\documentclass[10pt]{article} 
\usepackage{epsfig,amssymb}
\usepackage{amsmath}
\usepackage{natbib}
\usepackage{graphics}
\usepackage{graphicx}
\usepackage{dcolumn}
\usepackage{bm}
\usepackage{epsfig}

\begin{document}
\begin{center}

\Large{\bf Science, Technology and Mission Design \\
for the Laser Astrometric Test Of Relativity}

\vspace{0.3in}

\normalsize
\bigskip 

Slava G. Turyshev,$^a$
Michael Shao,$^a$
and 
Kenneth L. Nordtvedt, Jr.$^b$

\normalsize
\vskip 15pt

{\it $^a$Jet Propulsion Laboratory, 
California Institute of Technology, \\
4800 Oak Grove Drive, Pasadena, CA 91109, USA \\
$^b$Northwest Analysis, 118 Sourdough Ridge Road, Bozeman, MT 59715, USA}

\vspace{0.25in}

\end{center}


\begin{abstract}
The Laser Astrometric Test Of Relativity (LATOR) is a Michelson-Morley-type experiment designed to test the  metric nature of gravitation -- a fundamental postulate of the Einstein's general theory of relativity. The key element of LATOR is a geometric redundancy provided by the long-baseline optical interferometry and interplanetary laser ranging. By using a combination of independent time-series of gravitational deflection of light in the immediate proximity to the Sun, along with measurements of the Shapiro time delay on interplanetary scales (to a precision respectively better than 0.1 picoradians and 1 cm), LATOR will significantly improve our knowledge of relativistic gravity and cosmology.  
The primary mission objective is i) to measure the key post-Newtonian Eddington parameter $\gamma$ with accuracy of a part in 10$^9$.  $\frac{1}{2}(1-\gamma)$ is a direct measure for presence of a new interaction in gravitational theory, and, in its search, LATOR goes a factor 30,000 beyond the present best result, Cassini's 2003 test.  Other mission objectives include: ii) first measurement of gravity's non-linear effects on light to $\sim$0.01\% accuracy; including both the traditional Eddington $\beta$  parameter and also the spatial metric's 2nd order potential contribution (never measured before);  iii) direct measurement of the solar quadrupole moment $J_2$ (currently unavailable) to accuracy of a part in 200 of its expected size of $\simeq10^{-7}$; iv) direct measurement of the ``frame-dragging'' effect on light due to the Sun's rotational gravitomagnetic field, to 0.1\% accuracy. 

LATOR's primary measurement pushes to unprecedented accuracy the search for cosmologically relevant scalar-tensor theories of gravity by looking for a remnant scalar field in today's solar system. We discuss the science objectives of the mission, its technology, mission and optical designs, as well as expected performance of this  experiment. LATOR will lead to very robust advances in the tests of fundamental physics: this mission could discover a violation or extension of general relativity and/or reveal the presence of an additional long range interaction in the physical law.  There are no analogs to LATOR; it is unique and is a natural culmination of solar system gravity experiments. 

\end{abstract}


{\small 
\tableofcontents
}
 
\section{Introduction}
\label{sec:intro}

After almost ninety years since general theory of relativity was born, the Einstein's gravitational theory has survived every test. Such longevity, of course, does not mean that this theory is absolutely correct, but it serves to motivate more accurate tests to determine the level of accuracy at which it is violated. General theory of relativity began with its empirical success in 1915 by explaining the anomalous perihelion precession of Mercury's orbit.  Shortly thereafter, Eddington's 1919 observations of star lines-of-sight during a solar eclipse confirmed the doubling of the deflection angles predicted by general relativity as compared to Newtonian-like and Equivalence Principle arguments.  This test of gravitational deflection of light made  general relativity an instant success. 

From these beginnings, general theory of relativity has been verified at ever higher accuracy. Thus, microwave ranging to the Viking Lander on Mars yielded a $\sim$0.2\% accuracy in the tests of general relativity \citep{viking_shapiro1,viking_shapiro2,viking_reasen}. Spacecraft and planetary radar observations reached an accuracy of $\sim$0.15\% \citep{Anderson_etal_02,Pitjeva05}.  The astrometric observations of quasars on the solar background performed with Very-Long Baseline Interferometry (VLBI) improved the accuracy of the tests of gravity to $\sim$0.045\% \citep{RoberstonCarter91,Lebach95,eubanks97,Shapiro_SS_etal_2004}. Lunar laser ranging,  a continuing legacy of the Apollo program, provided $\sim$0.011\% verification of general relativity via precision measurements of the lunar orbit \citep{Ken_LLR68,Ken_LLR91,Ken_LLR98,Ken_LLR30years99,Ken_LLR_PPNprobe03,JimSkipJean96,Williams_etal_2001,LLR_beta_2004,pescara05}. Finally, the recent experiments with the Cassini spacecraft improved the accuracy of the tests to $\sim$0.0023\% \citep{iess_etal_1999,cassini_ber}. (See Section~\ref{sec:sci_mot} and Figure~\ref{fig:ppn}.)  As a result, today general relativity is the standard theory of gravity when astrometry and spacecraft navigation are concerned.  

Considering gravitation and fundamental physics, our solar system is the laboratory that still offers many opportunities to improve the tests of relativistic gravity. A carefully designed gravitational experiment that utilizes the strongest gravity potential available in the solar system, that provided by the sun itself, also has the advantage to conduct tests in a controlled and  well-understood environment. Indeed, compared to terrestrial conditions, the sun offers a factor of $M_\odot/M_\oplus\sim3.3\times 10^5$ increase in the strength of gravitational effects, the fact, that was recognized in a number of experiments proposed over the years (see discussion \citep{Will_book93,Will_review_2005}). Most of these proposals rely on sending an ensemble of ultra stable clocks to a close proximity to the sun, typically to distances of four solar radii \citep{anderson_solar_probe77,Spallicci_97,SpaceTime}. An approach, alternative to sending spacecraft on a highly eccentric trajectory into the challenging near-solar environment, would be to send a laser light instead.  This is due to the fact that optical technologies (i.e. long-baseline interferometry, laser ranging, etc.) have recently  demonstrated a very significant progress achieving the level of maturity needed for a major improvement of the accuracy of gravitational experiments in space.  The use of these technologies allows one to probe the strongest gravity in the solar system while still being separated by a safe distance from the sun; below we will develop this idea further.

This paper discusses the Laser Astrometric Test of Relativity (LATOR), the space-based experiment that is designed to significantly improve the tests of relativistic gravity in the solar system. The test will be performed in the solar gravity field using optical interferometry between two micro-spacecraft.  Precise measurements of the angular position of the spacecraft will be made using a fiber coupled optical interferometer on the ISS with a 100 m baseline. The primary objective of the LATOR mission will be to measure the gravitational deflection of light by the solar gravity to accuracy of 0.1 picoradians (prad), which corresponds to $\sim$10 picometers (pm) on a 100 m interferometric baseline. A combination of laser ranging among the spacecraft and direct interferometric measurements will allow LATOR to measure deflection of light in the solar gravity by a factor of $\sim$30,000 better than had recently been accomplished with the Cassini spacecraft. In particular, LATOR will not only measure the key PPN parameter $\gamma$ to unprecedented levels of accuracy of one part in 10$^9$; it will also reach ability to measure the next post-Newtonian order ($\propto G^2$) of light deflection resulting from gravity's intrinsic non-linearity. As a result, LATOR will measure values of other PPN parameters (see Eq.~(\ref{eq:metric})) such as parameter $\delta$ to 1 part in $10^4$ (never measured before), the solar quadrupole moment parameter $J_2$ to 1 part in 200, and the frame dragging effects on light due to the solar angular momentum to a precision of 1 parts in $10^3$.  

The paper is organized as follows: Section \ref{sec:sci_mot} discusses the theoretical framework and science motivation for the precision gravity tests in the solar system; it also presents the science objectives for the LATOR experiment. Section \ref{sec:lator_description} provides an overview for the LATOR experiment including basic elements of the current mission and optical designs. 
Section \ref{sec:interferometry} addresses design of the LATOR long-baseline optical interferometer inlcuding the laser metrology system. Section \ref{sec:lator_current} discusses the current design for the LATOR flight system and presents a preliminary design for LATOR optical receivers. 
In Section \ref{sec:model} we discuss modeling of the mission observables and address observational logic of LATOR measurements.  
In Section \ref{sec:error_bud} we present major constituents of the mission's error budget and discuss the expected mission performance. Section \ref{sec:conc} compares LATOR with other proposed gravity experiments and also discusses the next steps that will be taken in the development of LATOR.

\section{Scientific Motivation}
\label{sec:sci_mot}

Recent remarkable progress in observational cosmology has again submitted general relativity to a test by suggesting a non-Einsteinian model of universe's evolution \citep{Riess_supernovae98,perlmutter99,acfc_2003}. From the theoretical standpoint, the challenge is even stronger---if the gravitational field is to be quantized, general relativity will have to be modified. This is why the recent advances in the scalar-tensor extensions of gravity, that are consistent with the current inflationary model of the Big Bang, have motivated new search for a very small deviation of from the Einstein's theory, at the level of accuracy of three to five orders of magnitude below the level tested by experiment. 

In this section we will consider the recent theoretical end experimental motivations for the high-accuracy gravitational tests. We will also present the scientific objectives of the LATOR experiment.

\subsection{The PPN Formalism}

Generalizing on a phenomenological parameterization of the gravitational metric tensor field which Eddington originally developed for a special case, a method called the parameterized post-Newtonian (PPN) metric has been developed \citep{Ken_EqPrinciple68a,Ken_EqPrinciple68b,Ken_LLR68,Ken_LLR91,Ken_2PPN_87,WillNordtvedt72,Will_book93}.
This method  represents the gravity tensor's potentials for slowly moving bodies and weak interbody gravity, and it is valid for a broad class of metric theories including general relativity as a unique case.  The several parameters in the PPN metric expansion vary from theory to theory, and they are individually associated with various symmetries and invariance properties of underlying theory.  Gravity experiments can be analyzed in terms of the PPN metric, and an ensemble of experiments will determine the unique value for these parameters, and hence the metric field, itself.

In locally Lorentz-invariant theories the expansion of the metric field for a single, slowly-rotating gravitational source in PPN coordinates is given by:
{}
\begin{eqnarray}
\label{eq:metric}
g_{00}&=&1-2\frac{GM}{c^2r}\Big(1-J_2\frac{R^2}{r^2}\frac{3\cos^2\theta-1}{2}\Big)+2\beta\Big(\frac{GM}{c^2r}\Big)^2+{\cal O}(c^{-5}),\nonumber\\
g_{0i}&=& 2(\gamma+1)\frac{G[\vec{J}\times \vec{r}]_i}{c^3r^3}+
{\cal O}(c^{-5}),\\\nonumber
g_{ij}&=&-\delta_{ij}\Big[1+2\gamma \frac{GM}{c^2r} \Big(1-J_2\frac{R^2}{r^2}\frac{3\cos^2\theta-1}{2}\Big)+\frac{3}{2}\delta \Big(\frac{GM}{c^2r}\Big)^2\Big]+{\cal O}(c^{-5}),   
\end{eqnarray}
\noindent where $M$ and $\vec J$ being the mass and angular momentum of the Sun, $J_2$ being the quadrupole moment of the Sun and $R$ being its radius.  $r$ is the distance between the observer and the center of the Sun.  $\beta, \gamma, \delta$ are the PPN parameters and in general relativity they are all equal to $1$. The $M/r$ term in the $g_{00}$ equation is the Newtonian limit; the terms multiplied by the post-Newtonian parameters $\beta, \gamma$,  are post-Newtonian terms. The term multiplied by the post-post-Newtonian parameter  $\delta$ also enters the calculation of the relativistic light deflection \citep{Ken_cqg96}.

This PPN expansion serves as a useful framework to test relativistic gravitation in the context of the LATOR mission. In the special case, when only two PPN parameters ($\gamma$, $\beta$) are considered, these parameters have clear physical meaning. Parameter $\gamma$  represents the measure of the curvature of the space-time created by a unit rest mass; parameter  $\beta$ is a measure of the non-linearity of the law of superposition of the gravitational fields in the theory of gravity. General relativity, which corresponds to  $\gamma = \beta$  = 1, is thus embedded in a two-dimensional space of theories. The Brans-Dicke is the best known theory among the alternative theories of gravity.  It contains, besides the metric tensor, a scalar field and an arbitrary coupling constant $\omega$, which yields the two PPN parameter values $\gamma = (1+ \omega)/(2+ \omega)$, and $\beta$  = 1.  More general scalar tensor theories yield values of $\beta$ different from one.

\subsubsection{Current Limits on the PPN parameters $\gamma$ and $\beta$}
\label{sec:limits}

The PPN formalism has proved to be a versatile method to plan gravitational experiments in the solar system and to analyze the data obtained 
\citep{Ken_LLR68,Ken_LLR91,Ken_2PPN_87,Ken_EqPrinciple68a,Ken_EqPrinciple68b,WillNordtvedt72,Ken_cqg96,ken_icarus95,Will_review_1990,Will_book93,Will_review_2005,anderson_mars96,Bender_LLR97,acfc_2003}.
Different experiments test different combinations of the PPN parameters (for more details, see \citep{Will_book93,Will_review_2005}). The most precise value for the PPN parameter  $\gamma$ is at present given by the Cassini mission as: $\gamma -1 = (2.1\pm2.3)\times10^{-5}$ \citep{cassini_ber}. (Note that the Cassini result constraints the Brans-Dicke scalar coupling constant at the level of $|\omega|\ge4.35\times10^4$.) The secular trend of Mercury's perihelion, when described in the PPN formalism, depends on another linear combination of the PPN parameters $\gamma$  and  $\beta$ and the quadrupole coefficient $J_{2}$ of the solar gravity field:  $\lambda_\odot = (2 + 2\gamma -\beta)/3 + 0.296\times J_{2}\times 10^4$. The combination of parameters $\lambda_\odot$, was obtained with Mercury ranging data as $\lambda_\odot = 0.9996\pm 0.0006$ \citep{Pitjeva93,Pitjeva05}. Analysis of planetary ranging data recently yielded an independent determination of parameter $\gamma$: $\gamma -1 = 0.0015 \pm 0.0021$; it also gave $\beta$ with accuracy at the level of $\beta -1 = -0.0010 \pm 0.0012$ \citep{Williams_etal_2001,Anderson_etal_02,AndersonWilliams01}. The astrometric observations of quasars on the solar background performed with VLBI further reduced the uncertainty in the knowledge of the PPN parameter $\gamma$ resulting in the limit of $\gamma= 0.99983\pm0.00045$ \citep{eubanks97,Shapiro_SS_etal_2004}.

\begin{figure}[t!]
\begin{center}
\vskip 5pt
\psfig{figure=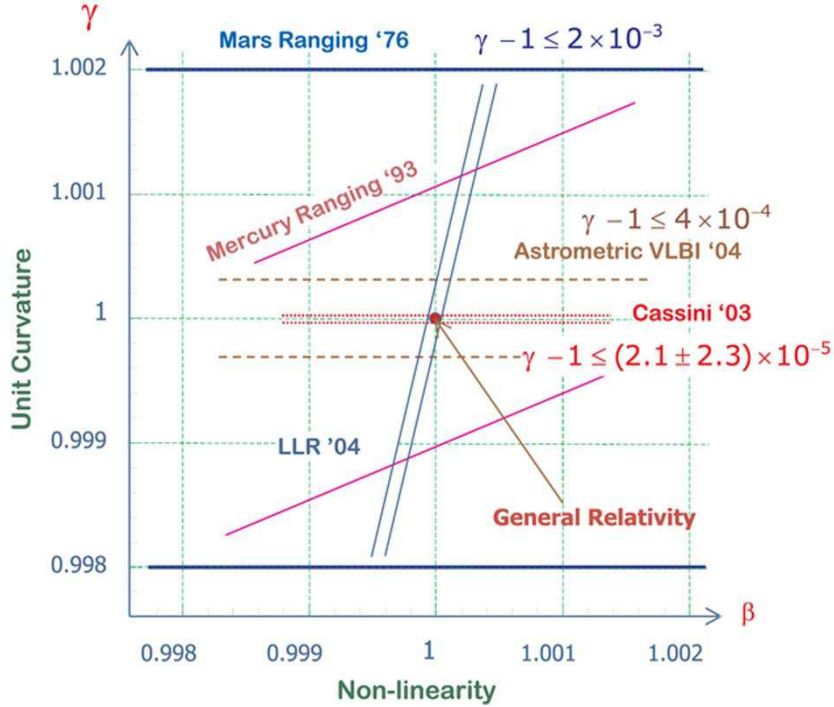,width=110mm}
\vskip -2pt 
\caption{\label{fig:ppn}
The progress in improving the knowledge of the PPN parameters $\gamma$ and $\beta$ for the last 30 years. So far, general theory of relativity survived every test; however, there are new compelling reasons to continue with the gravitational experiments  in the solar system at a significantly  improved accuracy. }
\end{center}
\vskip -4pt 
\end{figure} 


The PPN formalism has provided a useful framework for testing the violation of the Strong Equivalence Principle (SEP) for gravitationally bound bodies.  In that formalism, the ratio of passive gravitational mass $M_G$ to inertial mass $M_I$ of the same body is given by $M_G/M_I = 1 -\eta U_G/(M_0c^2)$, where $M_0$ is the rest mass of this body and $U_G$ is the gravitational self-energy. The SEP violation is quantified by the parameter $\eta$, which is expressed in terms of the basic set of PPN parameters by the relation $\eta = 4\beta-\gamma-3$. 
Additionally, with LLR finding that Earth and Moon fall toward the Sun at rates equal to 1.5 parts in 10$^{13}$, even in a conservative scenario, where a composition dependence of acceleration rates masks a gravitational self-energy dependence, $\eta$ is constrained to be less than 0.0008 \citep{AndersonWilliams01}; without such accidental cancelation the $\eta$ constraint improves to 0.0003. Using the recent Cassini result \citep{cassini_ber} on the PPN  $\gamma$, the parameter $\beta$ was measured as $\beta-1=(0.9\pm1.1)\times 10^{-4}$ from LLR \citep{Williams_etal_2004,LLR_beta_2004,pescara05}. (See Figure~\ref{fig:ppn}.) The next order PPN parameter $\delta$ has not yet been measured though its value can be inferred from other measurements.

Over the recent decade, the technology has advanced to the point that one can consider carrying out direct tests in a weak field to second order in the field strength parameter ($\propto G^2$). Although any measured anomalies in first or second order metric gravity potentials will not determine strong field gravity, they would signal that modifications in the strong field domain will exist.  The converse is perhaps more interesting:  if to high precision no anomalies are found in the lowest order metric potentials, and this is reinforced by finding no anomalies at the next order, then it follows that any anomalies in the strong gravity environment are correspondingly quenched under all but exceptional circumstances.\footnote{For example, a mechanism of a ``spontaneous-scalarization'' that, under certain circumstances, may exist in tensor-scalar theories \citep{Damour_EFarese93}.} 

We shall now discuss the recent motivations for the precision gravity experiments.

\subsection{Motivations for Precision Gravity Experiments}
\label{sec:mot} 

The continued inability to merge gravity with quantum mechanics, and recent cosmological observations indicate that the pure tensor gravity of general relativity needs modification.  The tensor-scalar theories of gravity, where the usual general relativity tensor field coexists with one or several long-range scalar fields, are believed to be the most promising extension of the theoretical foundation of modern gravitational theory. The superstring, many-dimensional Kaluza-Klein and inflationary cosmology theories have revived interest in the so-called ``dilaton fields,'' i.e. neutral scalar fields whose background values determine the strength of the coupling constants in the effective four-dimensional theory.  The importance of such theories is that they provide a possible route to the quantization of gravity and the unification of physical laws. 

Although the scalar fields naturally appear in the theory, their inclusion predicts different relativistic corrections to Newtonian motions in gravitating systems. These deviations from general relativity lead to a violation of the Equivalence Principle (either weak or strong or both), modification of large-scale gravitational phenomena, and generally lead to space and time variation of physical ``constants.'' As a result, this progress has provided new strong motivation for high precision relativistic gravity tests.

\subsubsection{Tensor-Scalar Extensions of General Relativity}
\label{sec:mot_theories}

\begin{figure}[h!]
\begin{center}
\vskip 5pt
\psfig{figure=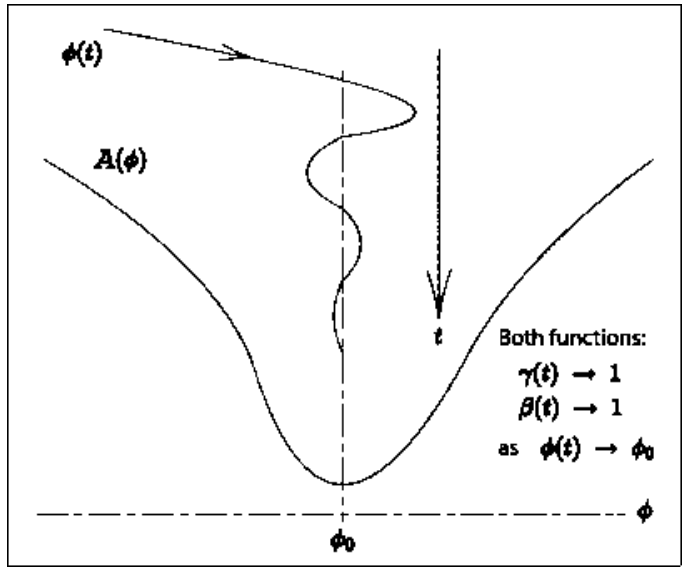,width=90mm}
\caption{\label{fig:attract}
Typical cosmological dynamics of a background scalar field is shown in the case when that field's coupling function to matter, $V(\phi)$, has an attracting point $\phi_0$. The strength of the scalar interaction's coupling to matter is proportional to the derivative (slope) of the coupling function, so it weakens as the attracting point is approached, and both the Eddington parameters $\gamma$ and $\beta$ (and all higher structure parameters as well)  approach their pure tensor gravity values in this limit \citep{Damour_EFarese96a,Damour_EFarese96b,Damour_Nordtvedt_93a,Damour_Nordtvedt_93b,DPV02a,DPV02b}.  But a small residual scalar gravity should remain today because this dynamical process is not complete, and that is what LATOR experiment seeks to find.
}
\end{center}
\end{figure}

Recent theoretical findings suggest that the present agreement between general relativity and experiment might be naturally compatible with the existence of a scalar contribution to gravity. In particular, Damour and Nordtvedt \citep{Damour_Nordtvedt_93a,Damour_Nordtvedt_93b} (see also \citep{DamourPolyakov94a,DamourPolyakov94b} for non-metric versions of this mechanism together with \citep{DPV02a,DPV02b} for the recent summary of a dilaton-runaway scenario) have found that a scalar-tensor theory of gravity may contain a ``built-in'' cosmological attractor mechanism toward general relativity.  These scenarios assume that the scalar coupling parameter $\frac{1}{2}(1-\gamma)$ was of order one in the early universe (say, before inflation), and show that it then evolves to be close to, but not exactly equal to, zero at the present time (Figure \ref{fig:attract} illustrates this mechanism in more details). 

The Eddington parameter $\gamma$, whose value in general relativity is unity, is perhaps the most fundamental PPN parameter, in that $\frac{1}{2}(1-\gamma)$ is a measure, for example, of the fractional strength of the scalar gravity interaction in scalar-tensor theories of gravity \citep{Damour_EFarese96a,Damour_EFarese96b}.  Within perturbation theory for such theories, all other PPN parameters to all relativistic orders collapse to their general relativistic values in proportion to $\frac{1}{2}(1-\gamma)$. This is why measurement of the first order light deflection effect at the level of accuracy comparable with the second-order contribution would provide the crucial information separating alternative scalar-tensor theories of gravity from general relativity \citep{Ken_2PPN_87} and also to probe possible ways for gravity quantization and to test modern theories of cosmological evolution \citep{Damour_Nordtvedt_93a,Damour_Nordtvedt_93b,DamourPolyakov94a,DamourPolyakov94b,DPV02a,DPV02b}. 
Under some assumptions (see e.g. \citep{Damour_Nordtvedt_93a,Damour_Nordtvedt_93b}) one can even estimate what is the likely order of magnitude of the left-over coupling strength at present time which, depending on the total mass density of the universe, can be given as $1-\gamma \sim 7.3 \times 10^{-7}(H_0/\Omega_0^3)^{1/2}$, where $\Omega_0$ is the ratio of the current density to the closure density and $H_0$ is the Hubble constant in units of 100 km/sec/Mpc. Compared to the cosmological constant, these scalar field models are consistent with the supernovae observations for a lower matter density, $\Omega_0\sim 0.2$, and a higher age, $(H_0 t_0) \approx 1$. If this is indeed the case, the level $(1-\gamma) \sim 10^{-6}-10^{-7}$ would be the lower bound for the present value of PPN parameter $\gamma$ \citep{Damour_Nordtvedt_93a,Damour_Nordtvedt_93b}. 

More recently, \cite{DPV02a,DPV02b} have estimated $\frac{1}{2}(1-\gamma)$, within the framework compatible with string theory and modern cosmology, which basically confirms the previous result \citep{Damour_Nordtvedt_93a,Damour_Nordtvedt_93b}. This recent analysis discusses a scenario when a composition-independent coupling of dilaton to hadronic matter produces detectable deviations from general relativity in high-accuracy light deflection experiments in the solar system. This work assumes only some general property of the coupling functions (for large values of the field, i.e. for an ``attractor at infinity'') and then only assume that $(1-\gamma)$ is of order of one at the beginning of the controllably classical part of inflation.
It was shown  in \citep{DPV02b} that one can relate the present value of $\frac{1}{2}(1-\gamma)$ to the cosmological density fluctuations. For the simplest inflationary potentials (favored by WMAP mission, i.e. $m^2 \chi^2$ \citep{[4c]}, see also WMAP technical papers at the mission's website: {\tt http://map.gsfc.nasa.gov/m$_-$mm/pub$_-$papers/}), \cite{DPV02a,DPV02b} found that the present value of $(1-\gamma)$ could be just below $10^{-7}$. In particular, within this framework $\frac{1}{2}(1-\gamma)\simeq\alpha^2_{\rm had}$, where $\alpha_{\rm had}$ is the dilaton coupling to hadronic matter; its value depends on the model taken for the inflation potential $V(\chi)\propto\chi^n$, with $\chi$ being the inflation field; the level of the expected deviations from general relativity is $\sim0.5\times10^{-7}$  for $n = 2$ \citep{DPV02b}. Note that these predictions are based on the work in scalar-tensor extensions of gravity which are consistent with, and indeed often part of, present cosmological models.  

Another example of recent theoretical progress is  the Dvali-Gabadadze-Porrati (DGP) brane-world model, which explores a possibility that  we live on a brane embedded in a large extra dimension, and where the strength of gravity in the bulk is substantially less than that on the brane \citep{dvali}. 
Although such theories can lead to perfectly conventional gravity on large scales, it is also possible to choose the dynamics in such a way that new effects show up exclusively in the far infrared providing a mechanism to explain the acceleration of the universe \citep{perlmutter99,Riess_supernovae98}.  It is interesting to note that DGP gravity and other modifications of GR hold out the possibility of having interesting and testable predictions that distinguish them from models of dynamical Dark Energy. One outcome of this work is that the physics of the accelerating universe may be deeply tied to the properties of gravity on relatively short scales, from millimeters to astronomical units \citep{Dvali_GP_2000,Dvali_GZ_2003,BP-2}.

To date general relativity and some other alternative gravitational theories are in good agreement with the experimental data collected from the relativistic celestial mechanical extremes provided by the relativistic motions in the binary millisecond pulsars.  At the same time, many modern theoretical models, which include general relativity as a standard gravity theory, are faced with the problem of the unavoidable appearance of space-time singularities. It is generally suspected that the classical description, provided by general relativity, breaks down in a domain where the curvature is large, and, hence, a proper understanding of such regions requires new physics. This is a reason why recently a considerable interest has been shown in the physical processes occurring in the strong gravitational field regime with relativistic pulsars providing a promising possibility to test gravity in this qualitatively different dynamical environment.  The general theoretical framework for pulsar tests of strong-field gravity was introduced in \citep{DamourTaylor92}; the observational data for the initial tests were obtained with PSR1534 \citep{Taylor_etal92}. An analysis of strong-field gravitational tests and their theoretical justification was presented in \citep{Damour_EFarese96a,Damour_EFarese96b,Damour_EFarese98}. The recent analysis of the pulsar data resulted in $\frac{1}{2}(1-\gamma)\simeq \alpha_{\rm had}^2\sim 4\times 10^{-4}$ at a 3$\sigma$ confidence level \citep{lange_etal2001}, with $\alpha_{\rm had}$ being the dilaton coupling to hadronic matter. While being a natural alternative to the weak-gravity tests, the pulsar tests of gravitation currently can not offer the accuracy at the level  that is presently available within the solar system. In fact, a carefully  designed dedicated experiment that utilizes the strongest gravitational potential available in the solar system, provided by the sun itself, offers a unique opportunity to test gravitation in a controlled and the well-understood environment. Therefore, despite the relative weakness of its gravitational field, the sun still has an advantage and offers an attractive opportunity to perform accurate tests of gravity.

The analyses discussed above not only motivate new searches for very small deviations of relativistic gravity in the solar system, they also predict that such deviations are currently present in the range from $10^{-5}$ to $5\times10^{-8}$ for $\frac{1}{2}(1-\gamma)$, i.e. for observable post-Newtonian deviations from general relativity predictions and, thus, should be easily detectable with LATOR. This would require measurement of the effects of the next post-Newtonian order ($\propto G^2$) of light deflection resulting from gravity's intrinsic non-linearity. An ability to measure the first order light deflection term at the accuracy comparable with the effects of the second order is of the utmost importance for gravitational theory and a major challenge for the 21st century fundamental physics.  

\subsubsection{Observational Motivations for Higher Accuracy Tests of Gravity}
\label{sec:mot_experiment}

Recent astrophysical measurements of the angular structure of the cosmic microwave background \citep{deBernardis_CMB2000}, the masses of large-scale structures \citep{Peacock_LargeScale01}, and the luminosity distances of type Ia supernovae \citep{perlmutter99,Riess_supernovae98} have placed stringent constraints on the cosmological constant $\Lambda$ and also have led to a revolutionary conclusion: the expansion of the universe is accelerating. The implication of these observations for cosmological models is that a classically evolving scalar field currently dominates the energy density of the universe. Such models have been shown to share the advantages of  $\Lambda$:  compatibility with the spatial flatness predicted inflation; a universe older than the standard Einstein-de Sitter model; and, combined with cold dark matter, predictions for large-scale structure formation in good agreement with data from galaxy surveys. Combined with the fact that scalar field models imprint distinctive signature on the cosmic microwave background (CMB) anisotropy, they remain currently viable and should be testable in the near future. This completely unexpected discovery demonstrates the importance of testing the important ideas about the nature of gravity. We are presently in the ``discovery'' phase of this new physics, and while there are many theoretical conjectures as to the origin of a non-zero $\Lambda$, it is essential that we exploit every available opportunity to elucidate the physics that is at the root of the observed phenomena.

There is now multiple evidence indicating that 70\% of the critical density of the universe is in the form of a ``negative-pressure'' dark energy component; there is no understanding as to its origin and nature. The fact that the expansion of the universe is currently undergoing a period of acceleration now seems rather well tested: it is directly measured from the light-curves of several hundred type Ia supernovae \citep{perlmutter99,Riess_supernovae98,[3c]}, and independently inferred from observations of CMB by the WMAP satellite \citep{[4c]} and other CMB experiments \citep{[6c],[5c]}. Cosmic speed-up can be accommodated within general relativity by invoking a mysterious cosmic fluid with large negative pressure, dubbed dark energy. The simplest possibility for dark energy is a cosmological constant; unfortunately, the smallest estimates for its value are 55 orders of magnitude too large (for reviews see \citep{Carroll_01,PeeblesRatra03}). Most of the theoretical studies operate in the shadow of the cosmological constant problem, the most embarrassing hierarchy problem in physics. This fact has motivated a host of other possibilities, most of which assume $\Lambda=0$, with the dynamical dark energy being associated with a new scalar field (see \citep{Carroll_etal_2004,Carroll_etal_2005} and references therein). However, none of these suggestions is compelling and most have serious drawbacks. Given the challenge of this problem, a number of authors considered the possibility that cosmic acceleration is not due to some kind of stuff, but rather arises from new gravitational physics (see discussion in 
\citep{Carroll_01,PeeblesRatra03,Carroll_HT_03,Carroll_etal_2004}).
In particular, certain extensions to general relativity in a low energy regime \citep{Carroll_etal_2004,Capozziello_Troisi_2005,Carroll_2005} were shown to predict an experimentally consistent universe evolution  without the need for dark energy (see discussion on the interplay between theory, experiment and observation in \citep{bean_carroll_troden_2005,bertolami_paramos_turyshev_2006}). These dynamical models are expected to explain the observed acceleration of the universe without dark energy, but may produce measurable gravitational effects on the scales of the solar system.  In particular, corresponding contribution to the parameter $\gamma$  in experiments conducted in the solar system are expected at the level of $1-\gamma \sim 10^{-7}-5\times10^{-9}$, thus further motivating the relativistic gravity research. Therefore, the PPN parameter $\gamma$ may be the only key parameter that holds the answer to most of the questions discussed above. Also an anomalous parameter $\delta$ will most likely be accompanied by a ``$\gamma$ mass'' of the Sun which differs from the gravitational mass of the Sun and therefore will show up as anomalous $\gamma$ (see discussion in \citep{Ken_LLR_PPNprobe03}).

Even in the solar system, general relativity still faces challenges. There is the long-standing problem of the size of the solar quadrupole moment and its possible effect on the relativistic perihelion precession of Mercury (see review in \citep{Will_book93}). The interest is in the study of the behavior of the solar quadrupole moment versus the radius and the heliographic latitudes. This solar parameter has been very often neglected in the past, because it was rather difficult to determine an accurate value. The improvement of the accuracy of our knowledge of $J_2$ is certainly due to the fact that, today, we are able to take into account the differential rotation with depth. In fact, the quadrupole moment plays an important role in the accurate computation of several astrophysical quantities, such as the ephemeris of the planets or general relativistic prediction for the precession of the perihelion of Mercury and minor planets such as Icarus.  Finally, it is necessary to accurately know the value of the quadrupole moment to determinate the shape of the Sun, that is to say its oblateness. Solar oblateness measurements by Dicke and others in the past gave conflicting results for $J_2$ (reviewed on p.~145 of \citep{CiufoliniWheeler_1995}). A measurement of solar oblateness with the balloon-borne Solar Disk Sextant gave a quadrupole moment on the order of $2\times 10^{-7}$ \citep{lydon96}. Helioseismic determinations using solar oscillation data have since implied a small value for $J_2$, on the order of $\sim 10^{-7}$, that is consistent with simple uniform rotation \citep{Will_book93,Brown,GoughToomre_InteriorSun91}. However, there exist uncertainties in the helioseismic determination for depths below roughly 0.4 $R_\odot$ which might permit a rapidly rotating core. (LATOR can measure $J_2$ with accuracy sufficient to put this issue to rest.) 

In summary, there are a number of theoretical and experimental reasons to question the validity of general relativity. Despite the success of modern gauge field theories in describing the electromagnetic, weak, and strong interactions, it is still not understood how gravity should be described at the quantum level. In theories that attempt to include gravity, new long-range forces can arise in addition to the Newtonian inverse-square law. Even at the purely classical level, and assuming the validity of the Equivalence Principle, Einstein's theory does not provide the most general way to generate the space-time metric. Regardless of whether the cosmological constant should be included, there are also important reasons to consider additional fields, especially scalar fields.   The LATOR mission is designed to  directly address theses challenges with an unprecedented accuracy; we shall now discuss LATOR in more details.

\section{Overview of LATOR}
\label{sec:lator_description}

\begin{figure*}[t!]
 \begin{center}
\noindent    
\psfig{figure=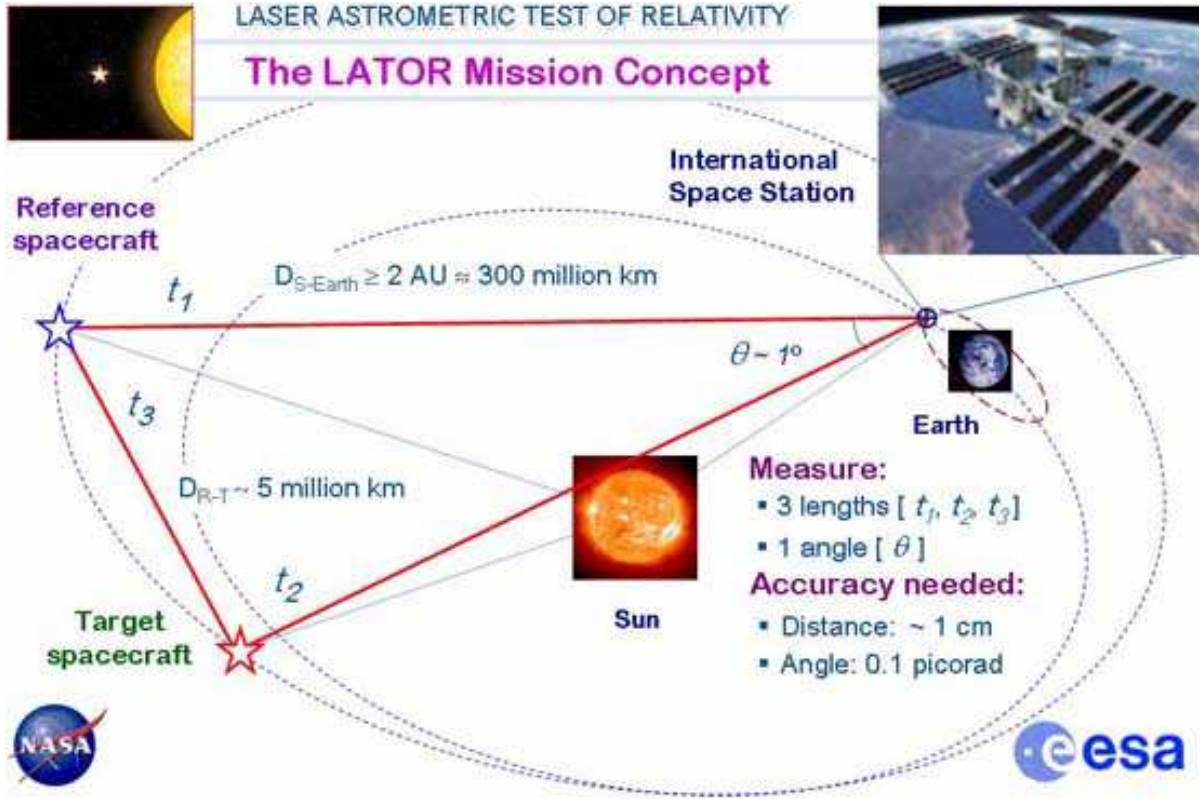,width=158mm}
\end{center}
\vskip -10pt 
  \caption{The overall geometry of the LATOR experiment.  
 \label{fig:lator}}
\end{figure*}

The LATOR experiment uses the standard technique of time-of-fight laser ranging (extended to interplanetary scales) between two micro-spacecraft whose lines of sight pass close by the Sun and also a long-baseline stellar optical interferometer (placed above the Earth's atmosphere) to accurately measure deflection of light by the solar gravitational field in the extreme proximity to the Sun \citep{lator_cqg_2004}.  Figure \ref{fig:lator} shows the general concept for the LATOR missions including the mission-related geometry, experiment details  and required accuracies.  

In this section we will consider the LATOR mission architecture in more detail.

\subsection{Evolving Light Triangle}
\label{sec:triangle}

The LATOR mission architecture uses an evolving light triangle formed by laser ranging between two spacecraft (placed in $\sim$1 AU heliocentric orbits) and a laser transceiver terminal on the International Space Station (ISS) (realized via European collaboration \citep{ESLAB2005_LATOR}).  The objective is to measure the gravitational deflection of laser light as it passes in extreme proximity to the Sun (see Figure \ref{fig:lator}).  To that extent, the long-baseline ($\sim$100 m) fiber-coupled optical interferometer on the ISS will perform differential astrometric measurements of the laser light sources on the two spacecraft as their lines-of-sight pass behind the Sun.  

As seen from the Earth, the two spacecraft will be separated by $\sim1^\circ$, which will be accomplished by a small maneuver immediately after their launch \citep{lator_cqg_2004,stanford_texas}. This separation would permit differential astrometric observations to an accuracy of $\sim$0.1 picorad needed to significantly improve measurements of gravitational deflection of light in the solar gravity.

The schematic of the LATOR experiment is quite simple and is given in Figure \ref{fig:lator}. Two spacecraft are injected into a heliocentric solar orbit on the opposite side of the Sun from the Earth. The triangle in the figure has three independent quantities but three arms are monitored with laser metrology. Each spacecraft equipped with a laser ranging system that enables a measurement of the arms of the triangle formed by the two spacecraft and the ISS. The uniqueness of this mission comes with its geometrically redundant architecture that enables LATOR to measure the departure from Euclidean geometry ($\sim8.48\times 10^{-6}$ rad) caused by the solar gravity field, to a very high accuracy \citep{lator_cqg_2004}.   This departure is shown as a difference between the calculated Euclidean value for an angle in the triangle and its value directly measured by the interferometer.  This discrepancy, which results from the curvature of the space-time around the Sun and can be computed for every alternative theory of gravity, constitutes LATOR's signal of interest. The precise  measurement of this departure constitutes the primary mission objective.

A version of LATOR with a ground-based receiver was proposed in \citep{Yu94}. Due to atmospheric turbulence and seismic vibrations that are not common mode to the receiver optics, a very long baseline interferometer (30 km) was proposed. This interferometer could only measure the differential light deflection to an accuracy of 0.1 $\mu$as, with a spacecraft separation of less than 1 arc minute. The shortening of the interferometric baseline (as compared to the previously studied version \citep{Yu94,Shao_1995,Shao96}) is achieved solely by going into space to avoid the atmospheric turbulence and Earth's seismic vibrations. On the space station, all vibrations can be made common mode for both ends of the interferometer by coupling them by an external laser truss. This relaxes the constraint on the separation between the spacecraft, allowing it to be as large as few degrees, as seen from the ISS. Additionally, the orbital motion of the ISS provides variability in the interferometer's baseline projection as needed to resolve the fringe ambiguity of the stable laser light detection by an interferometer.

We shall now consider the basic elements of the LATOR optical design. 

\subsection{General Approach in Optical Design}
\label{sec:optical_design}

\begin{figure*}[t!]
 \begin{center}
\noindent    
\epsfig{figure=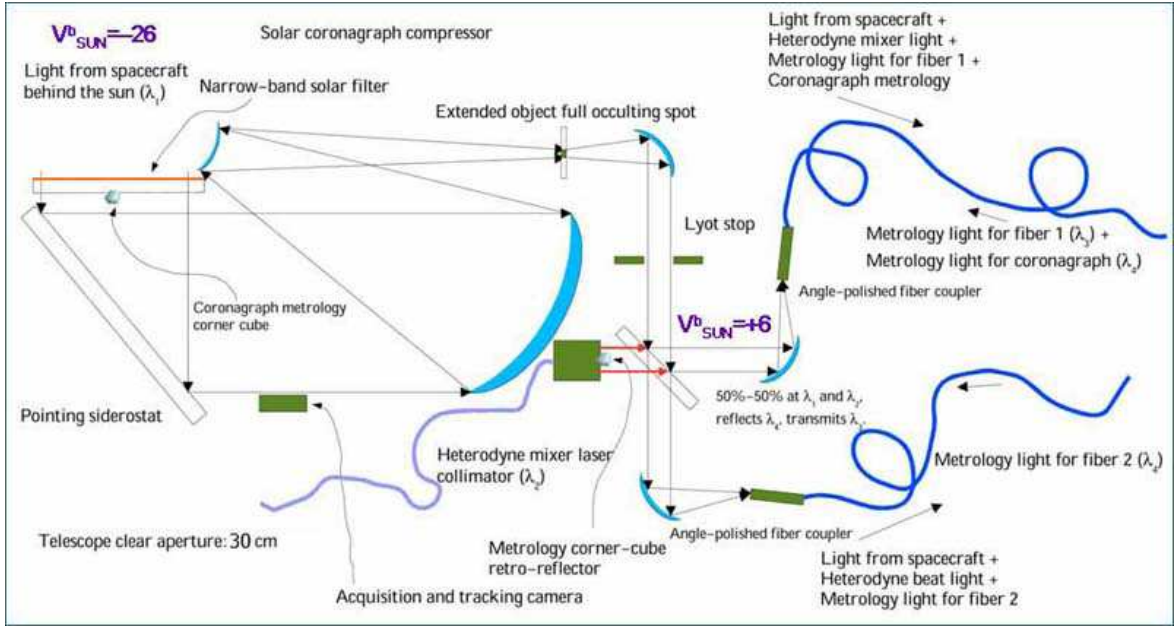,width=155mm}
\end{center}
\vskip -10pt 
  \caption{Basic elements of optical design for the LATOR interferometer: The laser light (together with the solar background) is going through a full aperture ($\sim 30$~cm) narrow band-pass filter with $5\times 10^{-5}~\mu$m bandwidth around wavelength of $\lambda=1064$~nm. The remaining light illuminates the baseline metrology corner cube and falls onto a steering flat mirror where it is reflected to an off-axis telescope with no central obscuration (needed for metrology). It then enters the solar coronograph compressor (with $1\times 10^{-5}$ suppression properties) by first going through a 1/2 plane focal plane occulter and then coming to a Lyot stop. At the Lyot stop, the background solar light is reduced by a factor of $10^{6}$. This combination of a narrow band-pass filter and coronograph enables the solar luminosity reduction from $V=-26$ to $V=4$ (as measured at the ISS), thus enabling the LATOR precision observations.
\label{fig:optical_design}}
\vskip -5pt 
\end{figure*}

A single aperture of the interferometer on the ISS consists of three 30 cm diameter telescopes (see Figure \ref{fig:optical_design} for a conceptual design). One of the telescopes with a very narrow bandwidth laser line filter in front and with an InGaAs camera at its focal plane, sensitive to the 1064 nm laser light, serves as the acquisition telescope to locate the spacecraft near the Sun.

The second telescope emits the directing beacon to the spacecraft. Both spacecraft are served out of one telescope by a pair of piezo controlled mirrors placed on the focal plane. The properly collimated laser light (1~W) is injected into the telescope focal plane and deflected in the right direction by the piezo-actuated mirrors. 

The third telescope is the laser light tracking interferometer input aperture, which can track both spacecraft at the same time. To eliminate beam walk on the critical elements of this telescope, two piezo-electric X-Y-Z stages are used to move two single-mode fiber tips on a spherical surface while maintaining focus and beam position on the fibers and other optics. Dithering at a few Hz is used to make the alignment to the fibers and the subsequent tracking of the two spacecraft completely automatic. The interferometric tracking telescopes are coupled together by a network of single-mode fibers whose relative length changes are measured internally by a heterodyne metrology system to an accuracy of less than 5~pm.

The spacecraft  are identical in construction and contain a relatively high powered (1~W), stable (2~MHz per hour $\sim$500 Hz per second), small cavity fiber-amplified laser at 1064~nm. The power of this laser is directed to the Earth through a 20~cm aperture telescope and its phase is tracked by the interferometer. With the available power and the beam divergence, there are enough photons to track the slowly drifting phase of the laser light. There is another 0.2~W laser operating at 780 nm, the power of which is transmitted through another telescope with small aperture of 5 cm, which points toward the other spacecraft. In addition to the two transmitting telescopes, each spacecraft has two receiving telescopes.  The receiving telescope, which points toward the area near the Sun, has laser line filters and a simple knife-edge coronagraph to suppress the Sun's light to 1 part in $10^5$ of the light level of the light received from the space station. The receiving telescope that points to the other spacecraft is free of the Sun light filter and the coronagraph.

In addition to the four telescopes they carry, the spacecraft also carry a tiny (2.5~cm) telescope with a CCD camera. This telescope is used to initially point the spacecraft directly toward the Sun so that their signal may be seen at the space station. One more of these small telescopes may also be installed at right angles to the first one, to determine the spacecraft attitude, using known, bright stars. The receiving telescope looking toward the other spacecraft may be used for this purpose part of the time, reducing hardware complexity. Star trackers with this construction were demonstrated many years ago and they are readily available. A small RF transponder with an omni-directional antenna is also included in the instrument package to track the spacecraft while they are on their way to assume the orbital position needed for the experiment. 

The LATOR experiment has a number of advantages over techniques that use radio waves to measure gravitational light deflection. Advances in optical communications technology, allow low bandwidth telecommunications with the LATOR spacecraft without having to deploy high gain radio antennae needed to communicate through the solar corona. The use of the monochromatic light enables the observation of the spacecraft almost at the limb of the Sun, as seen from the ISS. The use of narrowband filters, coronagraph optics and heterodyne detection will suppress background light to a level where the solar background is no longer the dominant noise source. In addition, the short wavelength allows much more efficient links with smaller apertures, thereby eliminating the need for a deployable antenna. Finally, the use of the ISS will allow conducting the test above the Earth's atmosphere---the major source of astrometric noise for any ground based interferometer. This fact justifies LATOR as a space mission.

\subsection{Science Objectives}
\label{sec:lator_science}

LATOR is a Michelson-Morley-type experiment designed to test the pure tensor metric nature of gravitation -- a fundamental postulate of Einstein's theory of general relativity \citep{lator_cqg_2004}.  With its focus on gravity's action on light propagation it complements other tests which rely on the gravitational dynamics of bodies.  The idea behind this experiment is to use a combination of independent time-series of highly accurate measurements of the gravitational deflection of light in the immediate proximity to the Sun along with measurements of the Shapiro time delay on the interplanetary scales (to a precision respectively better than 0.1 prad and 1 cm).  Such a combination of observables is unique and enables LATOR to significantly improve tests of relativistic gravity.  

\begin{table*}[t!]
\caption{Comparable sizes of various light deflection effects in the solar gravity field. The value of deflection angle is calculated on the limb of the Sun ($b=R_\odot$); the corresponding delay is given for a $b=100$~m interferometric baseline proposed for LATOR. \label{tab:eff}}
\vskip 5pt
\begin{center}
\begin{tabular}{|c|c|c|c|} \hline  
    & & Deflection & Delay for\\
Effect  & Analytical Form & angle, $\mu$as & $b=100$~m, pm \\ 
\hline \hline
&&&\\[-8pt]
   First  Order  &
   $2(1+\gamma )\frac{M}{b}$ & 
   $1.75$ arcsec   & 
   $0.849 $~mm     \\  
&&&\\[-8pt] \hline 
&&&\\[-8pt]
   Second Order   &
   $[\big(2(1+  \gamma)-\beta+\frac{3}{4} \delta\big)\pi- 
2(1+\gamma)^2]\frac{M^2}{b^2} $
   &  3.5   
   & 1697      \\ 
&&&\\[-8pt] \hline 
&&&\\[-8pt]
   Frame-Dragging   &
   $\pm 2(1+\gamma)\frac{J}{b^2}$ & 
   $\  \pm0.7$   &     $\pm339$ \\ 
&&&\\[-8pt] \hline 
&&&\\[-8pt]
   Solar Quadrupole  &
   $2(1+\gamma )J_2\frac{M}{b^3}$ & 
     0.2   &   97  \\[4pt] \hline  
\end{tabular} 
\end{center}\vskip-10pt
\end{table*}

The LATOR's primary mission objective is to measure the key post-Newtonian Eddington parameter $\gamma$ with an accuracy of a part in 10$^9$.  
When the light deflection in solar gravity is concerned, the magnitude of the first order effect as predicted by general relativity for the light ray just grazing the limb of the Sun is $\sim1.75$ arcsecond (asec) (for more details see Table \ref{tab:eff}). (Note that 1 arcsec $\simeq5~\mu$rad; when convenient, below we will use the units of radians and arcseconds interchangeably.) The effect varies inversely with the impact parameter. The second order term is almost six orders of magnitude smaller resulting in  $\sim 3.5$ microarcseconds ($\mu$as) light deflection effect, and which falls off inversely as the square of the light ray's impact parameter \citep{epstein_shapiro_80,FishbachFreeman80,RichterMatzner82a,RichterMatzner82b,Ken_2PPN_87,lator_cqg_2004}. The relativistic frame-dragging term\footnote{Gravitomagnetic frame dragging is the effect in which both the orientation and trajectory of objects in orbit around a body are altered by the gravity of the body's rotation.  It was studied by Lense and Thirring in 1918.} is $\pm 0.7 ~\mu$as, and contribution of the solar quadrupole moment, $J_2$, is sized as 0.2 $\mu$as (using theoretical value of the solar quadrupole moment $J_2\simeq10^{-7}$). The small magnitudes of the effects emphasize the fact that, among the four forces of nature, gravitation is the weakest interaction; it acts at very long distances and controls the large-scale structure of the universe, thus, making the precision tests of gravity a very challenging task.

\def\reff{\vskip 4pt \par \hangindent 12pt \noindent}
\begin{table*}[t!]
\caption{LATOR Mission Summary: Science Objectives 
\label{tab:summ_science}}
\vskip 5pt
\begin{center}
\begin{tabular}{m{14.25cm}} \hline \hline\\[-8pt] 

{\it Qualitative Objectives: }

\reff $\bullet$\hskip6pt
To test the metric nature of the Einstein's general theory of relativity in the most intense gravitational environment available in the solar system -- the extreme proximity to the Sun;

\reff $\bullet$\hskip6pt
To test alternative theories of gravity and cosmology, notably scalar-tensor theories, by searching for cosmological remnants of scalar field in the solar system;

\reff $\bullet$\hskip6pt
To verify the models of light propagation and motion of the gravitationally-bounded systems at the second post-Newtonian order (i.e. including effects $\propto G^2$).\\[4pt] 

{\it Quantitative Objectives: }

\reff $\bullet$\hskip6pt
To measure the key Eddington PPN parameter $\gamma$ with accuracy of 1 part in 10$^{9}$ -- a factor of 30,000 improvement in the tests of gravitational deflection of light;

\reff $\bullet$\hskip6pt
To provide direct and independent measurement of the Eddington PPN parameter $\beta$ via gravity effect on light to $\sim0.01$\% accuracy;

\reff $\bullet$\hskip6pt
To measure effect of the 2-nd order gravitational deflection of light with accuracy of 1 part in $10^{4}$, including first ever measurement of the post-PPN parameter $\delta$; 

\reff $\bullet$\hskip6pt
To directly measure the frame dragging effect on light (first such observation and also first direct measurement of solar spin) with accuracy of 1 part in $10^{3}$;

\reff $\bullet$\hskip6pt
To measure the solar quadrupole moment $J_2$ (using the theoretical value of $J_2 \simeq 10^{-7}$, currently unavailable) to 1 part in 200. 
\\[2pt]\hline 
\end{tabular}
\end{center}
\vskip-10pt
\end{table*}
 

If the Eddington's 1919 experiment was performed to confirm the general theory of relativity, LATOR is motivated to search for physics beyond the Einstein's theory of gravity with an unprecedented accuracy \citep{lator_cqg_2004}.  In fact, this mission is designed to address the questions of fundamental importance to modern physics. In particular, this solar system scale experiment would search for a cosmologically-evolved scalar field that is predicted by modern theories of quantum gravity and cosmology, and also by superstring and brane-world models \citep{dvali,BP-2,bertolami_paramos_turyshev_2006}. LATOR will also test the cosmologically motivated theories that attempt to explain the small acceleration rate of the Universe (so-called ``dark energy'') via modification of gravity at very large, horizon or super-horizon distances. 

By studying the effect of gravity on light and measuring the Eddington parameter $\gamma$, this mission will tests the presently viable alternative theories of gravity, namely the scalar-tensor theories. The value of the parameter $\gamma$ may hold the key to the solution of the most fundamental questions concerning the evolution of the universe.  In the low energy approximation suitable for the solar system, a number of modern theories of gravity and cosmology studied as methods for gravity quantization or proposed as an explanation to the recent cosmological puzzles, predict measurable contributions to the parameter $\gamma$ at the level of $\frac{1}{2}(1-\gamma)\sim 10^{-6}-10^{-8}$; detecting this deviation is LATOR's primary objective.  With the accuracy of one part in $10^{9}$, this mission could discover a violation or extension of general relativity, and/or reveal the presence of any additional long range interaction. 

We now outline the basic elements of the LATOR trajectory.

\subsection{Spacecraft Trajectory: a 3:2 Earth Resonant Orbit}
\label{sec:3:2orbit}

\begin{figure*}[ht!]
\hskip -10pt 
\begin{minipage}[b]{.46\linewidth}
\centering \psfig{figure=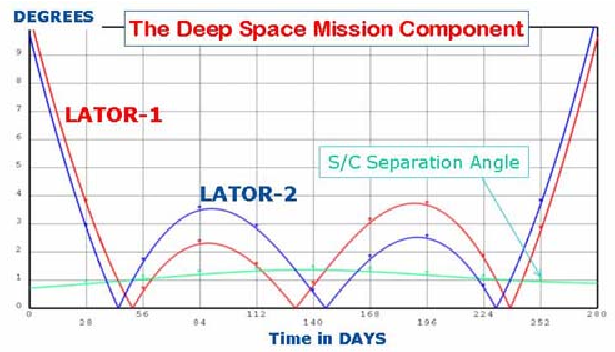,width=112mm}
\end{minipage}
\hfill  
\begin{minipage}[b]{.32\linewidth}
\centering 
\vbox{\psfig{figure=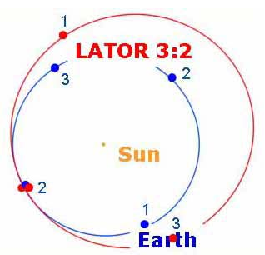,width=5.2cm}\\\vskip26pt}
\end{minipage}
\caption{Left: The Sun-Earth-Probe angle during the period of 3 occultations (two periodic curves) and the angular separation of the spacecraft as seen from the Earth (lower smooth line). Time shown is days from the moment when one of the spacecraft are at 10º distance from the Sun. Right: View from the North Ecliptic of the LATOR spacecraft in a 3:2 resonance. The epoch is taken near the first occultation.  
 \label{fig:lator_sep_angle2}}
\vskip -5pt 
\end{figure*}

To enable the primary objective, LATOR will place two spacecraft into a heliocentric orbit, to provide conditions for observing the spacecraft when they are behind the Sun as viewed from the ISS (see Figures~\ref{fig:iss_config},\ref{fig:lator_sep_angle2}).  With the help of the JPL Advanced Project Design Team (Team X), we recently conducted detailed mission design studies \citep{teamx}. In particular, we analyzed various trajectory options for the deep-space flight segment of LATOR, using both the Orbit Determination Program (ODP) and Satellite Orbit Analysis Program (SOAP)---the two standard JPL navigation software packages. 

One trajectory option would be to use a Venus flyby to place the spacecraft in a 1 yr orbit (perihelion at Venus orbit $\sim0.73$~AU and aphelion $\sim1.27$~AU). One complication of this approach is that the Venus orbit is inclined about 3.4$^\circ$ with respect to the ecliptic and the out-of-plane position of Venus at the time of the flyby determines the orbit inclination \cite{teamx}. The LATOR observations require that the spacecraft pass directly behind the Sun, i.e., with essentially no orbit inclination.  In order to minimize the orbit inclination, the Venus' flyby would need to occur near the time of Venus nodal crossing. An approach with a type IV trajectory and a single Venus flyby requires a powered Venus flyby with about 500 to 900 m/s.  However, a type I trajectory to Venus with two Venus gravity assists would get LATOR into a desirable 1 year orbit at Earth's opposition.  This option requires no velocity change, called $\Delta v$,  and provides repeated opportunities for the desired science observations. ($\Delta v$ is a desired spacecraft velocity change, which is typically enabled by either the on-board propulsion system or a planetary fly-by.) At the same time this orbit has a short launch window $\sim$17 days which motivated us to look for an alternative.

An orbit with a 3:2 resonance with the Earth\footnote{The 3:2 resonance occurs when the Earth does 3 revolutions around the Sun while the spacecraft does exactly 2 revolutions on a 1.5 year orbit. The exact period of the orbit may vary slightly, $<$1\%, from a 3:2 resonance depending on the time of launch.} was found to uniquely satisfy the LATOR orbital requirements \citep{teamx,lator_cqg_2004}. 
For this orbit, 13 months after the launch, the spacecraft are within $\sim10^\circ$ of the Sun with first occultation occurring 15 months after launch \citep{lator_cqg_2004}.  At this point, LATOR is orbiting at a slower speed than the Earth, but as LATOR approaches its perihelion, its motion in the sky begins to reverse and the spacecraft is again occulted by the Sun 18 months after launch.  As the spacecraft slows down and moves out toward aphelion, its motion in the sky reverses again, and it is occulted by the Sun for the third and final time 21 months after launch.

This entire process will again repeat itself in about 3 years after the initial occultation, however, there may be a small maneuver required to allow for more occultations.  Therefore, to allow for more  occultations in the future, there may be a need for an extra few tens of m/s of $\Delta v$. The energy required for launch, $C_3$, will vary between $\sim(10.6 - 11.4) ~{\rm km}^2/{\rm s}^2$ depending on the time of launch, but it is suitable for a Delta II launch vehicle.   The desirable $\sim1^\circ$ spacecraft separation  (as seen from the Earth) is achieved by performing a 30 m/s maneuver after the launch.  This results in the second spacecraft being within $\sim(0.6-1.4)^\circ$ separation during the entire period of 3 occultations by the Sun.

Figure \ref{fig:lator_sep_angle2} shows the  trajectory  and the occultations in more details.  The figure on the right is the spacecraft position in the solar system showing the Earth's and LATOR's orbits (in the 3:2 resonance) relative to the Sun.  The epoch of this figure shows the spacecraft passing behind the Sun as viewed from the Earth.  The  figure on the left shows the trajectory when the spacecraft would be within 10$^\circ$ of the Sun as viewed from the Earth.  This period of 280 days will occur once every 3 years, provided the proper maneuvers are performed.  Two similar periodic curves give the Sun-Earth-Probe angles for the two spacecraft while the lower smooth curve gives their angular separation as seen from the Earth.

As a baseline design for the LATOR orbit,\footnote{In addition to this 3:2 Earth resonant orbit, here is also an option to have both spacecraft move in opposite directions during each of the solar conjunctions \citep{Ken_lator05}.  In this option, the two LATOR spacecraft move either towards to or away from each other, as seen from the Earth. At the beginning of each conjunction the two craft are on the opposite sides from the Sun, moving towards each other in a such a way so that not only their impact parameters equal (i.e. $p_1=-p_2$), but also the rates of change of these impact parameters are also equal (i.e. $\dot p_1=-\dot p_2$). This option would increase the amount of $\Delta v$ which LATOR spacecraft should carry on-board, but it significantly reduces the experiment's dependence on the accuracy of the determination of the solar impact parameter. This particular option is currently being investigated and results will be reported elsewhere.} both spacecraft will be launched on the same launch vehicle. Almost immediately after the launch there will be a 30 m/s maneuver that separates the two spacecraft on their 3:2 Earth resonant orbits (see Figure \ref{fig:lator_sep_angle2}).  The sequence of events that occurs during each observation period will be initiated at the beginning of each orbit of the ISS. It assumed that bore sighting of the spacecraft attitude with the spacecraft transmitters and receivers have already been accomplished. This sequence of operations is focused on establishing the ISS to spacecraft link. The interspacecraft link is assumed to be continuously established after deployment, since the spacecraft never lose line of sight with one another (for more details consult Section~\ref{sec:operations}).

The 3:2 Earth resonant orbit provides an almost ideal trajectory for the LATOR mission, specifically i) it imposes no restrictions on the time of launch; ii) with a small propulsion maneuver after launch, it places the two LATOR spacecraft at the distance of less then 3.5$^\circ$ (or $\sim 14 ~R\odot$) for the entire duration of the experiment (or $\sim$8 months); iii) it provides three solar conjunctions even during the nominal mission lifetime of 22 months, all within a 7 month period; iv) at a cost of an extra maneuver, it offers a possibility of achieving  small orbital inclinations (to enable measurements at different solar latitudes); and, finally, v) it offers a very slow change in the Sun-Earth-Probe (SEP) angle of about $\sim R_\odot$ in 4 days. Furthermore, such an orbit provides three observing sessions during the initial 21 months after the launch, with the first session starting in 15 months \citep{lator_cqg_2004}. As such, this orbit represents a very attractive choice for LATOR. We intend to further study this 3:2 Earth resonant trajectory as the baseline option for the mission. 

In the next section we will discuss the preliminary design of the LATOR interferometer on the ISS.

\section{LATOR Interferometry}
\label{sec:interferometry}

In this section, we describe the process of how the LATOR interferometer will be measuring angles. Since the spacecraft will carry lasers that are monochromatic sources, the interferometer can efficiently use heterodyne detection to measure the phase of the incoming signal. To this extent, we first present a simplified explanation of heterodyne interferometry proposed for the LATOR interferometer. We then describe the interferometric design of the LATOR station on the ISS. 

\subsection{Heterodyne Interferometry}
\label{sec:heterod_interf}

\begin{figure*}[h!]
 \begin{center}
\noindent  \vskip -0pt   
\psfig{figure=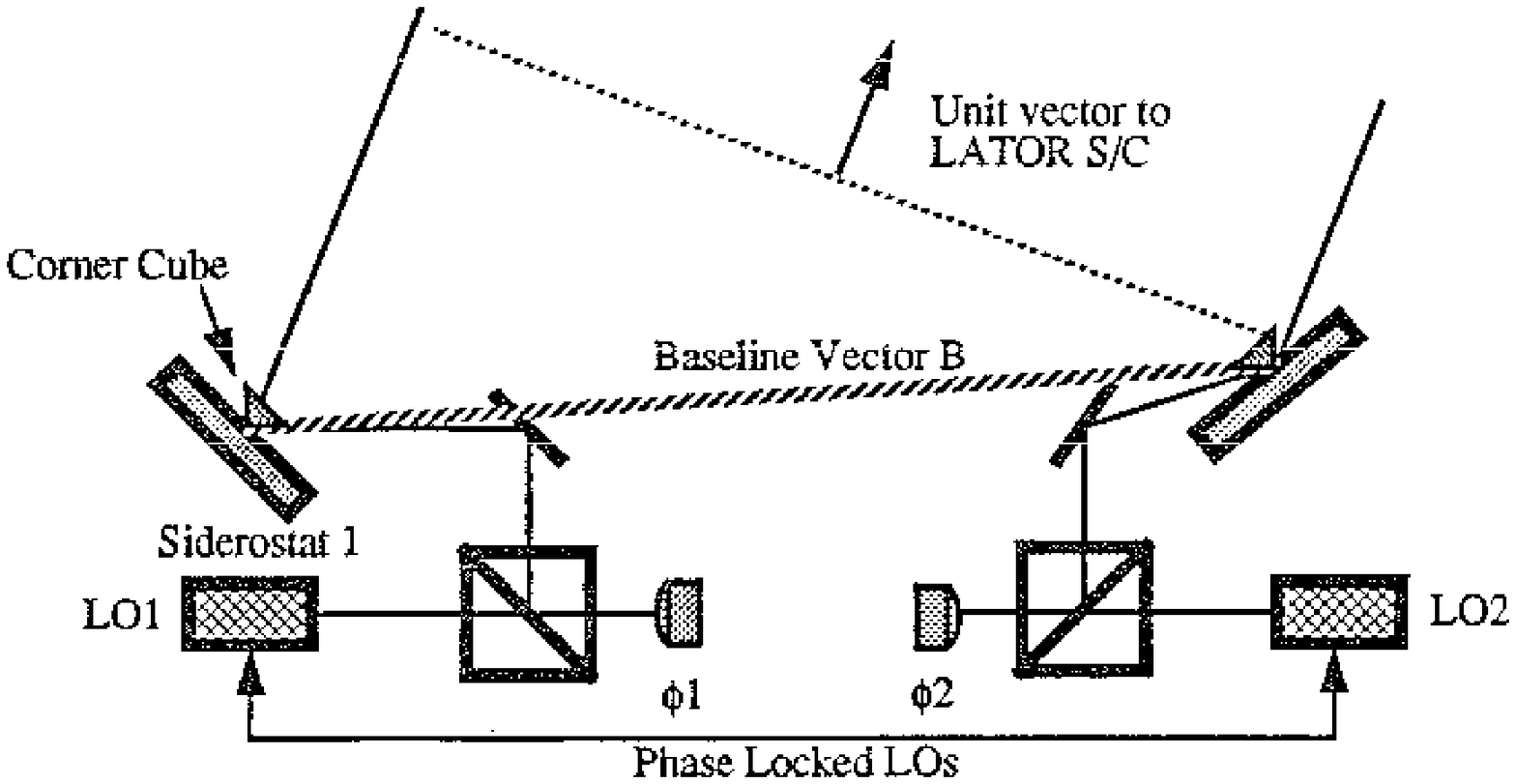,width=95mm}
\end{center}
\vskip -15pt 
  \caption{Heterodyne interferometry on 1 spacecraft with phase locked local oscillator.
 \label{fig:heterodyne1}}
%
\vskip 15pt \hskip -2pt 
\begin{minipage}[b]{.46\linewidth}
\centering \psfig{file=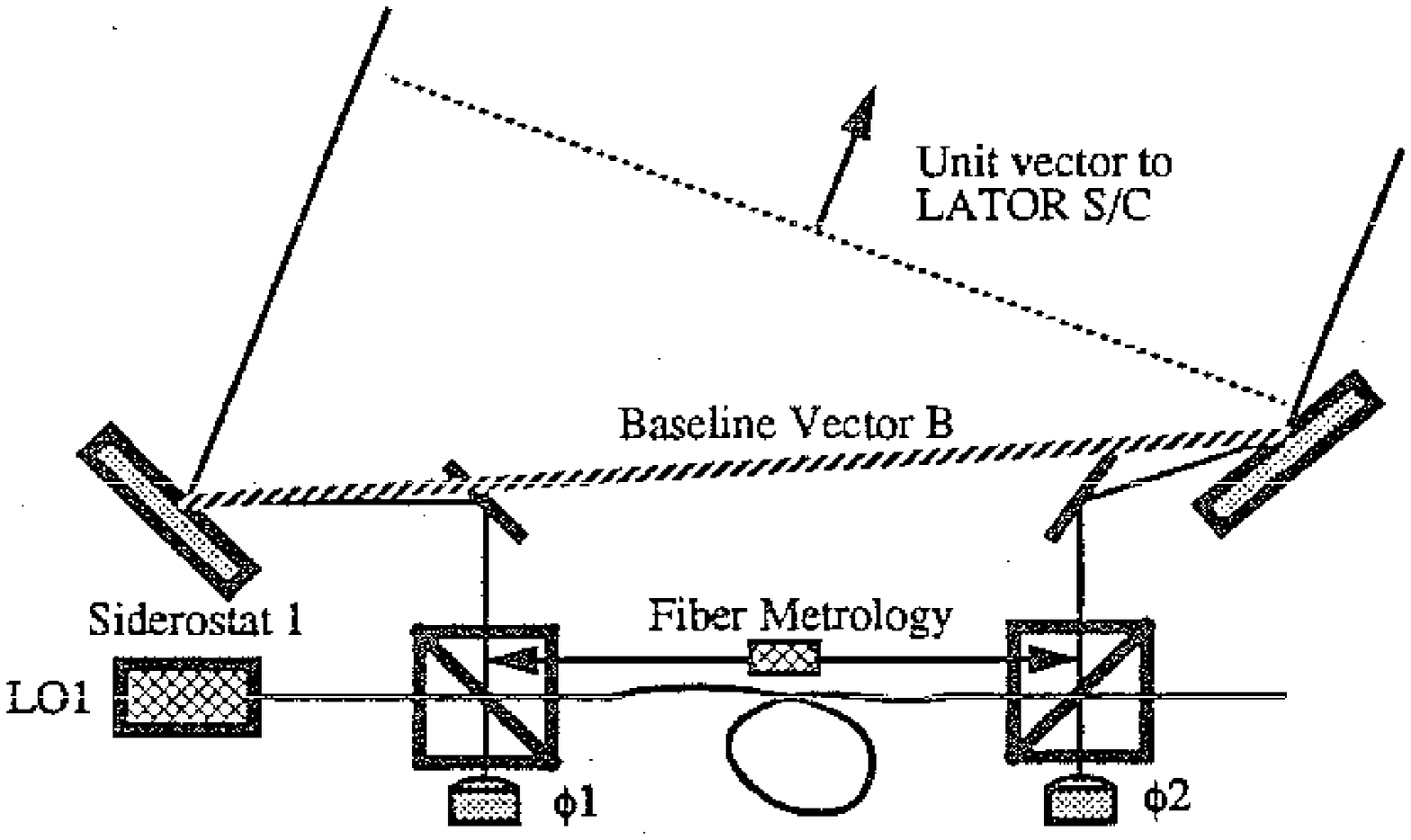, width=81.1mm}
\end{minipage} 
\hskip 15pt
\begin{minipage}[b]{.46\linewidth}
\vbox{\centering \psfig{file=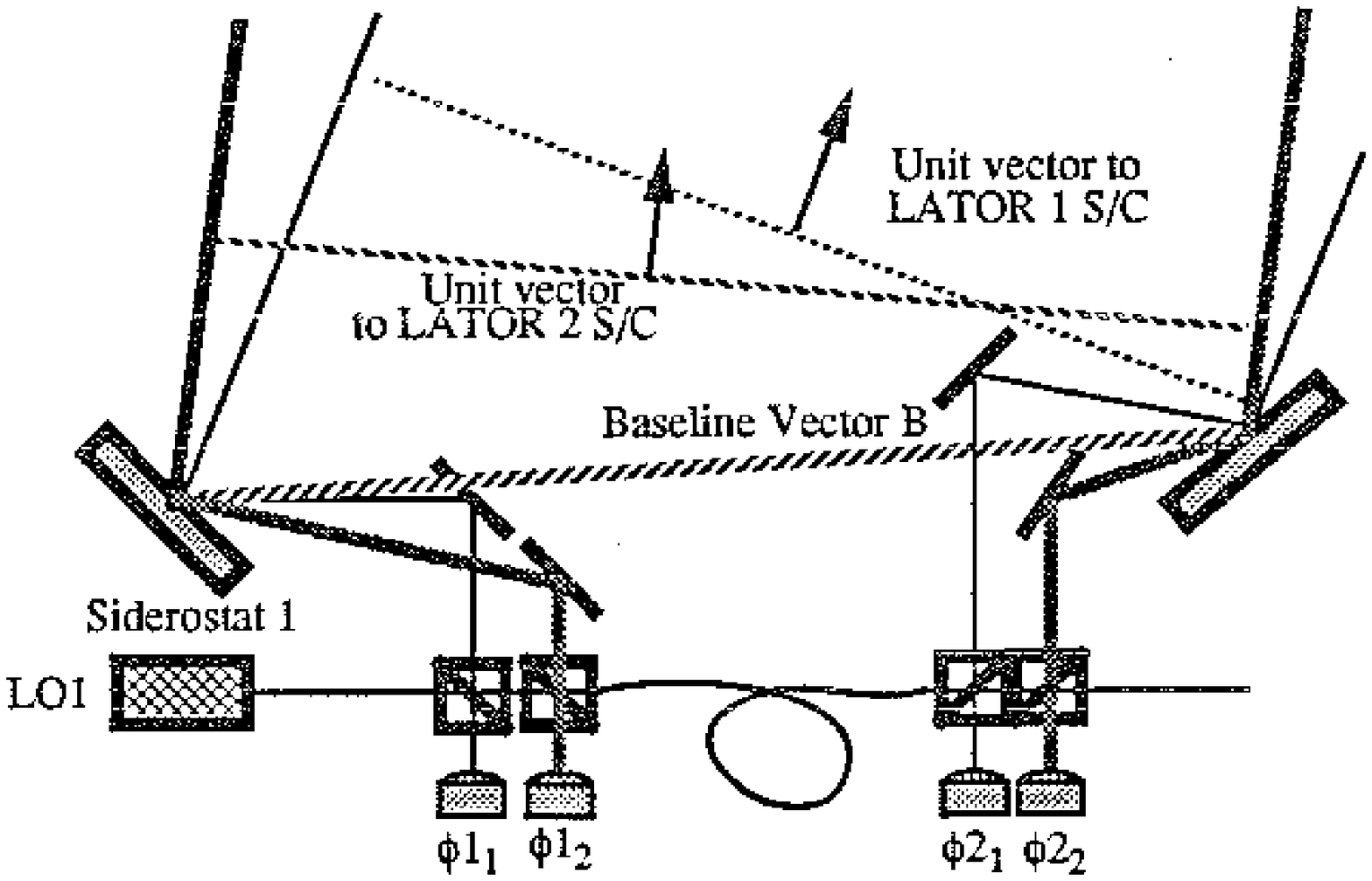, width=81.1mm} \\\vskip-8pt}
\end{minipage}
\caption{Left: Fiber-linked heterodyne interferometry and fiber metrology system. Right: Fiber-linked heterodyne interferometry on 2 spacecraft.}
 \label{fig:heterodyne2-3}
\vskip -0pt 
\end{figure*}

Figures \ref{fig:heterodyne1}--\ref{fig:heterodyne2-3} show a simplified schematic of how angles are measured using a heterodyne interferometer. In Figure \ref{fig:heterodyne1}, two siderostats are pointed at a target. Two fiducials, shown as corner cubes, define the end points of the interferometer baseline. The light from each of the two arms is interfered with stable local oscillators (LOs) and the phase difference recorded. If the LOs in each arm were phase locked, the angles of the target with respect to the baseline normal is
{}
\begin{equation}
\theta=\arcsin\Big[\frac{(2\pi n+\phi_1-\phi_2)\lambda}{2\pi b}\Big],
\end{equation}

\noindent 
where $\lambda$ is the wavelength of the downlink laser, $n$ is an unknown integer arising from the fringe ambiguity and $b$ is the baseline length. In order to resolve this ambiguity, multiple baselines were used in the previous mission design (this is discussed in greater detail in \citep{Yu94}). In reality, it is difficult to phase lock the two LOs over the long baseline lengths. The left graphics in Figure \ref{fig:heterodyne2-3} shows how a single LO can be used and transmitted to both siderostats, using a single mode fiber. In this configuration, a metrology system is used to monitor changes in the path length as seen by the LO as it propagates through the fiber. The metrology system measures the distance from one beamsplitter to the other. In this case, the angle is given by
\begin{equation}
\theta=\arcsin\Big[\frac{(2\pi n+\phi_1-\phi_2 +m_1)\lambda}{2\pi b}\Big],
\end{equation}
\noindent
where $m_1$ is the phase variations introduced by changes in the optical path of the fiber as measured by the metrology system. 

Now consider the angle measurement between two spacecraft (shown on the right in Figure \ref{fig:heterodyne2-3}). In this case the phase variations due to changes in the path through the fiber are common to both spacecraft. The differential angle is
 \begin{eqnarray}
\theta&=&\arcsin\Big[\frac{(2\pi (n_1-n_2)+(m_1-m_2))\lambda}{2\pi b}+
\frac{((\phi1_1-\phi1_2)-(\phi2_1-\phi2_2))\lambda}{2\pi b}\Big].
\end{eqnarray}
\noindent
Since the spacecraft are monochromatic sources, the interferometer can efficiently use heterodyne detection to measure the phase of the incoming signal.  Note that because of the fact that this is a differential measurement, it is independent of the any changes in the fiber length. In reality, the interferometer will have optical paths that are different between the two spacecraft signal paths. These paths must be monitored accurately with a metrology system to correct for phase changes in the optical system due to thermal variations. However, this metrology must only measure path lengths in each interferometric station and not along the entire length of the fiber.  

\begin{figure*}[t!]
 \begin{center}
\noindent    
\psfig{figure=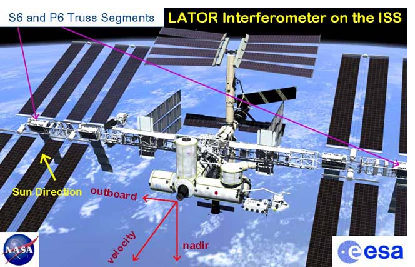,width=115mm}
\psfig{figure=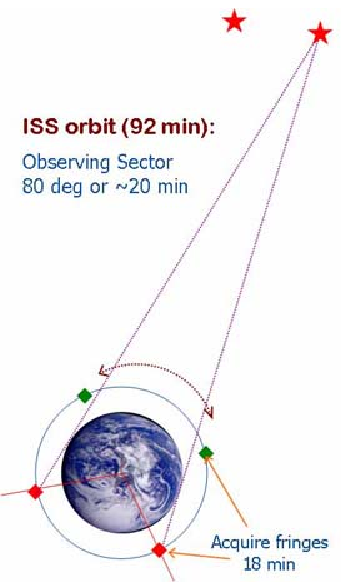,width=4.4cm}
\vskip -5pt 
  \caption{Left: Location of the LATOR interferometer on the ISS. To utilize the inherent ISS Sun-tracking capability, the LATOR optical packages will be located on the outboard truss segments P6 and S6 outwards. Right: Signal acquisition for each orbit of the ISS. Note that variation of the baseline's projection allows to successfully solve the issue of the monochromatic fringe ambiguity.  
 \label{fig:iss_config}}
 \end{center}
\end{figure*}

The use of multiple interferometers is a standard solution to resolve the fringe ambiguity resulting from the interferometric detection of monochromatic light \citep{lator_cqg_2004}. The current LATOR mission proposal is immune from the fringe ambiguity problem, as the orbit of the ISS provides enough variability (at least $\sim 30$\%) in the baseline projection (see Figure~\ref{fig:iss_config} for general description of the geometry of the interferometer on the ISS and its orbit). This variablity enables one to take multiple measurements during one orbit, in order to uniquely resolve the baseline orientation for each ISS orbit, which successfully solves the fringe ambiguity issue for LATOR.

%

\subsection{Long-Baseline Optical Interferometer on the ISS}
\label{sec:interf-ISS}

\begin{figure}[h!]
 \begin{center}
\noindent  \vskip -0pt   
\psfig{figure=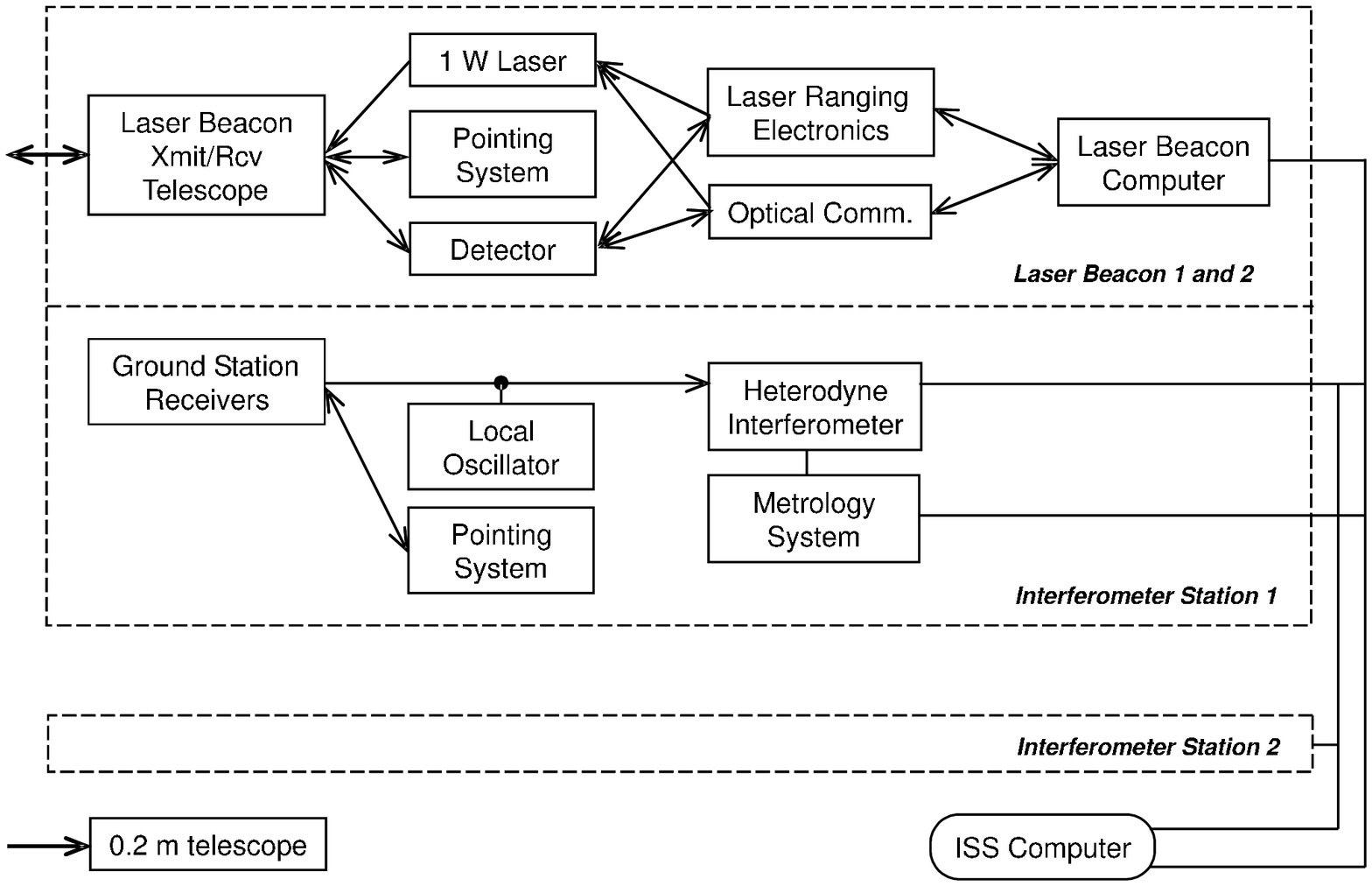,width=117mm}
\end{center}
\vskip -15pt 
  \caption{A block diagram of the LATOR station on the ISS.
 \label{fig:iss_block_giagram}}
\end{figure}

The LATOR station on the ISS is used to interferometrically measure the angle between the two spacecraft and also to transmit and receive the laser ranging signals to each of the spacecraft. A block-diagram of the laser station is shown in Figure~\ref{fig:iss_block_giagram} and is described in more detail below. The station on the ISS is composed of a two laser beacon stations that perform communications and laser ranging to the spacecraft and two interferometer stations that collect the downlink signal for the astrometric measurement. This station also uses a fiber optic link to transmit the common local oscillator to the interferometer station.

\subsubsection{General Description}

The interferometer on the ISS will be formed by two optical packages (or laser beacon stations) with approximate dimensions of (0.6~m $\times$ 0.6~m $\times$ 0.6~m) for each package. The mass of each telescope assembly $\sim$120~kg. Both laser beacon stations must be physically located and integrated with the ISS infrastructure (see description of the ISS in \citep{ISS-2000}). Their location must provide a straight-line separation of $\sim$100 m between the two stations and have a clear line-of-sight (LOS) path between the two transceivers during the observation periods. Both packages must have clear LOS to both spacecraft during pre-defined measurement periods. Location on the ISS should maximize the inherent ISS sun-tracking capability.  Both telescope assemblies will have to be able to point toward the Sun during each observing period which can be achieved by locating these payloads on the ISS outboard truss segments (P6 and S6 outward, see Figure~\ref{fig:iss_config}). In addition, a limited degree of automatic Sun-tracking capability is afforded by the $\alpha$-gimbals on the ISS.

The minimum unobstructed LOS time duration between each transponder on the ISS and the transponders and their respective spacecraft will be 58 minutes per the 92 minute orbit of the ISS.   	
The pointing error of each transceivers to its corresponding spacecraft will be no greater than 1 $\mu$rad for control,  1 $\mu$rad for knowledge, with a stability of 0.1 $\mu$rad/sec, provided by a combination of the standard GPS link available on the ISS and $\mu$-g accelerometers. 

\subsubsection{Laser Beacon Station}

The laser beacon stations provide the uplink signals to the LATOR spacecraft and detect their downlink signals (see Figure~\ref{fig:iss-beacon-station}). The transmitter laser signal is modulated for laser ranging and to provide optical communications. Separate transmitters are used for each spacecraft each using a 1~W laser at 1064 nm as the source for each laser beacon. The laser beam is expanded to a diameter of 30~cm and is directed toward the spacecraft using a siderostat mirror. Fine pointing is accomplished with a fast steering mirror in the optical train.

During initial acquisition, the optical system of the laser beacon is modified to produce a beam with a 10~arcsec divergence. This angular spread is necessary to guarantee a link with the spacecraft, albeit a weak one, in the presence of pointing uncertainties. After the acquisition sequence is complete, the beam is narrowed to a diffraction limited beam, thereby increasing the signal strength.

The downlink laser signal at 1064 nm, is detected using a $12\times 12$ ($10 \times 10$ arcsec) array of Germanium detectors. In order to suppress the solar background, the signal is heterodyned with a local oscillator and detected within a narrow 1 MHz bandwidth. In the initial acquisition mode, the detection system searches over a 300 MHz bandwidth and uses a spiral search over a 30~arcsec angular field to find the downlink signal. Upon acquisition, the search bandwidth is decreased to 1~MHz and a quad-cell subarray is used to point the siderostat and fast steering mirrors of the beacon.

\subsubsection{Interferometer Station}

The interferometer stations collect the laser signal from both spacecraft to perform the heterodyne measurements needed for the interferometric angle measurement. There are a total of five receivers to make the four angular measurements needed to resolve fringe ambiguity.

\begin{figure}[h!]
 \begin{center}
\noindent  \vskip -5pt   
\psfig{figure=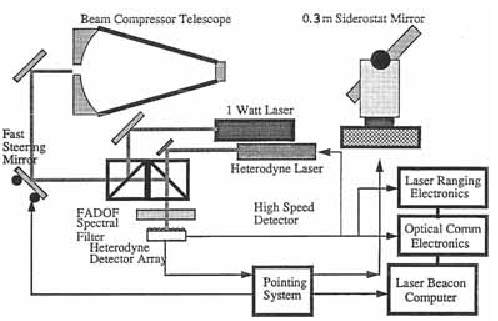,width=120mm}
\end{center}
\vskip -15pt 
  \caption{Schematic of the laser beacon station on the ISS.
 \label{fig:iss-beacon-station}}
 \begin{center}
\noindent  \vskip -5pt   
\psfig{figure=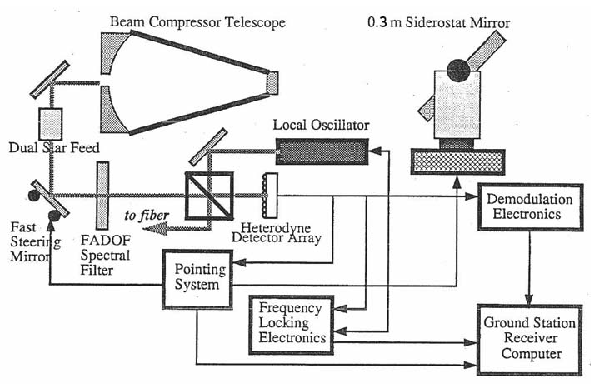,width=125mm}
\end{center}
\vskip -15pt 
  \caption{Component description of the receiver on the ISS.
 \label{fig:iss-receiver}}
\end{figure} 

\begin{figure}[h!]
 \begin{center}
\noindent  \vskip -5pt   
\psfig{figure=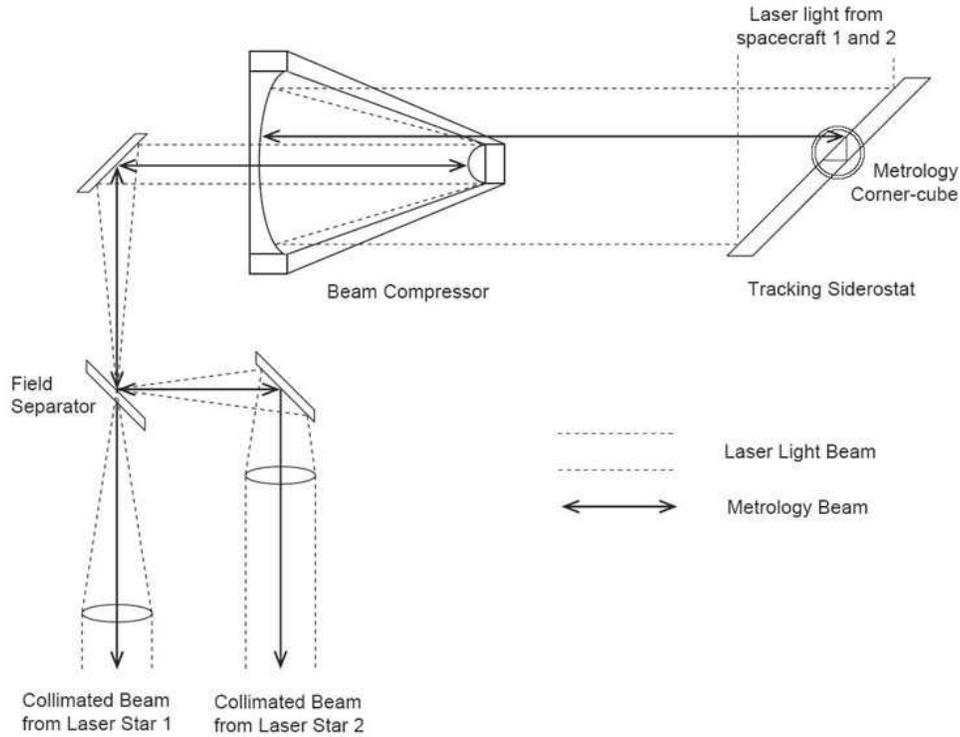,width=129mm}
\end{center}
\vskip -15pt 
  \caption{
Principal diagram for the dual feed optical system. The diagram describes propagation of laser starlight and metrology beams through the optical system (note metrology corner-cube, beam compressor, and field separator).
 \label{fig:dual-feed}}
\end{figure}

Figure~\ref{fig:iss-receiver} shows a schematic of an interferometer station. The detection and tracking system is basically similar to the receiver arm of the laser beacon described in the previous section. Light is collected by a 0.3~m siderostat mirror and compressed with a telescope to a manageable beam size. The light from each of the spacecraft is separated using a dual feed optical system as shown in Figure~\ref{fig:dual-feed}. A fast steering mirror is used for high bandwidth pointing of the receiver. In addition a combination of a wideband interference filter and a narrow band Faraday Anomalous Dispersion Optical Filter (FADOF) will used to reject light outside a 0.05 nm band around the laser line. Each spacecraft signal is interfered with a local oscillator and the phase measurement time tagged and recorded. A $6\times6$~Ge array ($5\times5$ arcsec FOV) is used to provide heterodyne acquisition and tracking of the LATOR spacecraft.

\subsubsection{Interferometer on the ISS}

\begin{figure*}[t!]
 \begin{center}\hskip -5pt
\epsfig{figure=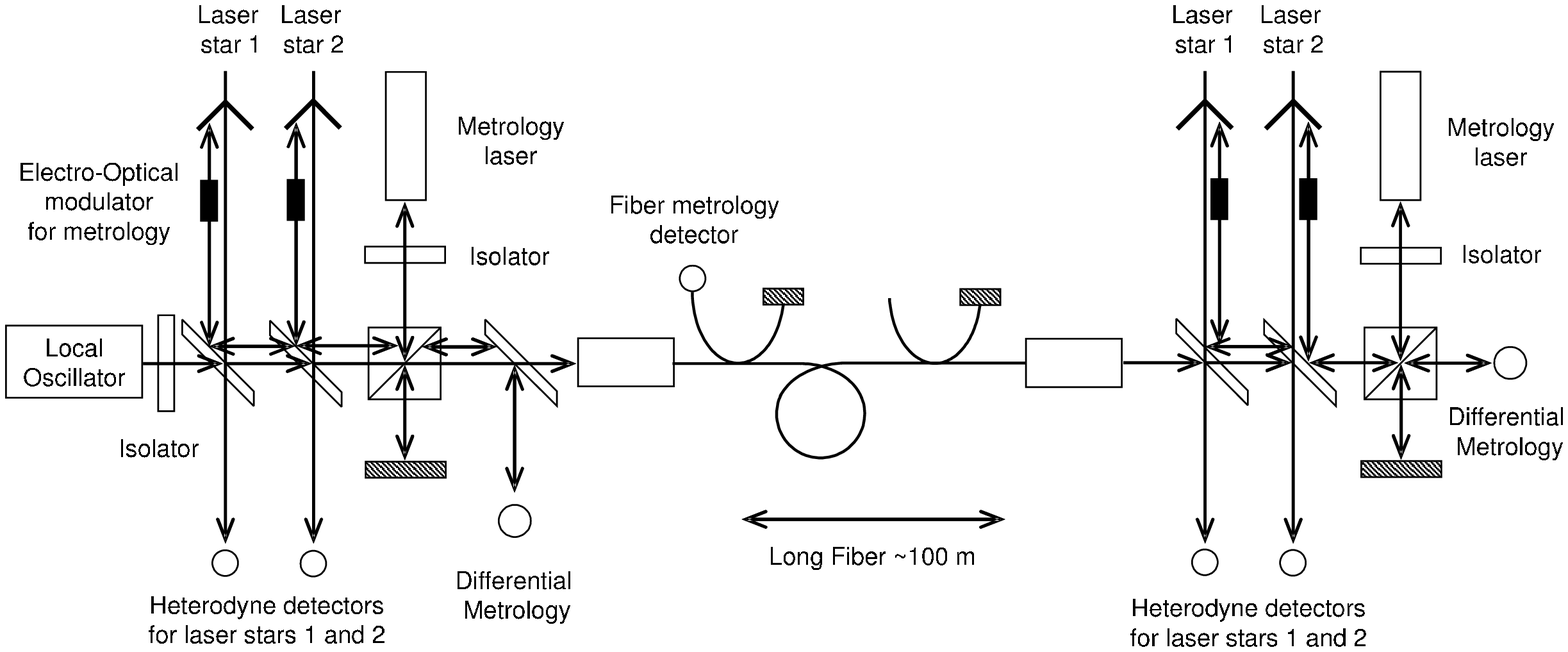,width=160mm}
\end{center}
\vskip -10pt 
  \caption{Component description of the ISS-based interferometer. 
 \label{fig:iss_interf}}
\end{figure*}

Figure \ref{fig:iss_interf} shows a schematic of the ISS-based fiber interferometer that will be used to perform the angular measurement between the two LATOR spacecraft. The interferometer includes the heterodyne detection of the downlink signals that have been described in the previous section. The local oscillator (LO) is generated in one of the ground station receivers and is frequency locked to the laser signal from one of the spacecraft. The LO is then broadcast to the other station on the ISS through a 100~m single mode polarization preserving fiber. The heterodyne signals from all the stations (2 stations, 2 signals each) are recorded and time tagged.

Figure~\ref{fig:iss_interf} also shows two metrology systems used in the interferometer. The first metrology system measures the difference in optical path between the two laser signal paths and is essential to proper processing of the heterodyne data. The second metrology system measures the changes in the optical path through the fiber. This measurement monitors the length of the fiber and is used in the post processing of the interferometer data. The internal path metrology system, shown in the figure, measures the paths from corner cube on the siderostat mirror (shown as two, really only one) to the metrology beamsplitter. It is essential that the laser metrology system be boresighted to the laser signal path so the correct distance is measured. A Michelson interferometer with a frequency shift in one arm measures changes in the length of each signal path. Both spacecraft signal paths are measured simultaneously. This is accomplished by using an electro-optic cell and modulating each beam at a different frequency. A He-Ne laser is used as the light source for this metrology system. Filters at the output of the detector are then used to separate the signals corresponding to each metrology beam.

\begin{figure}[h!]
 \begin{center}
\noindent  \vskip -5pt   
\psfig{figure=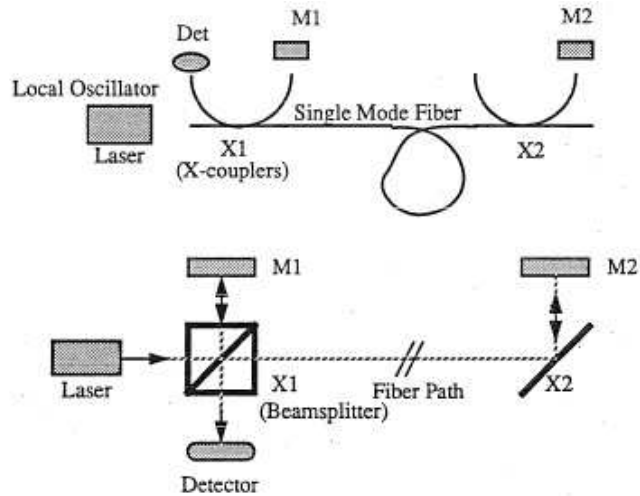,width=90mm}
\end{center}
\vskip -15pt 
  \caption{Fiber metrology system.
 \label{fig:fiber_metr}}
\end{figure} 

The fiber metrology system measures changes in the optical path through the fiber. This system uses local oscillator signal in a Michelson configuration. Figure~\ref{fig:fiber_metr} shows the correspondence between a standard Michelson interferometer and the fiber metrology system. The two X couplers serve as the beam splitters. Reflectors at the ends of the fiber couplers serve as the reference and signal mirrors. One of these reflectors is dithered to frequency shift the output signal. The phase measurement at the detector measures changes in the path length between points X1 and X2, if Ml-X1 and M2-X2 are held constant. This is accomplished by placing the X couplers and mirrors at each end of the fiber on a single thermally stable optical breadboard.

\subsection{Laser Metrology Transceiver Subsystem} 

The metrology transceiver consists of the laser, frequency modulators, optics, and frequency stabilizer. The laser light is first frequency-stabilized to better than 1 part in 10$^{10}$, this is done in order to make the measurements.  The laser light is then frequency-modulated in order to produce the heterodyne signal and distinguish between incoming and outgoing beams.  Finally, light is collimated and injected into the beam launcher optics. The incoming metrology signal is received by the beam launcher optics and is interfered with the local laser.  A cat's eye retroreflector serves as the spacecraft fiducial and is common to all three beam launchers.  Below we discuss these elements in more details.
  
\subsubsection{Laser} 
A 1~W Nd:YAG laser operating at 1064 nm is used to transmit the metrology signals to the other spacecraft. The laser will be thermally tunable over a range of several GHz. Two lasers are used in each spacecraft for redundancy.

\subsubsection{Frequency Stabilization}
\label{sec:freq-stab}

The source laser is stabilized to 1 part in 10$^{10}$ long term using a temperature controlled Fabry Perot etalon. A Pound-Drever scheme is used to servo the frequency of the laser to one of the longitudinal modes of the cavity. Control of the $\sim$3 cm cavity to 10 mK will achieve the required stability. Calibration of the cavity length on the ground will be done by injecting a second laser locked onto an adjacent longitudinal cavity mode and beating the two signals together. For the 3 cm cavity, the 5 GHz beat frequency must be known to 10$^{-10}$. Temperature control of the cavity will allow fine tuning of the laser frequencies between the spacecraft so that the heterodyne signal between two lasers lies below 2 MHz. This will require knowledge of the spacecraft relative velocity to 1 m/s, which is easily achievable.

\subsubsection{Frequency Modulators}

The laser frequency is modulated in order to distinguish between the various transmit and receive beams used in the LATOR measurements. In addition, the relative velocity between the spacecraft can reach as high as 100 m/s. This will produce a Doppler frequency of up to 200 MHz between lasers from two spacecraft. The frequency of the modulator will be tuned to slightly offset from the Doppler frequency to minimize the bandwidth at which the data needs to be recorded.

Acousto-optic modulators (AOM) with fiber-coupled input and output are used. For a single metrology channel 3 different frequencies are needed for the reference, and two unknown signals. One implementation is a fiber-fed modulator which uses a bulk AOM and is insensitive to alignment errors. Other implementations for the AOMs will also be studied. These include integrated optic AOMs and multi-channel Bragg cells, both of which will be capable of generating the multiple
signals at much lower mass.

The metrology system will also need to phase lock the outgoing laser with the incoming laser. The AOM provides the phase modulation to the laser beam. The incoming signal and the laser output from the AOM are interfered on a high frequency detector. This signal is then used to servo the frequency of the AOM to null. This will produce a phase
locked signal whose phase error is determined by the level of the null. In reality, because of the AOM, center frequency, the interfered signal will be upshifted by a stable local oscillator and the servoing done in RF. The stability of this local oscillator is the same as the required stability of the phase locked loop, 10$^{-10}$ (discussed in Sec.~\ref{sec:freq-stab}).

\subsubsection{Beam Launcher and Receiver Optics}

In the current instrument design, the modulated laser beam is injected using a polarization preserving single mode fiber and expanded to a 0.5 cm beam.  A cat's eye retroreflector is one of several devices that can be used as the metrology fiducial and is common to the three metrology beams. The cat's eye uses two optically contacted concentric hemispheres with radius of $\sim$ 10 cm and $\sim$ 20 cm. The cat's eye is sized many times larger than the beam in order to minimize the effect of spherical aberration. 

The beam is then expanded to a 5 cm beam using a refractive telescope. A refractive design was chosen because changes in the optical path are relatively insensitive to changes in the position and orientation of the optical elements.  

In order to measure the path length to better $\sim5$~pm, errors due to thermal effects on the beam launcher optics must be controlled. For example a change in the temperature of 1 milli-Kelvin on a 0.5 cm beamsplitter would produce a path length error of $\sim5$~pm; consequently an active thermal controller would be used on the beamsplitters, and telescope optics. Furthermore, baffles on the optics will be used to prevent external radiation from affecting the temperature of the instrument. The metrology optics will be mounted on a GrEp bench for thermal stability.

\subsubsection{Acquisition Camera Subsystem}
The acquisition camera  will be used as the sensor for pointing the metrology beam.  A $512\times512$ camera may be used to detect the position of the incoming laser beam to 0.5 arcsec over a 1$^\circ$ field by interpolating the centroid of the spot to 0.1 pixel. Three cameras will be used to track each of the incoming metrology beams. The outgoing laser beam will be retroreflected from the alignment corner cube to produce a spot on the acquisition camera on which to servo the pointing gimbal.  The direction of the outgoing beam is set to the position of the target spacecraft, taking into account the point ahead angle.

\subsubsection{Pointing Subsystem}
 In the current instrument design the entire beam launcher optical assembly is gimbaled to point the metrology beam to the target spacecraft.  The 2-axis gimbal has a center of rotation at the center of the cat's eye retro reflector.  This optical arrangement measures the distance between the optical fiducials and is not sensitive to slight misalignments to the first order.  The gimbal will have a range of $1^\circ$ and a pointing resolution of 0.5 arcsec. 

\subsubsection{Laser Ranging Subsystem}
The laser ranging system is used to determine the positions of the spacecraft with respect to the ISS. This is required to determine the impact parameter of the laser beam grazing the sun as well as the co-planarity of the three spacecraft. A time of flight laser ranging system is used to triangulate the spacecraft positions. A laser transponder system on the spacecraft is used to increase the SNR of the return pulse. 

Laser ranging will be performed with an accuracy of $\sim$1 cm by integrating over a number of laser pulses. If the system were capable of instantaneously detecting delays of 100 ps (3 cm), at a 1 kHz repetition rate, it would take under 1 second to reach the desired accuracy. Assuming this level of ranging and using baseline 100 m will result in an accuracy in the transverse direction of 1 m at LATOR's orbit with spacecraft separated as much as 2 AU (see discussion of inter-spacecraft laser ranging operations in Section~\ref{sec:X-sc-rec/trans}).

In the next section we turn our attention to the mission flight system.

\section{LATOR Flight System}
\label{sec:lator_current}

The LATOR flight system consists of two major components: the deep-space component that will be used to transmit and receive the laser signals needed to make science measurements and the interferometer on the ISS that will be used to interferometrically measure the angle between the two spacecraft and to transmit and receive the laser ranging signals to each of the spacecraft. 

\begin{figure}[t!]
 \begin{center}
\noindent  \vskip -5pt   
\psfig{figure=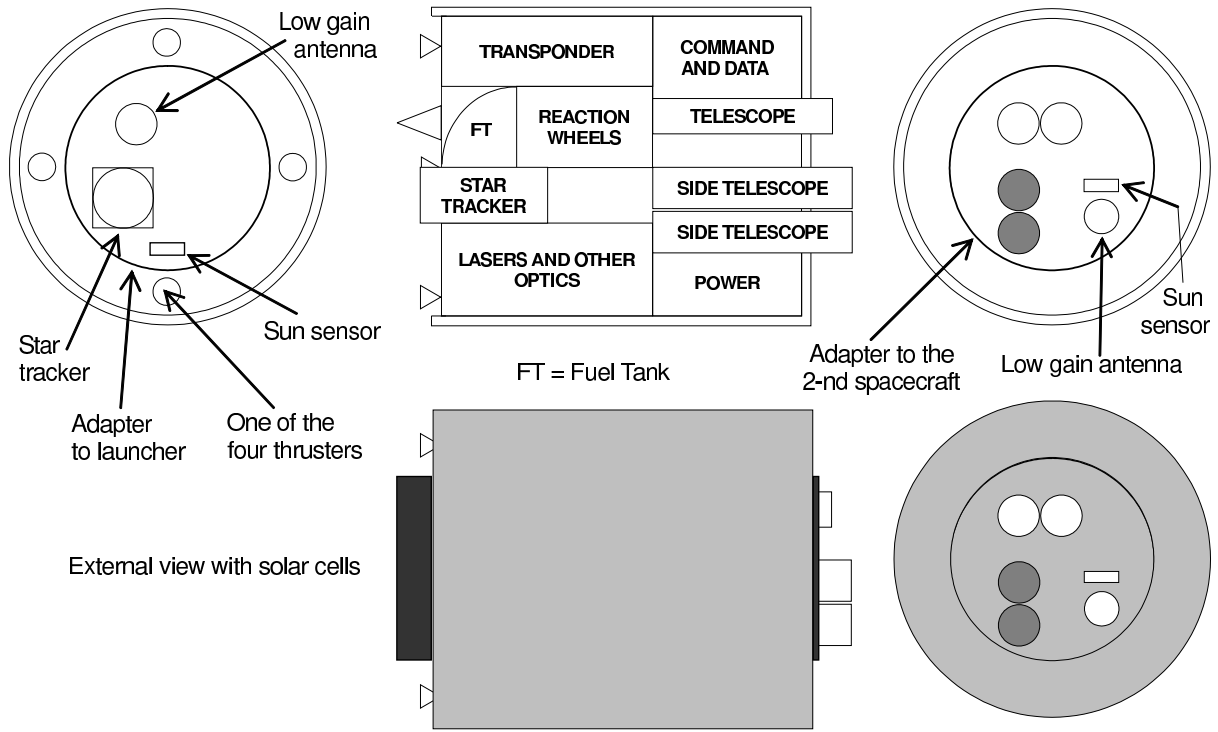,width=160mm}
\end{center}
\vskip -15pt 
  \caption{LATOR spacecraft concept.
 \label{fig:sc_concept}}
\end{figure}

\begin{figure*}[h!]
\begin{minipage}[b]{.40\linewidth}
\centering \psfig{figure=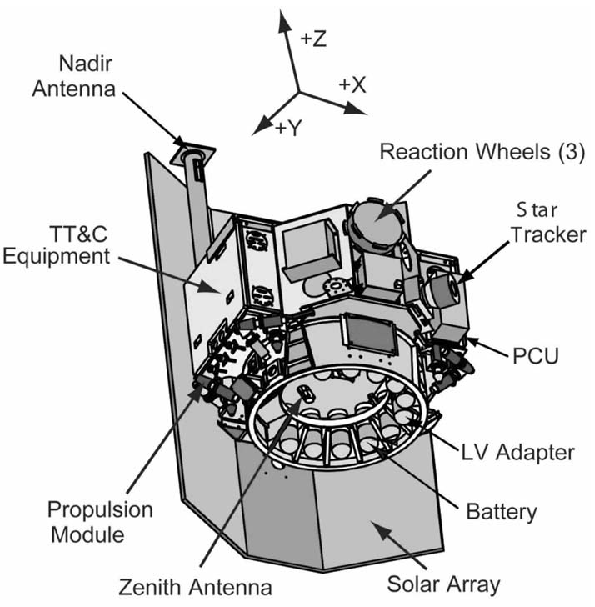,width=76mm}
\end{minipage}
\hskip 32pt
\begin{minipage}[b]{.40\linewidth}
\centering 
\vbox{\psfig{figure=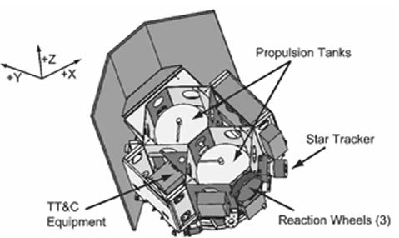,width=83mm}\\\vskip26pt}
\end{minipage}
\caption{A typical Spectrum Astro SA-200S/B bus. With minor modifications this configuration may be adopted for the deep-space component of the LATOR mission.
 \label{fig:SA200S_config2}}
\vskip -5pt 
\end{figure*}

There are two LATOR spacecraft in the deep-space component of the mission, which will be used to transmit and receive the laser signals needed to make the science measurements. Figure \ref{fig:sc_concept} shows a schematic of the flight system without the solar cell array. The flight system is subdivided into the instrument payload and the spacecraft bus (note that SA200S spacecraft built by SpectrumAstro already has the needed capabilities, see Figure \ref{fig:SA200S_config2}). The instrument includes the laser ranging and communications hardware and is described in more detail in the following section. The spacecraft contains the remainder of the flight hardware which includes solar cells, attitude control, and the spacecraft structure.

In this Section we will discuss the design of these components in more detail.

\subsection{LATOR Instrument}

\begin{figure}[h!]
 \begin{center}
\noindent  \vskip -0pt   
\psfig{figure=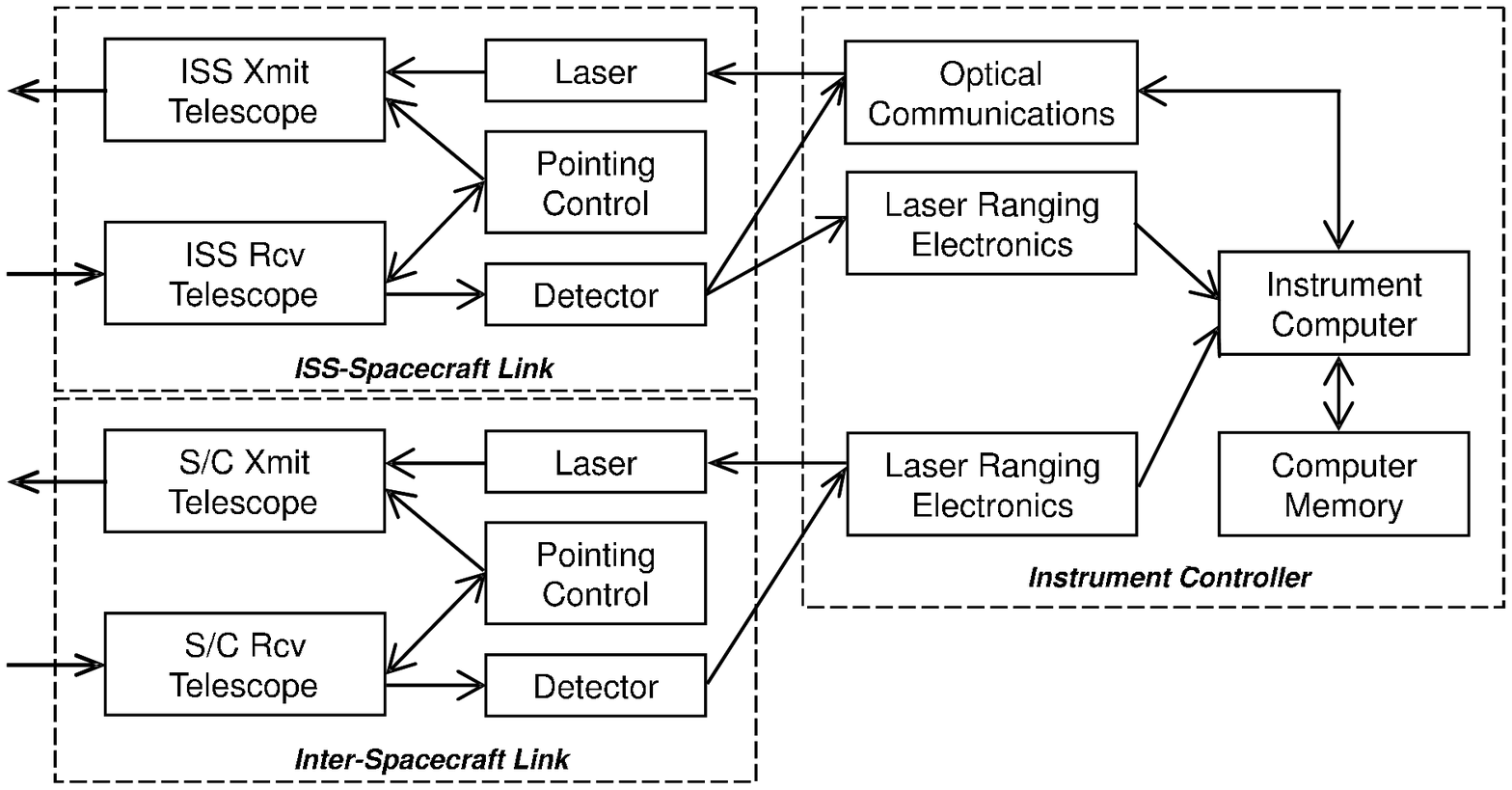,width=127mm}
\end{center}
\vskip -15pt 
  \caption{LATOR instrument subsystem block diagram.
 \label{fig:instrument}}
\end{figure}

The LATOR instrument in each of the two spacecraft consists of three laser metrology transmitters and receivers that can be gimbaled to point at the other spacecraft, and a camera system to acquire the incoming laser signals and to control the pointing of the outgoing beams.  In addition, the instrument contains a laser ranging transponder in order to determine the spacecraft position from the ground. The LATOR instrument is used to perform laser ranging between the two spacecraft; it is also used (the second set) for laser ranging and optical communications between the spacecraft and the ISS. Figure \ref{fig:instrument} shows a block diagram of the instrument subsystems, which we describe in more detail below.

\subsubsection{ISS-to-Spacecraft Receiver and Transmitter} 

\begin{figure}[h!]
 \begin{center}
\noindent  \vskip -0pt   
\psfig{figure=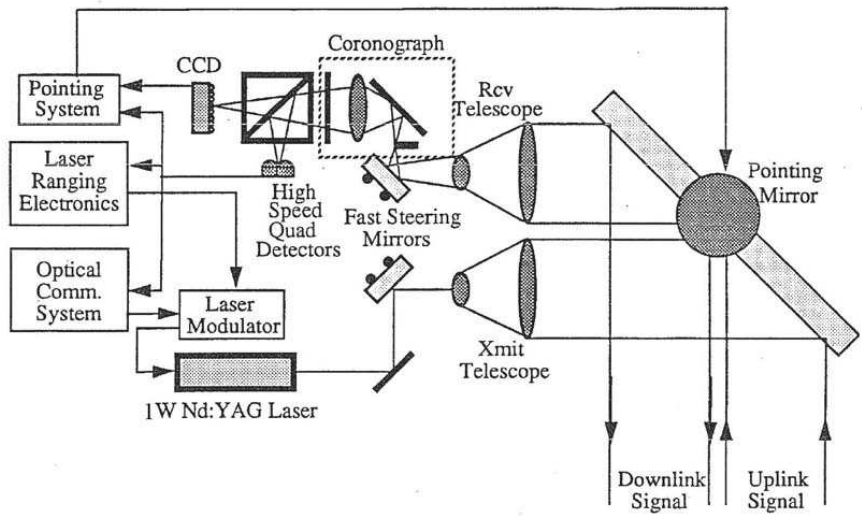,width=115mm}
\end{center}
\vskip -15pt 
  \caption{Spacecraft transmitter and receiver for the ISS--Spacecraft link.
 \label{fig:transm_receiver}}
\end{figure}

The ISS-to-spacecraft receiver performs the acquisition, tracking, and detection of the signals from the ISS (Figure \ref{fig:transm_receiver}). This uplinked signal will be sent at 1064 nm and will contain modulation both to perform laser ranging and to send control signals to the spacecraft. The signals from the ISS are detected by a telescope with a collecting aperture of 20~cm. A coronograph will be used to suppress stray light from the Sun. In addition a combination of a wideband interference filter and a narrow band FADOF filter will used to reject light outside a 0.05 nm band around the laser line. The incoming signal is subdivided with one portion going to a high bandwidth detector and the other to an acquisition and tracking CCD array. Using a $64 \times 64$ CCD array with pixels sized to a diffraction limited spot, this array will have a 5 arcmin field of view, which is greater than the pointing knowledge of the attitude control system and the point ahead angle (30 arcsec). After acquisition of the ISS beacon, a $2\times 2$ element subarray of the CCD will be used as a quad cell to control the ISS--S/C two axis steering mirror. This pointing mirror is common to both the receiver and transmitter channel, to minimize misalignments between the two optical systems due to thermal variations. The pointing mirror will have 10 arcmin throw and a pointing accuracy of 0.5 arcsec, which will enable placement of the uplink signal on the high bandwidth detector.

The ISS-to-spacecraft transmitter sends a laser signal to both the interferometer collectors and the beacon receivers. The signal will be encoded for both ranging and communication information. In particular, the transmitted signal will include the inter-spacecraft ranging measurements. The transmitter uses a 1~W frequency stabilized Nd:YAG laser at 1064 nm. A 5~kHz line width is required to simplify heterodyne detection at the ground station. A 20~cm telescope is used to transmit the laser beam and a steering mirror is used for pointing. The mirror uses information from the attitude control system, the quad-cell detector in the receiver, and the point ahead information from the instrument controller to determine the transmit direction. A fast steering mirror is used to maintain high bandwidth pointing control for both the transmitter and receiver. 

\begin{figure}[h!]
 \begin{center}
\noindent  \vskip -5pt   
\psfig{figure=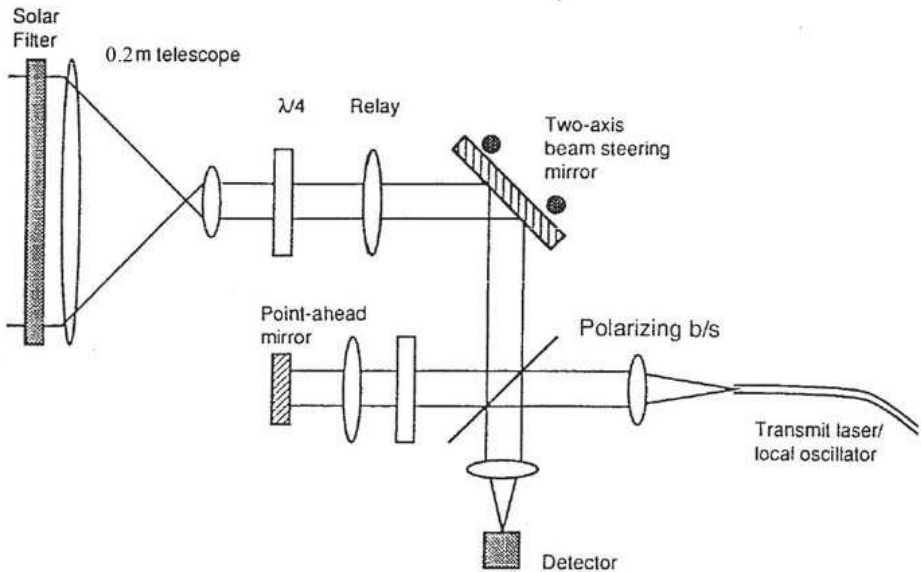,width=125mm}
\end{center}
\vskip -15pt 
  \caption{ISS-to-spacecraft link with common optics: spiral scanning spatial acquisition; open loop point ahead control with piezo actuators; fiber-coupled, frequency stabilized transmitter; pupil planes at the steering mirror and mixing apertures.
 \label{fig:optics}}
\end{figure} 

We have also considering the possibility of using a common optical system for both the transmitter and receiver. Figure \ref{fig:optics} shows a schematic of such a transmitter/receiver system. Because of the difference in the receive and transmit wavelengths, dichroic beam splitters and filters are used to minimize losses from the optics and leakage into the detectors. In this scheme a point-ahead mirror is used maintain a constant angular offset between the received and transmitted beams. Because of the common optical elements, this system is more tolerant to misalignments than the previous configuration.

\subsubsection{Inter-Spacecraft Receiver and Transmitter} 
\label{sec:X-sc-rec/trans}

\begin{figure}[h!]
 \begin{center}
\noindent  \vskip -5pt   
\psfig{figure=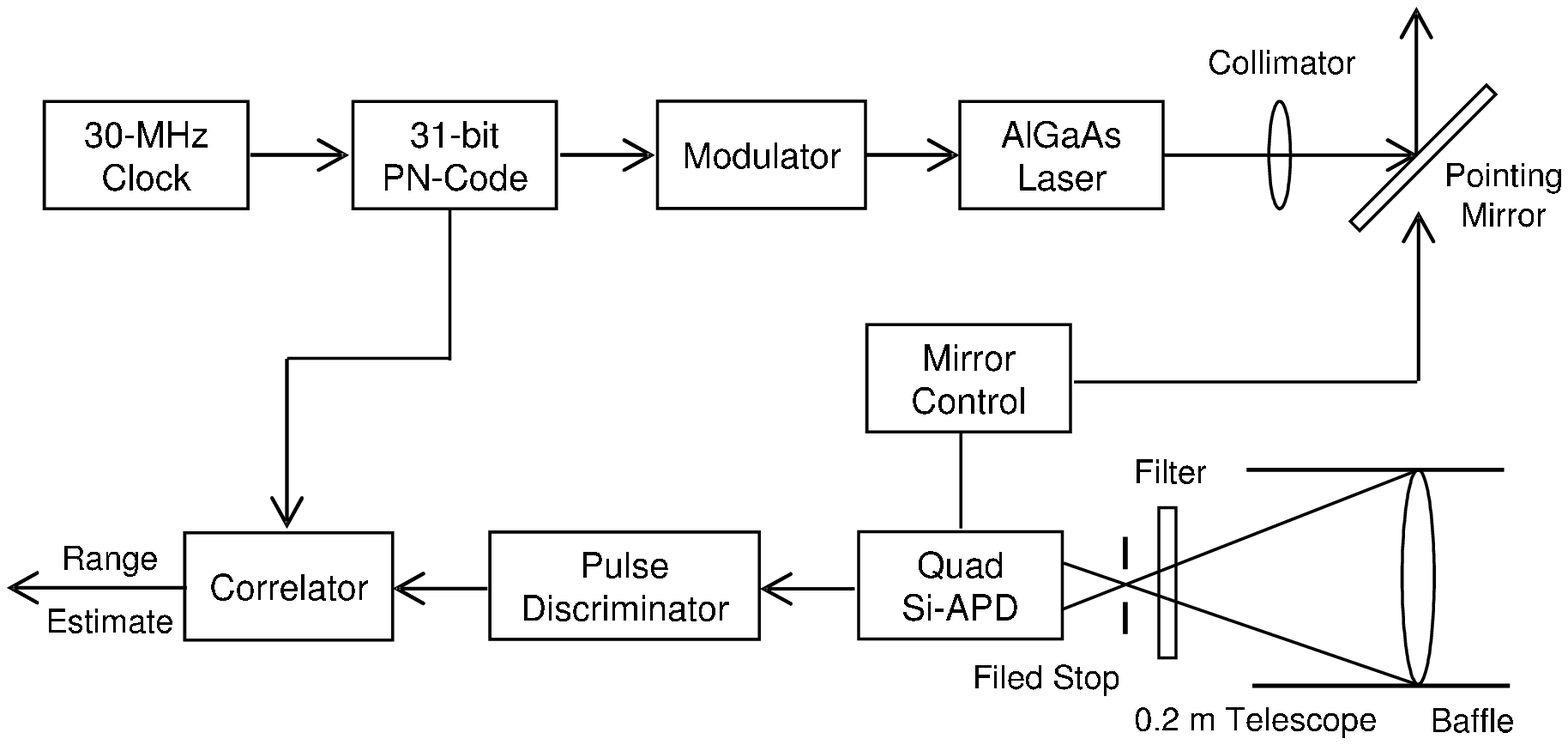,width=125mm}
\end{center}
\vskip -15pt 
  \caption{A block diagram of the inter-spacecraft transmitter and receiver.
 \label{fig:inter_sc}}
\end{figure}

The inter-spacecraft receiver/transmitter uses two separate optical systems. The receiver detects the laser ranging signal from the other spacecraft (shown in Figure~\ref{fig:inter_sc}). The receiver is similar in design to the ISS--Spacecraft receiver subsystem. Since there is no solar background contribution, the coronograph and FADOF filter have been removed. Detection of the signal is accomplished using a CCD for acquisition and a quad cell subarray for tracking. The tracking signal is also used to control the pointing of the transmitter minor. A separate high bandwidth detector is used for detecting the laser ranging signal.

The inter-spacecraft transmitter sends the laser ranging signal to the other spacecraft. The transmitter uses a 780 nm laser with an output power of 0.2~W (alternatively it may use a small fraction of the laser light that is used to establish spacecraft-to-ISS link). The transmitter and receiver telescopes have an aperture of 5~cm diameter. Because of the proximity of the LATOR spacecraft, thermal drifts that cause misalignments between the transmitter and receiver optical systems can be sensed and corrected rapidly. In addition, the LATOR geometry requires minimal point-ahead, since the transverse velocity between spacecraft is nearly zero.

\subsubsection{Instrument Controller} 
The instrument controller subsystem contains the remainder of the instrument hardware. This includes the electronics needed for the laser ranging and optical communications as well as the computer used to control the instrument. The instrument computer will take information from the attitude control system and receiver subsystems in order to control the pointing of the transmit subsystems and the modulation of their laser signals.


\subsection{LATOR Spacecraft}

The LATOR spacecraft, like most spacecraft, will be composed of the following subsystems: thermal, structural, attitude control, power, command and data handling, telecommunications, and propulsion, that will be discussed below. 

\subsubsection{Thermal Subsystem} 
The basic thermal design will similar to that of the SA-200B, with modifications to account for the variation in range. This design uses basically passive thermal control elements with electric heaters/thermostats.  The thermal control flight elements are multilayer insulation, thermal surfaces, thermal conduction control, and sensors.  The active elements are minimized and will be only electric heaters/thermostats.  To minimize heater power thermal louvers may be used. The current design assumes that the spacecraft uses passive thermal control with heaters/thermostats, because it is basically designed for Earth orbit (alternative thermal designs that utilize active elements are also being studied.) 

\subsubsection{Structural Subsystem} 
The current best estimate for the total dry mass is 115~kg including a set of required modifications to the standard SA-200B bus (i.e., a small propulsion system, a 0.5~m HGA for deep-space telecom, etc.) The design calls for launching the two spacecraft on a custom carrier structure, as they should easily fit into the fairing (for instance, Delta II 2425-9.5). The total launch mass for the two spacecraft will be 552~kg. (This estimate may be further reduced, given more time to develop a point design.) 

\subsubsection{Attitude Control Subsystem} 
An attitude control system may be required to have pointing accuracy of 6 $\mu$rad and a pointing knowledge of 3 $\mu$rad. This may be achieved using a star tracker and sun sensor combination to determine attitude together with reaction wheels (RW) to control attitude. Cold-gas jets may be used to desaturate RWs. A Spectrum Astro SA-200B 3-axis stabilized bus with RWs for fine pointing and thrusters for RW desaturation is a good platform \citep{teamx}. 
For the current experiment design it is sufficient to utilize a pointing architecture with the following performance (3$\sigma$, per axis): control 6 $\mu$rad; knowledge 3 $\mu$rad; stability 0.1 $\mu$rad/sec.  The SA-200B readily accommodates these requirements.

\subsubsection{Power Subsystem} 
The flight system will require $\sim$50 W of power. This may be supplied by a 1 square meter GaAs solar cell array. To maintain a constant attitude with respect to the Sun, the solar cells must be deployed away from the body of the spacecraft. This will allow the cells to radiate away its heat to maintain the cells within their operating temperature range.

\subsubsection{Telecommunications Subsystem} 
The telecommunications subsystem will be a hybrid that utilizes the optical communications capability of the instrument as the primary means of transmitting and receiving commands and data. In addition, a small low gain antenna for low data rate radio communications will be used for emergency purposes. This system will use a 15~W transmitter and 10~dB gain antenna. Using X band this system can operate with a 5 bit per second (bps) data rate. The design calls for an SDST X-Band transponder, operating at 15~W; X-Band SSPA; a 0.5~m HGA; two X-Band LGAs pointed opposite each other; a duplexer; two switches; and coax cabling -- the standard options of present day spacecraft design.

\subsubsection{Propulsion Subsystem}	
The propulsion subsystem for SA-200S bus may be used as is. This will ensure a minimal amount of engineering is required. The system is a blowdown monopropellant system with eight 5-N thrusters and two 32~cm tanks each with 22~kg propellant capacity. The system exists and was tested in many applications.


We shall now consider the basic elements of the LATOR optical receiver system.  While we focus on the optics for the two spacecraft, the interferometer has essentially similar optical architecture.

\subsection{Optical Receiver System}
 

\begin{table*}[t!]
\begin{center}
\caption{Summary of design parameters for the LATOR spacecraft optical receiver system.
\label{table:requirements}} \vskip 8pt
\begin{tabular}{rl} \hline\hline
Parameters/Requirements   & Value/Description \\\hline
 & \\[-10pt]
Aperture &  200 mm, unobstructed \\[3pt]
Wavelength & 1064 nm \\[3pt]
Narrow bandpass Filter & 0.05 nm FWHM over full aperture \\[3pt] 
Focal Planes & APD Data \& CCD Acquisition/Tracking \\[3pt]
APD Field of View & Airy disk field stop (pinhole) in front of APD\\[3pt]
APD Field Stop (pinhole) & Approximately 0.009 mm in diameter \\[3pt] 
APD Detector Size & TBD (a little larger than 0.009 mm)\\[3pt] 
CCD Field of View & 5 arc minutes \\[3pt] 
CCD Detector Size & 640 $\times$ 480 pixels (9.6 mm $\times$ 7.2 mm)\\[3pt]
CCD Detector Pixel Size & 15 $\mu$m\\[3pt] 
Beamsplitter Ratio (APD/CCD) & 90/10\\[3pt] 
Field Stop & `D'-shaped at primary mirror focus \\[3pt] 
Lyot Stop & Circular aperture located at telescope exit pupil\\[3pt] 
\hline\hline
\end{tabular} 
\end{center} 
\end{table*}

\begin{figure*}[t!]
 \begin{center}
\noindent    
\psfig{figure=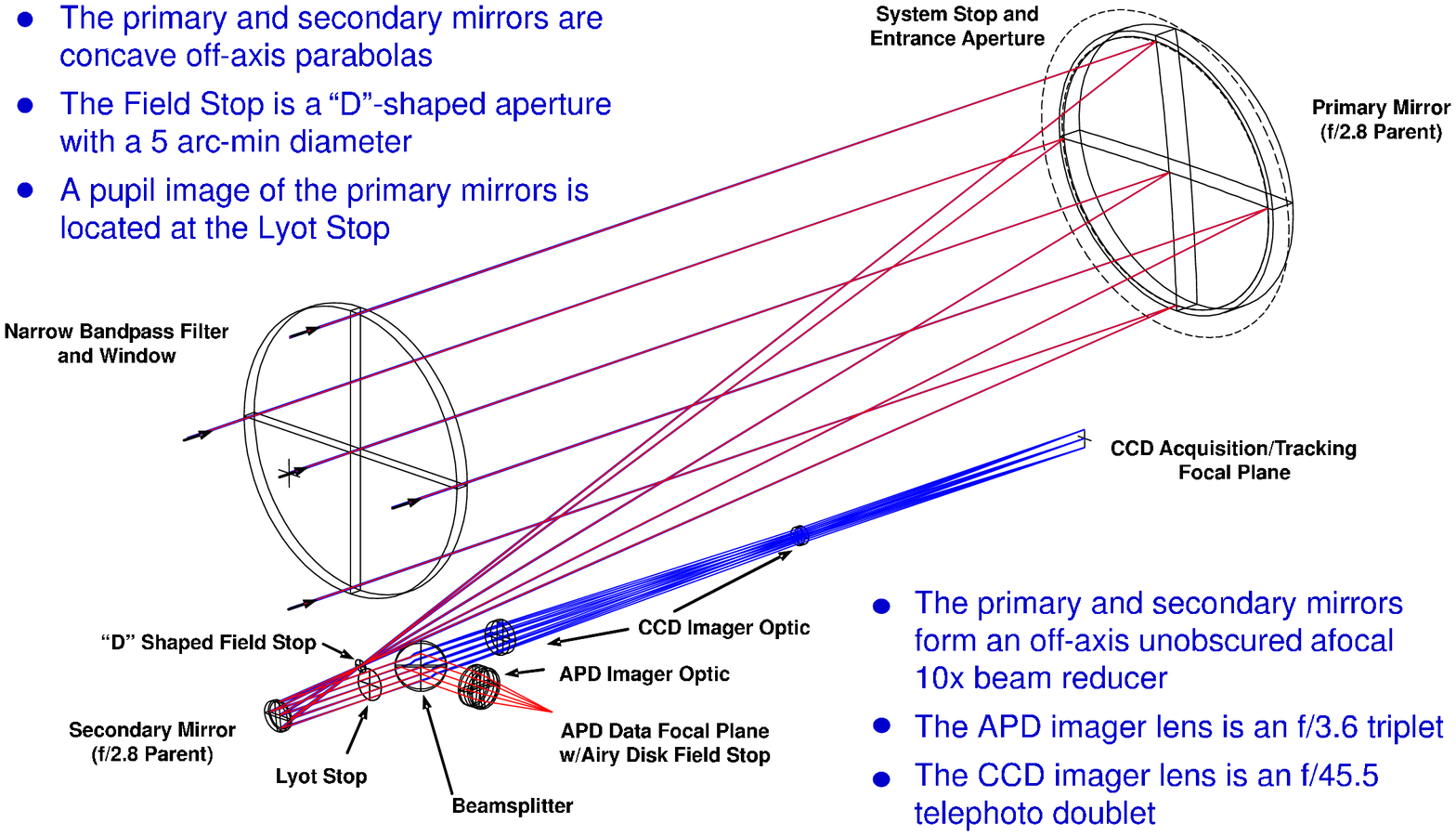,width=164mm}
\end{center}
\vskip -10pt 
  \caption{Layout for LATOR optical receiver system.  
 \label{fig:lator_receiver}}
\end{figure*} 
\begin{figure*}[t!]
 \begin{center}
\noindent    
\psfig{figure=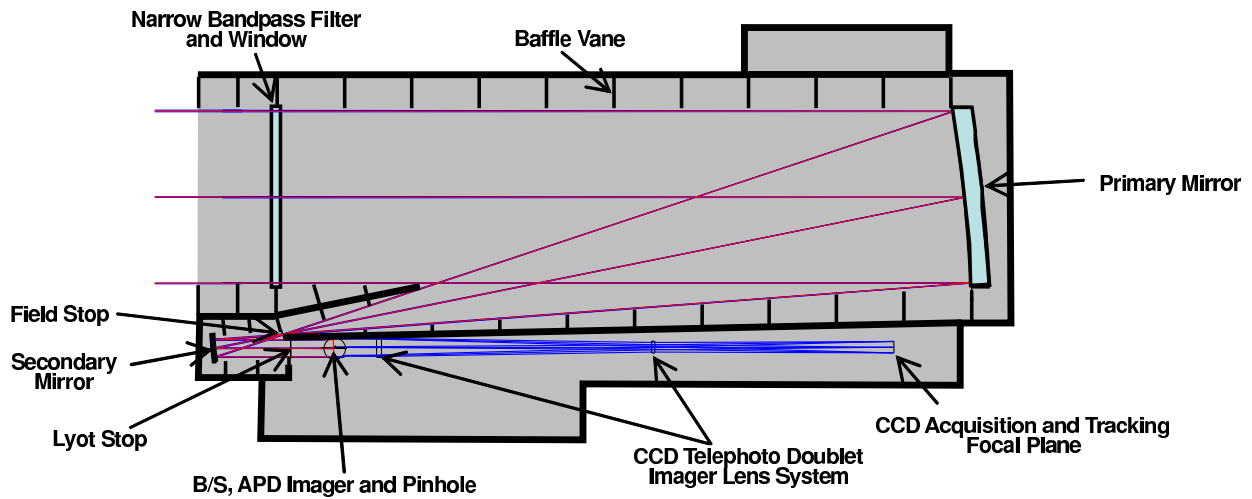,width=164mm}
\end{center}
\vskip -10pt 
  \caption{Preliminary baffle design for LATOR optical receiver system.  
 \label{fig:lator_buffle}}
\end{figure*} 
\begin{figure}[t!]
 \begin{center}
\noindent    
\psfig{figure=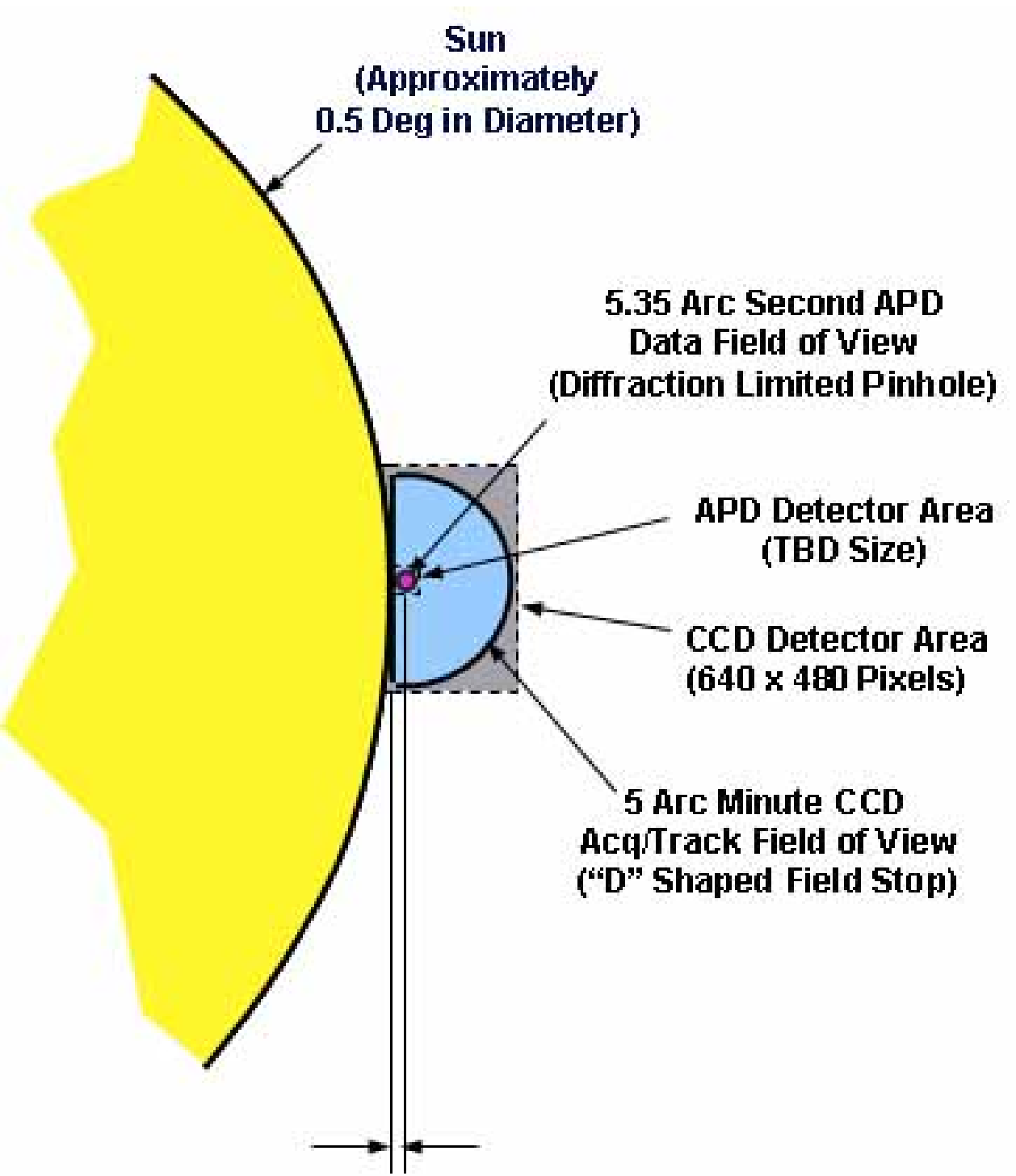,width=82mm}
\end{center}
\vskip -10pt 
  \caption{LATOR focal plane mapping (the diagram not to scale).  
 \label{fig:lator_focal}}
\end{figure} 

The LATOR 200 mm receiver optical system is a part of a proposed experiment. This system is located at each of two separate spacecraft placed on heliocentric orbits, as shown in Figure \ref{fig:lator}. The receiver optical system captures optical communication signals form a transmitter on the ISS, which orbits the Earth. To support the primary mission objective, this system must be able to receive the optical communication signal from the uplink system at the ISS that passes through the solar corona at the immediate proximity of the solar limb (at a distance of no more than 5 Airy disks).

Our recent analysis of the LATOR receiver optical system successfully satisfied all the configuration and performance requirements (shown in Table \ref{table:requirements}) \citep{stanford_ijmpd,stanford_texas,hellings_2005}. We have also performed a conceptual design (see Figure \ref{fig:lator_receiver}), which was validated with a ray-trace analysis. The ray-trace performance of the designed instrument is diffraction limited in both the APD and CCD channels over the specified field of view at 1064 nm. The design incorporated the required field stop and Layot stop. A preliminary baffle design has been developed for controlling the stray light. 

The optical housing is estimated to have very accommodating dimensions; it measures (500 mm $\times$ 220 mm $\times$ 250 mm). The housing could be made even shorter by reducing the focal length of the primary and secondary mirrors, which may impose some fabrication difficulties. These design opportunities are being currently investigated. 

\subsubsection{Preliminary Baffle Design}

Figure \ref{fig:lator_buffle} shows the LATOR preliminary baffle design. The out-of-field solar radiation falls on the narrow band pass filter and primary mirror; the scattering from these optical surfaces puts some solar radiation into the FOV of the two focal planes. This imposes some requirements on the instrument design.  
Thus, the narrow band pass filter and primary mirror optical surfaces must be optically smooth to minimize narrow angle scattering. This may be difficult for the relatively steep parabolic aspheric primary mirror surface. However, the field stop will eliminate direct out-of-field solar radiation at the two focal planes, but it will not eliminate narrow angle scattering for the filter and primary mirror.  Finally, the Lyot stop will eliminate out-of-field diffracted solar radiation at the two focal planes. Additional baffle vanes may be needed at several places in the optical system. 

This design will be further investigated in series of trade-off studies by also focusing on the issue of stray light analysis. Figure \ref{fig:lator_focal} shows the design of the focal plane capping. The straight edge of the ``D''-shaped CCD field stop is tangent to the limb of the Sun and it is also tangent to the edge of APD field stop. There is a 2.68 arcsecond offset between the straight edge and the concentric point for the circular edge of the CCD field stop.  The results of the analysis of APD and CCD channels point spread functions can be found in \citep{stanford_ijmpd,stanford_texas}. 

\begin{figure*}[t!]
 \begin{center}
\noindent    
\psfig{figure=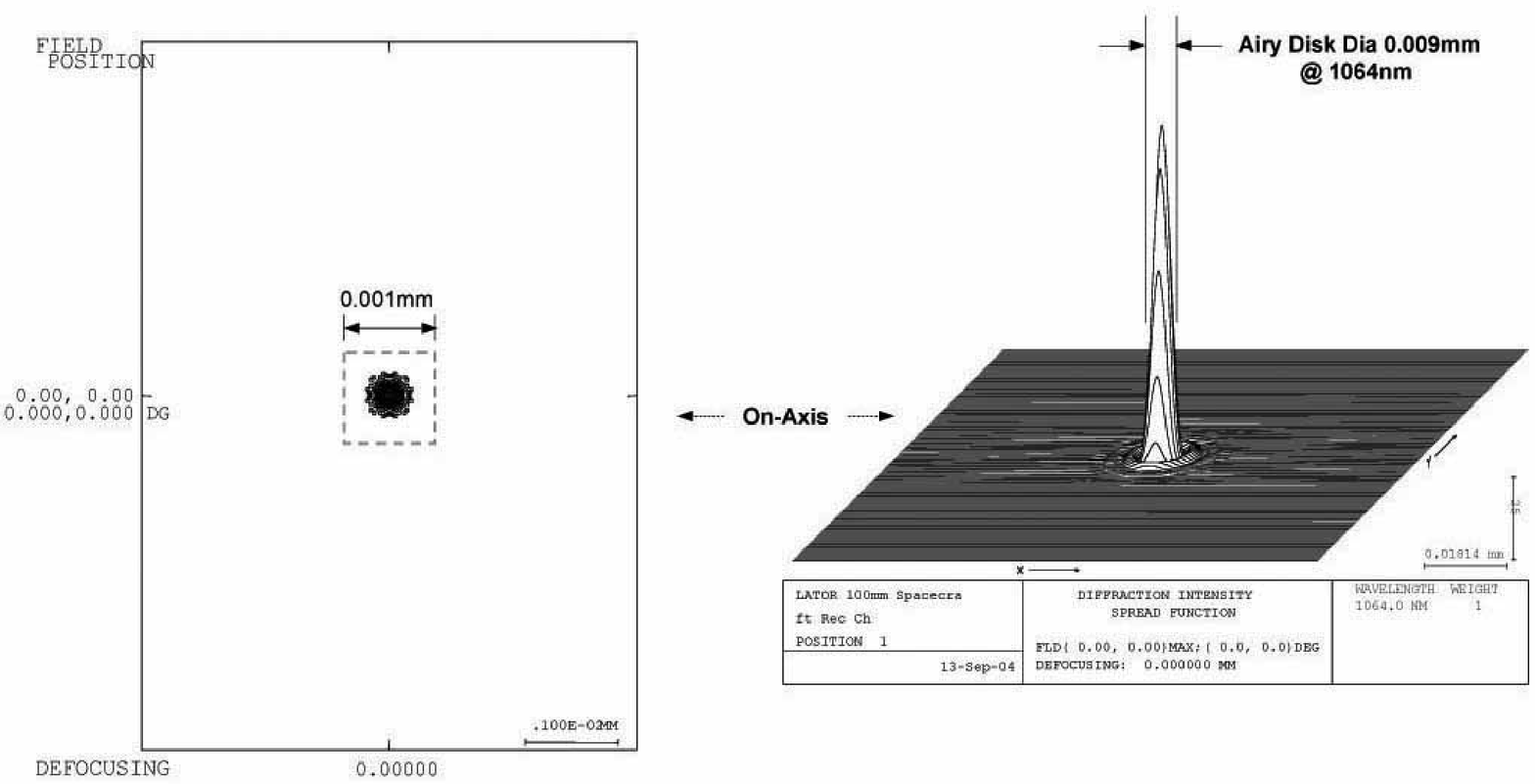,width=158mm}
\end{center}
\vskip -10pt 
  \caption{APD channel geometric (left) and diffraction (right) PSF.  
 \label{fig:lator_apd}}
 \begin{center}
\noindent    
\psfig{figure=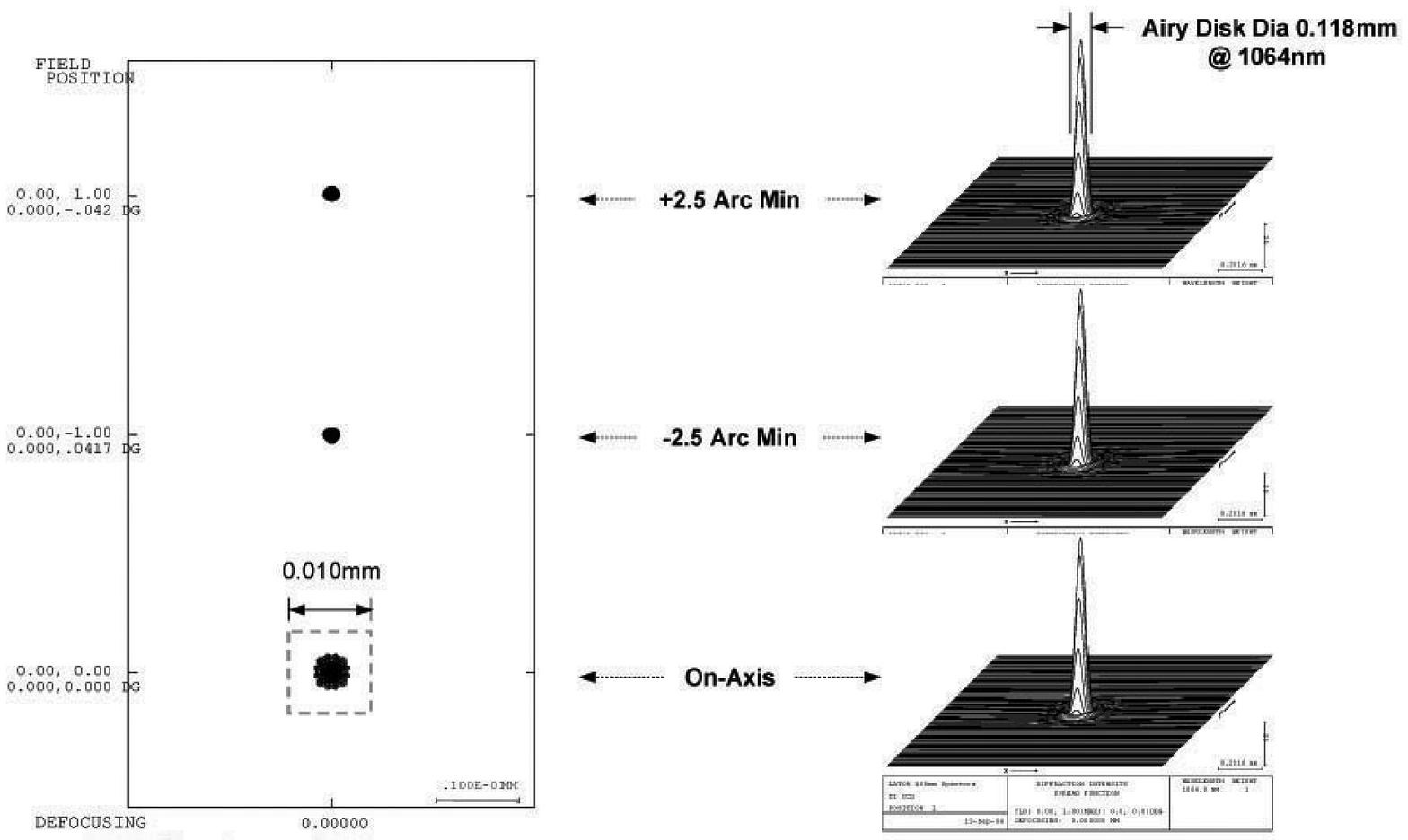,width=158mm}
\end{center}
\vskip -10pt 
  \caption{CCD channel geometric (left) and diffraction (right) PSF.  
 \label{fig:lator_ccd}}
\end{figure*} 

\subsubsection{Focal Plane Mapping}

Figure \ref{fig:lator_focal} shows the design of the focal plane capping. The straight edge of the D-shaped CCD field stop is tangent to the limb of the Sun and it is also tangent to the edge of APD field stop (pinhole). There is a 2.68 arcsecond offset between the straight edge and the concentric point for the circular edge of the CCD field stop (D-shaped aperture). In addition, the APD field of view and the CCD field of view circular edges are concentric with each other. Depending on the spacecraft orientation and pointing ability, the D-shaped CCD field stop aperture may need to be able to be rotated to bring the straight edge into a tangent position relative to the limb of the Sun. 
The results of the analysis of APD and CCD channels point spread functions (PSF) are shown in Figures \ref{fig:lator_apd} and \ref{fig:lator_ccd}.

\subsection{LATOR Coronograph}
\label{sec:coronograph}

\begin{figure}[h!]
 \begin{center}
\noindent  \vskip -5pt   
\psfig{figure=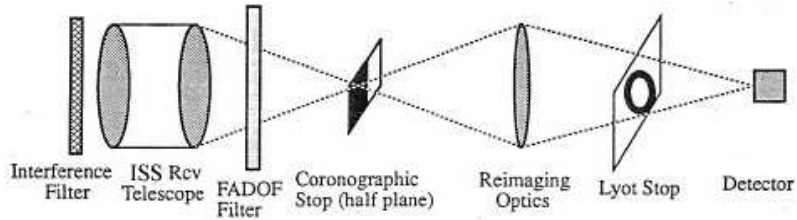,width=105mm}
\end{center}
\vskip -15pt 
  \caption{LATOR coronograph system.
 \label{fig:coronograph}}
\end{figure}

In order to have adequate rejection of the solar background surrounding the laser uplink from Earth, the spacecraft optical system must include a coronagraph. Figure \ref{fig:coronograph} shows a schematic of the coronagraph. A 20 cm telescope forms an image on the chronographic stop. This stop consists of a knife-edge mask placed 6 arcseconds beyond the solar limb. The transmitted light is then reimaged onto a Lyot stop, which transmits 88\% of the incident intensity. Finally, the light is reimaged onto the tracking detector.
 
\begin{figure}[h!]
 \begin{center}
\noindent  \vskip -4pt   
\psfig{figure=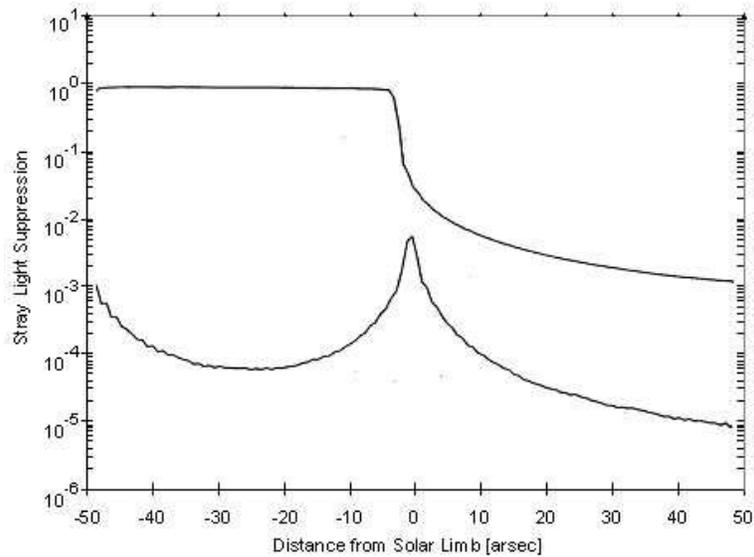,width=100mm}
\end{center}
\vskip -15pt 
  \caption{Results of coronograph performance simulation.
 \label{fig:coronograph_sim}}
\end{figure} 
 
The results of a simulated coronograph showing the stray light rejection as a function of the distance from the solar limb is shown in 
Figure~\ref{fig:coronograph_sim}. The solar surface has been approximated as a vertical edge extending along the entire length of a $256 \times 256$ array. The upper curve shows the stray light levels for an optical system without a coronograph. In this case, the flux from the solar surface has only been decreased by a factor of 100. With the coronograph, however, a further factor of 100 rejection can be achieved. In addition to decreasing the stray solar radiation, the coronograph will decrease the transmission of the laser signal by 78\% (for a signal 12 arcsec from limb) due to coronographic transmission and broadening of the point spread function. At these levels of solar rejection, it is possible for the spectral filter to reject enough starlight to acquire the laser beacon. It is interesting to note that without the coronograph, the stray light from the Sun, decreases proportionally to the distance from the limb. In contrast, with the use of the coronograph, the stray light decreases as the square of the distance from the limb.

\subsection{LATOR Observing Sequence}
\label{sec:operations}

It is important to discuss the sequence of events that will lead to the signal acquisition and that occur during each observation period (i.e., every orbit of the ISS).  This sequence will be initiated at the beginning of the experiment period, after ISS emergence from the Earth's shadow (see Figure~\ref{fig:iss_config}). It assumes that boresighting of the spacecraft attitude with the spacecraft transmitters and receivers have already been accomplished. This sequence of operations is focused on establishing the ISS to spacecraft link. The interspacecraft link is assumed to be continuously established after final deployment (at $\sim15^\circ$ off the Sun), since the spacecraft never lose line of sight with one another.

\begin{table*}[h!]
\begin{center}
\caption{Analysis of various links between ISS and spacecraft in  observation and acquisition modes. \label{tab:link-sc-ISS}}
\vskip 5pt
\begin{tabular}{|l|c|c|c|c|} \hline 
& \multicolumn{2}{c}{}&\multicolumn{2}{|c|}{}\\[-8pt] 
& \multicolumn{2}{c}{Spacecraft-to-ISS Link}
& \multicolumn{2}{|c|}{ISS-to-Spacecraft Link}\\ 
 ~~~~Optical Link Parameters & \multicolumn{2}{c}{spacecraft=Xmit, ISS=Rcv}&\multicolumn{2}{|c|}{ISS=Xmit, spacecraft=Rcv}\\\cline{2-5}
 & &&&\\[-8pt]
 & Acquisition & Observation & Acquisition &Observation \\  \hline \hline 
  &&&&\\[-8pt]
{\bf Transmitter Parameters} &  & & & \\  
~~~~Laser power, W	& 1 &1 &1 & 1\\
~~~~Wavelength, $\mu$m 	& 1.064& 1.064 & 1.064& 1.064 \\
~~~~Xmit telescope diameter, m  & 0.2& 0.2 & 0.3& 0.3 \\
~~~~Beam divergence, $\mu$rad  & 10 asec & 5.32 & 10 asec & 3.55   \\
~~~~Distance ($L=2$~AU), m	     & $3.00 \times 10^{11}$ & $3.00 \times 10^{11}$ & $3.00 \times 10^{11}$ & $3.00 \times 10^{11}$ \\
~~~~Footprint diameter at $L$, m & $1.45 \times 10^{7}$ & $1.59 \times 10^{6}$ & $1.45 \times 10^{7}$ &  $1.06 \times 10^{6}$\\
~~~~Optics efficiency	   & 0.7& 0.7 & 0.7& 0.7 \\  
~~~~Pointing efficiency	   & 0.9& 0.9 & 0.9& 0.9 \\[4pt] 
\hline 
  &&&&\\[-8pt]
{\bf Receiver Parameters} &  &  &&\\  
~~~~Rcv telescope diameter, m   & 0.3 & 0.3& 0.2 & 0.2 \\
~~~~Rcv optics efficiency    & 0.7& 0.7 & 0.7& 0.7 \\
~~~~Detector quantum efficiency	     & 0.9& 0.9& 0.9& 0.9 \\
~~~~Power received, W  & $1.70 \times 10^{-16}$ & $1.41 \times 10^{-14}$  & $7.55 \times 10^{-17}$ &  $1.41 \times 10^{-14}$\\
~~~~Photon flux received, ph/s & $9.10\times 10^2$ & $7.55 \times 10^4$ & $4.04 \times 10^2$ & $7.55 \times 10^4$ \\[4pt] \hline 
  &&&&\\[-8pt]
{\bf Solar Background Parameters} &   & &&\\  
~~~~Solar irradiance, ph/s/m$^{2}$/sr/$\mu$m 
& $4.64 \times 10^{25}$ & $4.64 \times 10^{25}$ 
& $4.64 \times 10^{25}$ & $4.64 \times 10^{25}$  \\
~~~~Rcv detector size, $\mu$rad &3.55 &3.55 & 5.32& 5.32 \\
~~~~Heterodyne spectral bandwidth &  300~MHz &1~MHz  & --- & ---\\
~~~~Narrow-bandpass filter, $\mu$m & --- & --- & $5 \times 10^{-5}$ & $5 \times 10^{-5}$ \\
~~~~Coronograph efficiency& $1 \times 10^{-5}$ & $1 \times 10^{-5}$ & $1 \times 10^{-5}$ & $1 \times 10^{-5}$\\
~~~~Solar photon flux received, ph/s	& $2.95 \times 10^2$  & $9.82 \times 10^{-1}$ & $1.30 \times 10^{4}$& $1.30 \times 10^4$ \\
  & & &&\\[-8pt] \hline  & & &&\\[-8pt] 
{\bf Signal to noise ratio}   &  22.5  &24.7  &  23.2 &21.4 \\
~~~~Integration time  &  1 sec & 10 msec &  60 sec & 10 msec\\\hline  
\end{tabular} 
\end{center}
\end{table*}
\begin{table*}[h!]
\begin{center}
\caption{Spacecraft-to-spacecraft link parameters. \label{tab:link-sc-sc-link}}
\vskip 5pt
\begin{tabular}{|l|c|} \hline  
  &\\[-8pt]
{\bf Transmitter Parameters} &   \\  
~~~~Laser power, W	             & 0.2 \\
~~~~Wavelength, $\mu$m 	 	     & 0.780 \\
~~~~Xmit telescope diameter, m  & 0.05 \\
~~~~Beam divergence, $\mu$rad  & 15.6  \\
~~~~Distance ($r$, 0.5$^\circ$ at 2 AU), m & $2.61 \times 10^{9}$ \\
~~~~Footprint at $r$, m	     & $4.07 \times 10^{4}$\\
~~~~Optics efficiency	     & 0.7\\
~~~~Pointing efficiency	     & 0.9 \\\hline
  &\\[-8pt]
{\bf Receiver Parameters} &    \\  
~~~~Rcv telescope diameter, m   & 0.05 \\
~~~~Rcv optics efficiency    & 0.7 \\
~~~~Detector quantum efficiency	     & 0.9 \\
~~~~Power at detector, W  & $1.20 \times 10^{-13}$ \\
~~~~Photon flux received, ph/s & $4.70 \times 10^5$\\ \hline 
  &\\[-8pt]
{\bf Signal to noise ratio} &   30.1  \\
~~~~Integration time &  10 msec \\\hline  
\end{tabular} 
\end{center}
\vskip -5pt
\end{table*}

The laser beacon transmitter at the ISS is expanded to have a beam divergence of 10 arcsec in order to guarantee illumination of the LATOR spacecraft (see Table~\ref{tab:link-sc-sc-link}). After re-emerging from the Earth's shadow this beam is transmitted to the craft and reaches them in about 18 minutes. At this point, the LATOR spacecraft acquire the expanded laser beacon signal. In this mode, a signal-to-noise ratio (SNR) of 23.2 can be achieved with 60 seconds of integration. With an attitude knowledge of 10 arcsec and an array field of view of 30 arcsec no spiral search is necessary. Upon signal acquisition, the receiver mirror on the spacecraft will center the signal and use only the center quad array for pointing control. Transition from acquisition to tracking should take about 1 minute. Because of the weak uplink intensity, at this point, tracking of the ISS station is done at a very low bandwidth. The pointing information is fed-forward to the spacecraft transmitter pointing system and the transmitter is turned on. The signal is then re-transmitted down to the ISS with a light-travel time of 18 minutes.

Each interferometer station and laser beacon station searches for the spacecraft laser signal. In acquisition mode, the return is heterodyned by using an expanded bandwidth of 300~MHz, to assure capture of the laser frequency. In this case, the solar background is the dominant source of noise, and an SNR of 22.5 is achieved with 1 second integration. Because of the small field of view of the array, a spiral search will take 30 seconds to cover a 30 arcsec field. Upon acquisition, the signal will be centered on the quad cell portion of the array and the local oscillator frequency locked to the spacecraft signal. The frequency band will then be narrowed to 1~MHz and the local oscillator frequency locked to the download laser. In this regime, the solar background is no longer the dominant noise source and an SNR of 24.7 can be achieved in only 10 msec of integration. 
The laser beacon transmitter will then narrow its beam to be diffraction limited ($\sim$1 arcsec) and to point toward the LATOR spacecraft. This completes the signal acquisition phase, and the entire architecture is in-lock and transmits scientific signal.  This procedure is re-established during each 92-minute orbit of the ISS.

The inter-spacecraft optical link budget is given in Table~\ref{tab:link-sc-sc-link}. In this case, the sun is not contributing to the signal to noise analysis, one has to account for the detector's noise contribution only.   

\subsection{Factors Affecting SNR Analysis}

In conducting the signal-to-noise analysis we pay significant attention to several important factors. In particular, we estimate what fraction of the transmitted signal power captured by the 20 cm receiver aperture and analyze the effect of the Gaussian beam divergence (estimated at $\sim 7 ~\mu$rad) of the 30 cm transmit aperture on the ISS. Given the fact that the distance between the transmitter and receiver is on the order of 2 AU, the amount captured is about $2.3\times 10^{-10}$ of the transmitted power.  

We also consider the amount of solar disk radiation scattered into the two receiver focal planes. In particular, the surface contamination, coating defects, optical roughness and substrate defects could scatter as much as $1\times10^{-4}$ or more (possibly $1\times 10^{-3}$)  of the solar energy incident on the receive aperture into the field of view.   These issues are being considered in our current analysis. We also study the amount of the solar corona spectrum within the receive field of view that is not blocked by the narrow band pass filter.  The factors we consider is the filter's FWHM band-pass is 0.05 nm, the filter will have 4.0 optical density blocking outside the 0.05 nm filter band pass from the X-ray region of 1200 nm; the filter efficiency within the band pass will be about 35\%, and the detector is probably sensitive from 300 nm to 1200 nm.  

Additionally, we consider the amount of out-of-field solar radiation scattered into the focal plane by the optical housing. This issue needs to be investigated in a stray light analysis which can be used to optimize the baffle design to minimize the stray light at the focal plane.  Finally, we study the effectiveness of the baffle design in suppressing stray light at the focal plane. Thus, in addition to the stray light analysis, the effectiveness of the final baffle design should be verified by building an engineering model that can be tested for stray light.  

Our recent conceptual design and a CODEV  ray-trace analysis met all the configuration and performance requirements (shown in Table \ref{table:requirements}).  The ray-trace performance of the resulted instrument is diffraction limited in both the APD and CCD channels over the specified field of view at 1064 nm. The design incorporated the required field stop and Layot stop. A preliminary baffle design has been developed for controlling the stray light.  
In the near future, we plan to perform a stray light analysis which should be performed to optimize the baffle design and calculate the amount of stray light that could be present at each of the two focal planes.  This stray light analysis will take into account the optical smoothness of the band-pass filter and primary mirror surfaces. Narrow angle scattering may be a problem at the two focal planes in the filter and primary mirror are not optically very smooth and, thus, it requires a more detailed study. Finally, a rigorous signal-to-noise analysis will be performed to validate the power required to achieve a high signal-to-noise ratio in detecting received beam signal in the presence of the expected focal beam stray light predicted by the stray light analysis and the engineering model stray light tests.

In the next section we will consider the modeling of the LATOR observables  and will discuss the logic of its measurements. 

\section{LATOR Preliminary Observational Model} 
\label{sec:model}

The goal of measuring deflection of light in solar gravity with accuracy of one part in $10^{9}$ requires serious consideration of systematic errors. One would have to properly identify the entire set of factors that may influence the mission accuracy at this level. Fortunately, we initiated this process aided by previous experience in the development of a number of instruments that require similar technology and a comparable level of accuracy \citep{lator_cqg_2004}, notably Space Inerferometry Mission, Keck and Palomar Testbed Interferometers. This experience comes with understanding various constituents of the error budget, expertise in developing appropriate instrument models; it is also supported by the extensive verification of the expected  performance with instrumental test-beds and existing flight hardware. Details of the LATOR error budget are being developed and will be published elsewhere, when fully analyzed. Recent covariance studies confirmed the expected mission performance and emphasized the significant potential of the mission \citep{Ken_lator05,hellings_2005}. 

In this Section we will discuss the LATOR observables, based on a simplified model that will be used to introduce the observational logic of this experiment. We first discuss the model for the relativistic delay of the laser signals as they transit between the nodes of the LATOR's light triangle.  We  then will introduce the model for differential astrometric interferometry to be implemented for the mission.  

\subsection{Relativistic Light-Time Model} 

In development of the mission's error budget we use a simple model to capture all error sources and their individual impact on the mission performance \citep{lator_cqg_2004}. 
The first step into a relativistic modeling of the light path consists of determining the direction of the incoming photon as measured by an observer as a function of the barycentric coordinate position of the light source. 

From a geometrical point of view the Sun, Earth, and other planets each curve space-time in their vicinity to varying degrees. The effect of this curvature is the increase of the round-trip travel time of a laser pulse. Effects of the gravitational monopole on light propagation are the largest among those in the solar system. To the first order in gravitational constant, the one-way relativistic light-time expression was derived in heliocentric form by \cite{Shapiro_1964}; in its most general form it was given by \cite{Tausner_1966} and independently by \cite{Holdridge_1967}. It was formulated in expanded solar-system barycentric form and incorporated into JPL orbit determination software by \cite{Moyer_1977,Moyer_2003}. 

The portion of the Moyer's formulation due to the Sun and Earth is
{}
\begin{equation}
(t_j-t_i) =\frac{r_{ij}^B}{c}+{(1+\gamma)} \Big[\frac{\mu_\odot}{c^3}\ln\Big(\frac{r_i^S+r_j^S+r_{ij}^S+(1+\gamma)\frac{\mu_S}{c^3}}
{r_i^S+r_j^S-r_{ij}^S+(1+\gamma)\frac{\mu_\odot}{c^2}}\Big)+
\frac{\mu_E}{c^3} \ln\Big(\frac{r_i^E+r_j^E+r_{ij}^E}
{r_i^E+r_j^E-r_{ij}^E}\Big)\Big].
\label{eq:shapiro}
\end{equation}
The first term on the right is the geometric travel time due to coordinate separation; the remaining two terms represent the gravitational curvature effects due to the Sun and Earth. The complete equation gives the elapsed coordinate time between two photon events, where an event is indicated by the subscript $i,j\in\{1,3\}$ (with subscript $i=3$ reserved for the ISS).  $\mu_\odot=GM_S$ and $\mu_E=GM_E$ are the solar and Earth's gravitational constants correspondingly.  A Latin superscript denotes the origin of a vector: $B$ is the solar-system barycenter, $S$ is the Sun, and $E$ is the Earth.  The use of the symbols in the equation is as follows. $r_i^S=|{\bf r}^S_i|$ is the magnitude of the vector from the Sun to photon event $i$ transmission (or reception) at coordinate time $t_i$. $\mathbf{r}_{ij}=\mathbf{r}^S_j-\mathbf{r}^S_i$ is the vector and $r_{ij}=|\mathbf{r}_{ij}|$ is the magnitude of the difference between the vector from the Sun to photon event $j$ at time $t_j$ and the vector from the Sun to photon event $i$ at time $t_i$. 

When the ray path is near the solar limb, Eq.~(\ref{eq:shapiro}) is greatly simplified taking the following standard \citep{Will_book93} form
\begin{eqnarray}
(t_j-t_i) =\frac{r_{ij}}{c}+(1 + \gamma) \frac{\mu_\odot}{c^3}\ln\Big(\frac{4r_ir_j}{p^2}\Big).
\label{eq:delay_short}
\end{eqnarray}
For the shortest arm of the triangle, $\ell_{12}$, the solar impact parameter is comparable to the distances involved, thus the relativistic delay in this arm is very small and, for the purposes of theis paper, it can be neglected; thus, we can write $(t_2-t_1)\simeq r_{12}/c$, which is sufficient for our purposes here. 

The obtained equations may be used to model the light paths, $\ell_{ij}=c(t_j-t_i)$, for the signals to transit between all three vortices of the LATOR triangle, as shown in Fig.~\ref{fig:lator}. Indeed, one can write the following approximate expressions:
\begin{equation}
\ell_{3j}={r_{3j}}+(1 + \gamma) \frac{\mu_\odot}{c^2}\ln\Big(\frac{4r_jr_3}{p_j^2}\Big),  \qquad \ell_{12}={r_{12}}.
\label{eq:path}
\end{equation}
\noindent Another observable that will be available to LATOR is the range-rate, ${\dot \ell}_{ij}$ that essentially is a time-derivative of ${\ell}_{ij}(p_j(t))$. The accuracy of the range-rate time series is expected to produce even more accurate results, similar to the situation with recent cassini experiment \citep{cassini_ber,cassini_and}. The impact of the high-accuracy range-rate data on the final accuracy of the experiment is being investigated and results will be reported elsewhere.   

A more rigorous analysis that includes effects of the order $G^2 v/c$ had been initiated at JPL.  The results of this analysis and the  corresponding covariance studies will be reported elsewhere. The recent covariance studies already show very interesting results \citep{hellings_2005,Ken_lator05}. 

\subsection{Interferometric Delay Model} 
\label{sec:grav_defl}

Because a light signal propagating in a gravitational potential is retarded relative to its travel-time in an field-free space, as predicted by general relativity, the computed value for the differential time of arrival of the signals at two telescopes forming an interferometer must be corrected for gravitational effects \citep{Sovers98}.  The LATOR interferometer is highly sensitive to these effects of gravity on light propagation. In this section we will derive the expression for the contribution of the relativistic gravity to the optical path difference (OPD) measured by an interferometer in solar orbit. 

Let us define ${\bf r}_{3(i)}(t_i)$ with $i=1,2$ to be the barycentric positions of the two telescopes of the interferometer, such that ${\bf b}= {\bf r}_{3(2)}-{\bf r}_{3(1)}$ is the interferometer's baseline. 
By keeping only the terms to first order in gravitational constant $G$, the expression Eq.~(\ref{eq:shapiro}) can be used to derive the optical path difference $d_j=\ell_{j3}(\mathbf{r}_3+\mathbf{b})-\ell_{j3}(\mathbf{r}_3)=c(t_{3(2)}-t_j)-(t_{3(1)}-t_j)$ registered by the intreferometer for the light received from the $j$-th source:
{}
\begin{equation}
d_j={r_{j3(2)}-r_{j3(1)}}+{(1+\gamma)}\frac{\mu_\odot}{c^2}\ln\Big[\Big(\frac{r_j+r_{3(2)}+r_{j3(2)}}{r_j+r_{3(1)}+r_{j3(1)}}\Big)\Big(\frac{r_j+r_{3(1)}-r_{j3(1)}}{r_j+r_{3(2)}-r_{j3(2)}}\Big)\Big].
\label{eq:inerf_delay}
\end{equation}
\noindent 
This is the required correction to coordinate time delay due to the solar gravitational monopole to the time of arrival of the light sent by $j$-th spacecraft and received by $(i)$-th telescope.

Eq.~(\ref{eq:inerf_delay}) is the differenced Shapiro time delay for the two telescopes separated by $\mathbf{b}$; it is appropriate for the most general geometry, in which $r_j\approx r_{3(i)} \approx r_{j3(i)}$.  For practical purposes, however, $({r_{j3(2)}-r_{j3(1)}})/{r_{j3(1)}}$ is a small quantity that allows further simplification of Eq.~(\ref{eq:inerf_delay}). Using relations $ r_{3(2)} \simeq r_{3(1)} +({\bf b}\cdot{\hat {\bf r}_{3(1)}})$, ~$r_{j3(2)} \simeq r_{j3(1)} +({\bf b}\cdot{\hat {\bf r}_{j3(1)}}),$ and expressing all quantities at $t_3\equiv t_{3(1)}$, we can present the interferometric delay, $ d_{j}$ as follows:
{}
\begin{equation}
d_j\simeq\frac{({\bf b}\cdot{\hat {\bf r}_{j3}})}
{1-({\bf v}_{3}\cdot{\hat {\bf r}_{j3}})/c}+
(1+\gamma)\frac{\mu_\odot}{c^2} \Big[\frac{r_j+r_{3}-r_{j3}}{2r_jr_{3}}\frac{ {\bf b}\cdot( {\hat {\bf r}_3}+{\hat{\bf r}_{j3}})}{1+(\hat {\bf r}_{j}\cdot\hat {\bf r}_{3})}-\frac{r_j+r_{3}+r_{j3}}{2r_jr_3}\frac{ {\bf b}\cdot( {\hat {\bf r}_3}-{\hat{\bf r}_{j3}})}{1+(\hat {\bf r}_{j}\cdot\hat {\bf r}_{3})}\Big],
\label{eq:delay}
\end{equation} 
\noindent where ${\bf r}_{j3}\equiv{ {\bf r}_{j3(1)}}$ and ${\bf r}_{3}\equiv{\bf r}_{3(1)}$, and ${\bf v}_{3}$ is the barycentric velocity of the interferometer. The obtained expression Eq.~(\ref{eq:delay}) is appropriate for the most general geometry, however, when the ray path is near the solar limb and also the corresponding impact parameters $p_j$ are small compare to the barycentric distances to emmiter and receiver, $p_j/r_j\ll1$, $p_j/r_3\ll1$, it is further simplified, taking the form:
{}
\begin{equation}
d_j\simeq \frac{bp_j}{r_3}
\frac{1}{1-({\bf v}_{3}\cdot{\hat {\bf r}_{j3}})/c}+
(1+\gamma)\frac{2\mu_\odot}{c^2} 
\frac{b}{p_j}\frac{ r_j}{r_3+r_j}.
\label{eq:delay_simple}
\end{equation} 
\noindent
The interferometric delay Eq.~(\ref{eq:delay_simple}) is for the geometry when both transmitter and receiver are at finite and comparable distances from the sun.  Note that, if transmitter is located at a distance far greater then that of the receiver, $r_j\gg r_3$, Eq.~(\ref{eq:delay_simple}) transforms to a typical expression for VLBI observations with sources being at infinity \citep{Will_book93,Sovers98}. Therefore, for a typical LATOR geometry with $r_j\approx r_3$, the magnitude of the relativistic delay (and corresponding angle) is approximately twice smaller when compared to the case when the light source is at infinity.

As we discussed in Section~\ref{sec:heterod_interf}, the LATOR interferometer reaches its highest accuracy in differential mode by accurately measuring differential OPD, $d_{12}=d_2-d_1$, which is given from Eq.~(\ref{eq:delay_simple}) as:
{}
\begin{equation}
d_{12}\simeq \frac{b}{r_3}\Big[
\frac{p_2}{1-({\bf v}_{3}\cdot{\hat {\bf r}_{23}})/c}-
\frac{p_1}{1-({\bf v}_{3}\cdot{\hat {\bf r}_{13}})/c}\Big]+(1+\gamma)\frac{2\mu_\odot}{c^2} 
\Big[\frac{b}{p_2}\frac{r_2}{r_3+r_2}-\frac{b}{p_1}\frac{r_1}{r_3+r_1}\Big].
\label{eq:diff_delay}
\end{equation}
Note that interferometric delay rate is another observable that will be available for LATOR. Having determined the expression for delay Eq.~(\ref{eq:diff_delay}), the delay rate, ${\dot d}_{12}$, it is simply given as time-derivative of the delay, $d_{12}(p_j(t))$. Currently, we are investigating the impact of this observable on the experiment and will incorporate ${\dot d}_{12}$ in our further studies. 

Eq.~(\ref{eq:diff_delay}) that captures the largest terms in the model of LATOR interferometric observations. The entire LATOR model accounts for a significant number of other effects, including those due to gravitational multipoles, second order deflection, angular momentum contribution, and etc. The work to develop a complete mission model work had being initiated; the results will be reported elsewhere. 

In the next section we will consider on a conceptual formulation of the LATOR observables.

\subsection{Logic of LATOR Observations} 
\label{sec:logic} 

In this section, we discuss the observational logic of the LATOR experiment.  This is done to only conceptually demonstrate the features of the mission design.

The range observations Eq.~(\ref{eq:path}) may be used to measure any angle between the three fiducials in the triangle. However, for observations in the solar gravity field, measuring the lengths do not give you a complete information to determine the angles, and some extra information is needed. This information is the mass of the Sun, and, at least one of the impact parameters. Nevertheless, noting that the paths $\underline{\ell}_{ij}$ correspond to the sides of the connected, but gravitationally distorted triangle, one can write $\underline{\ell}_{12}+\underline{\ell}_{23}+\underline{\ell}_{31}=0$, where $\underline{\ell}_{ij}$ is the null geodesic path for light moving between the two points $i$ and $j$. This leads to an expression for the angle between the two spacecraft computed from the range measurements; using Eqs.~(\ref{eq:path}) this quantity may be given as
{}
\begin{equation}
\sin(\angle {\underline\ell}_{31}{\underline\ell}_{32})\simeq\sin\alpha_3-(1+\gamma)\frac{\mu_\odot}{c^2} \frac{r_{12}}{r_{31}r_{32}} 
\Big[\cos\alpha_1\ln\frac{4r_3r_1}{p_1^2} +\cos\alpha_2\ln\frac{4r_3r_2}{p_2^2}\Big]\cot\alpha_3,
\label{eq:angle_range}
\end{equation}
where the three Euclidian angles within the LATOR  light triangle are computed from the laser ranging measurements of the three arms of the triangle, $r_{12}, r_{31}, r_{32}$ using usual formulae $\cos\alpha_3=(r_{32}^2+r_{31}^2-r_{12}^2)/(2 r_{32} r_{31})$, 
$\cos\alpha_1=(r_{31}^2+r_{12}^2-r_{32}^2)/(2 r_{31} r_{12})$, and 
$\cos\alpha_2=(r_{32}^2+r_{12}^2-r_{31}^2)/(2 r_{32} r_{12})$ and $\sin\alpha_3$ is given as below:
{}
\begin{equation}
\sin\alpha_3=\frac{1}{2 r_{31} r_{32}}
\Big[2r_{12}^2(r_{31}^2+r_{32}^2)-(r_{32}^2-r_{31}^2)^2-r_{12}^4\Big]^{1/2}.
\label{eq:sin_alpha_3}
\end{equation}
LATOR will provide highly-accurate time-series of $r_{12}, r_{31}, r_{32}$, that can be used to simulate measurement of $\sin(\angle {\underline\ell}_{31}{\underline\ell}_{32})$. For a typical LATOR configuration, Eq.~(\ref{eq:angle_range}) can be approximated as:
{}
\begin{equation}
\sin(\angle {\underline\ell}_{31}{\underline\ell}_{32})\simeq \frac{r_{12}}{2r_{3}}-(1+\gamma)\frac{\mu_\odot}{c^2} \frac{r_{12}}{4r_3^2} \ln\frac{4r_3^2}{p_1p_2}\approx0.01745-\frac{1}{2}(1+\gamma)~1.82\times 10^{-9},
\label{eq:angle_range_lator_0}
\end{equation}
\noindent where we evaluated the expression for a typical set of numerical values suitable for LATOR with one spacecraft is at the solar limb: $r_1\sim r_2\sim r_3\sim 1$~AU, $p_1=R_\odot\equiv 0.2667^\circ,$ $\alpha_3=1^\circ$, $p_2=2r_3\sin(\angle p_1+\alpha_3)$, and $r_{12}=2r_3\sin \alpha_3$.  

Eq.~(\ref{eq:angle_range_lator_0}) has an immediate impact on the logic of the LATOR observations. In particular, it follows from this equation that the smallness of the opening angle $\alpha_3$ making it very difficult to measure $\gamma$ to a sufficient accuracy using only laser ranging observables. Thus, if one desires to achieve accuracy of  $\sigma_\gamma\sim10^{-8}$ in measuring the PPN $\gamma$, one needs to know  $r_{12}=5\times 10^6$ km to accuracy of $\Delta r_{12}\sim 2~\mu$m. This ranging accuracy is certainly achievable with modern technologies, but would rely on drag-free spacecraft and very precise clocks. Notably, in the next decade LISA (i.e. Laser Interferometer Space Antenna) would be able to achieve the accuracy of few pm over the same distance by utilizing a complex scheme of coherent detection of light.  However, this is not what LATOR is going to do. The experiment uses time-of-flight laser ranging to measure distances between the nodes of the light triangle.  These observations are then used to directly compute the opening angle $\angle {\underline\ell}_{31}{\underline\ell}_{32}$ for further input in the astrometric observations.  Below we shall discuss the interferometric component of the mission model. 

Differential astrometric observations Eq.~(\ref{eq:diff_delay}) will be used to obtain another measurement of the same angle $\angle {\underline\ell}_{31}{\underline\ell}_{32}$  between the two spacecraft. The LATOR interferometer will perform differential observations between the two sources of laser light, measuring the differential delay $d_{12}=d_2-d_1$ to the high accuracy. Note that, for a typical LATOR configuration, $r_1\sim r_2\sim r_3$ and $p_2\simeq p_1+r_3\sin\alpha_3$ and one can benefit from the following approximate relation:
{}
\begin{equation}
\sin(\angle {\underline\ell}_{31}{\underline\ell}_{32})\simeq \sin\alpha_3\simeq \frac{p_2-p_1}{r_3}
\simeq \frac{r_{12}}{2r_{3}}.
\label{eq:angle_range_lator_1}
\end{equation}
This is exactly the angle that will be measured interferometrically.  However, the obtained relation (\ref{eq:angle_range_lator_1}) may be used to provide a direct geometric measurement of the opening angle $\angle {\underline\ell}_{31}{\underline\ell}_{32}$ to the required accuracy, which would aid the LATOR astrometric interferometry.

We will use result (\ref{eq:angle_range_lator_1}) to further simplify the expression for the differential delay, Eq.~(\ref{eq:diff_delay}) which for a typical LATOR geometry and circular motion of the interferometer takes the following form:
\begin{equation}
d_{12}\simeq b \Big[\frac{r_{12}}{2r_{3}}-\frac{v_3}{c}\frac{p_3^2-p_1^2}{r_3^2}+(1+\gamma)\frac{\mu_\odot}{c^2} 
\big(\frac{1}{p_2}-\frac{1}{p_1}\big)\Big].
\label{eq:diff_delay_lator}
\end{equation}

The LATOR interferometer is designed to provide highly-accurate time-series of the interferometric delay $d_{12}$, while laser ranging will determine the light travel times between the nodes of the triangle,  $\Delta t_{ij}= t_j-t_i$ (that will be used to compute $r_{ij}=c\Delta t_{ij}$).  Evaluating expression Eq.~(\ref{eq:diff_delay_lator}) for a typical LATOR configuration with one spacecraft is at the solar limb and $v_3=30$~km/s, one finds the characteristic sizes of the effects: 
{}
\begin{equation}
d_{12}\simeq \big(1.745 -4.675\times 10^{-6} -
\frac{1}{2}(1+\gamma)~3.352\times 10^{-4}\big)~{\rm m}.
\label{eq:diff_delay_quant}
\end{equation}
Optical interferometry is a mature technology that may achieve very high accuracy in measuring the quantities involved in Eq.~(\ref{eq:diff_delay_quant}). 
Thus, a measurement of the delay  $d_{12}$ with uncertainty of 5 pm results in the accuracy in the parameter $\gamma$ of $\Delta \gamma=2\times 10^{-8}$. Currently we are able to measure  delays on large interferometric baselines with accuracy $\sim1$~pm which certainly benefit LATOR by satisfying its primary objective.   

As is evident from Figure \ref{fig:lator}, the key element of the LATOR experiment is a redundant-geometry optical truss to measure the effects of gravity on the laser signal trajectories.  LATOR will generate four time series of measurements: one for the optical range of each side of the triangle, plus the angle between light signals arriving at one vertex of the light triangle.  Within the context of a moving Euclidean light triangle, these measurements are redundant.  From a combination of these four times series of data, the several effects of gravity on the light propagations can be precisely and separately determined.  For example: the first and second order gravity monopole deflections go as $p^{-1}$ and $p^{-2}$ while the solar quadrupole deflection goes as $p^{-3}$, with $p(t)$ being a laser signal's evolving impact parameter; the quadrupole moment's deflection has further latitude dependence if spacecraft lines of sight are so located.  

The data will be taken over periods in which the laser light's impact parameters $p(t)$ vary from one to ten solar radii, producing time signatures in the data which permits both the separation of the several gravitational effects and the determination of key spacecraft location coordinates needed to calibrate the deflection signals. In our mission simulations we use the complete set of LATOR observables, including range and range-rates and also delay and delay-rates. These additional data types account for the temporal evolution of the entire LATOR light triangle and further improve the mission accuracy. To demonstrate our design considerations, in the discussion below we will use only equations for range and interferometric delay given by Eqs.~(\ref{eq:path}) and (\ref{eq:diff_delay})  correspondingly.  

In the next section we will discuss the expected performance of the experiment and will present criteria for the mission design. 

\section{LATOR Astrometric Performance}
\label{sec:error_bud}

It is convenient to present error sources in three broad categories: i)~the ones that are related to mission architecture, ii) those that are external to the triangle and have an astrophysical origin, and iii) those that  originated within the instrument itself. 
Typical mission-related errors are those that result from the uncertainties in the orbits of the spacecraft and the ISS,  chosen mission design and observing scenario and, in general, those errors that result from the  geometry of the experiment and affect the range and angle determination. 
The astrophysical errors are those that are external to the instrument and are due to various phenomena that influence both  mission planning and observing scenario. Such errors are due to the gravity effects of planets and largest asteroids,  unmodeled motion of the fiducial stars, optical properties of the Sun and solar corona near the limb, etc. 
Example of the instrumental errors include effects of the long term laser stability, errors in pointing of the laser beams, instrumental drifts and other systematic and random errors originating within the instrument itself. (By instrument we understand the experimental hardware situated at all three vortices of the triangle.)

We consider $\Delta \gamma=2\times 10^{-8}$ to be the accuracy of determining the PPN parameter $\gamma$ in a single measurement. This design accuracy drives the flow-down of mission requirements that we will discuss in this section. 
Thus, based on the Eq.~(\ref{eq:diff_delay_lator}), such a design accuracy results in the following requirement on the accuracy of the OPD measurements:  
{}
\begin{equation}
\Delta d_{12}\simeq \Delta\gamma \,\frac{\mu_\odot}{c^2}\frac{b(p_2-p_1)}{p_1p_2}= 5~{\rm pm} \,\Big(\frac{\Delta\gamma}{2\times 10^{-8}}\Big) \Big(\frac{b}{100~{\rm m}}\Big)\Big(\frac{R_\odot}{p_1}\Big)\Big(1-\frac{p_1}{p_2}\Big).
\label{eq:acc_opd}
\end{equation}
Therefore, in our design considerations we take $\Delta d_{12}=5$~pm to be the target accuracy for the interferometric measurements on the LATOR's 100 m baseline. 

Here we discuss a preliminary astrometric error budget for the LATOR experiment and  present design considerations that enable the desirable instrument performance. (A more detailed model to the second order in gravitational effects is available and is being used in simulations to verify the expected mission performance.) Our goal for this section is to present a comparative analysis of the accuracy that mission needs to achieve in order to satisfy its science requirements. 

\subsection{Trajectory Measurement Accuracy}
\label{sec:error_bud_traj}

In this section we will discuss the constituents of LATOR error budget for the measurement of the angular deflection light. 
The error budget is subdivided into three components -- range and interferometer measurements, and spacecraft stability which are described in this Section. 

\subsubsection{Range Measurement} 
 
This component describes the angular errors due to uncertainties in the distance between the spacecraft and the ISS as they determined by laser ranging. The angular uncertainty due to an inter-spacecraft ranging error, $\Delta r_{12}$, is
{}
\begin{equation}
\Delta r_{12} \simeq \Delta\gamma \,\frac{\mu_\odot}{c^2}\frac{2r_{3}(p_2-p_1)}{p_1p_2}=1~{\rm cm} \,\Big(\frac{\Delta\gamma}{2\times 10^{-8}} \Big)\Big(\frac{r_3}{\rm AU}\Big)\Big(\frac{R_\odot}{p_1}\Big)\Big(1-\frac{p_1}{p_2}\Big).
\label{eq:acc_r12}
\end{equation}

\noindent Therefore, the experiment will require a  spacecraft-to-spacecraft laser ranging accuracy of 1 cm, which for spacecraft separated by 1$^\circ$ at 2~AU distance,  results in an angular error of 0.03 prad (i.e. corresponding to a delay uncertainty of $\sigma_d=3$~pm). 

The angular error due to an ISS-to-spacecraft ranging error is given by
{}
\begin{equation}
\Delta r_{31} \simeq \Delta\gamma \,\frac{\mu_\odot}{c^2}
\frac{4r_{3}^2}{r_{12}}\frac{(p_2-p_1)}{p_1p_2}=60~{\rm cm} \,\Big(\frac{\Delta\gamma}{2\times 10^{-8}} \Big)\Big(\frac{\sin 1^\circ}{\sin\alpha_3}\Big)\Big(\frac{r_3}{\rm AU}\Big)\Big(\frac{R_\odot}{p_1}\Big)\Big(1-\frac{p_1}{p_2}\Big).
\label{eq:acc_r31}
\end{equation}
 
\noindent Therefore, we have allocated a 60 cm range uncertainty for each of the two ISS-to-spacecraft laser links, results in an angular uncertainty of 0.035 prad (i.e. $\sigma_d=3.5$~pm). The total error budget for the laser ranging distance measurements is $4.6$~pm. 

\subsubsection{Knowledge of the Baseline}
 
Experiment uncertainties in the ISS interferometer measurement contribute additional terms to the overall error budget.   The baseline design for the instrument is a 100~m baseline with a 10 s integration time. Using Eq.~(\ref{eq:diff_delay_lator}), one obtains the requirements on the accuracy of the  baseline estimation:
{}
\begin{equation}
\Delta b \simeq \Delta\gamma \,\frac{\mu_\odot}{c^2}\frac{2r_{3}}{r_{12}}\frac{b(p_2-p_1)}{p_1p_2}= 0.2~{\rm nm}\,\Big(\frac{\Delta\gamma}{2\times 10^{-8}} \Big)\Big(\frac{\sin 1^\circ}{\sin\alpha_3}\Big)\Big(\frac{b}{100~{\rm m}}\Big)\Big(\frac{R_\odot}{p_1}\Big)\Big(1-\frac{p_1}{p_2}\Big).
\label{eq:acc_b}
\end{equation}
Based on theoretical predictions for narrow angle measurements, we require an angular error of 0.025 prad (i.e. $\sigma_d=2.5$~pm), limited by long-term instrument systematics. 

The current requirement for systematic errors in the instrument has been set at 0.05 prad. This corresponds to measurement of the laser fringe phase to 1 part in $2.9\times10^5$ ($\lambda =1.064 ~\mu$m, $b = 100$~m). This term includes errors in the metrology and fringe detection of the interferometer, as well as the effect of photon noise. 

\subsubsection{Orbit Stability of the Spacecraft}

In order to determine the first and second order terms of gravitational deflection, LATOR will make a number of measurements at different spacecraft separations and various impact parameters.  During the period between measurements, it is assumed that the impact parameter is known. An error in this assumption will cause an equivalent error in the computation of the deflection term. Thus, it is important to set a requirement on the knowledge of the impact parameter, which is given by
{}
\begin{equation}
\Delta p_1 \simeq \Delta\gamma \,\frac{1}{2}\frac{p_1p_2}{p_1+p_2}=5.75~{\rm m} \,\Big(\frac{\Delta\gamma}{2\times 10^{-8}}\Big)\Big(\frac{p_1}{R_\odot}\Big)\Big(\frac{1}{1+p_1/p_2}\Big).
\label{eq:acc_p1}
\end{equation}

\noindent This is a highly requirement if one conducts only a static measurement of the light deflection. In our case, the delay-rate observable and smooth motion of the spacecraft significantly reducing the sensitivity of the experiment on the absolute knowledge of the impact parameter $p_1=p_1(t_0)+{\dot p}_j (t-t_0)$. In fact, LATOR will have a very good orbit determined by the combination of the laser ranging and conventional radio-metric navigation, that will provide ${\dot p}_j$ to a high accuracy. This allows one can solve for the initial impact parameter $p_1(t_0)$ in the numerical analysis. 

Therefore, one might require that the spacecraft be stable to 4 $\mu$as, which corresponds to a drift in the transverse distance of 5.8 m and results in an angular error of 0.04 prad.

Similarly, the separation between the spacecraft should also be stable to keep the knowledge of the impact parameters. This is quantified by the allowable uncertainty in the difference between impact parameters $\delta p=p_2-p_1$ that is given by  
{}
\begin{equation}
\Delta \delta p \simeq \Delta\gamma \,\frac{p_2}{2p_1}\,(p_2-p_1)=124.0~{\rm m}\,\Big(\frac{\Delta\gamma}{2\times 10^{-8}}\Big)\Big(\frac{p_2}{p_1}\Big)\Big(\frac{r_3}{\rm AU}\Big)\Big(\frac{\sin\alpha_3}{\sin1^\circ}\Big).
\label{eq:acc_Dp12}
\end{equation}
\noindent Therefore, we require that the spacecraft be stable to 0.05 mas, which results in an angular error of 0.035 prad (i.e. $\sigma_d=3.5$~pm of corresponding delay uncertainty).

\subsubsection{Orbit Stability of the ISS} 

In addition to the spacecraft error, the ISS's orbital error will also produce contribution to the angular measurement. Most of the errors on the ISS can be made common-mode; therefore, their influence on the differential astrometry with LATOR interferometer will be either negligible  or it will be small and well modeled. However, there are some errors that would still produce measurable contribution to the differential delay, if not properly addressed; notably, the accuracy of the ISS orbit.  The current mission design calls for an enhancement of the ISS orbit solution by utilizing GPS receivers at the location of each optical (see Section~\ref{sec:operations}). This will also help to address the issue of the extended structure low-frequency vibrations of the ISS. As we mentioned above, the effect of these vibrations will be addressed by using $\mu$-g level accelerometers, that will be integrated within both optical packages on the ISS. A combination of the GPS receivers and $\mu$-g accelerometers will provide information needed to improve the ISS attitude information; this improvement will be done for each cornercube fiducial (needed for the interferometric baseline determination). 

Our current error budget for the differential observations with the LATOR interferometer allocates $\sim 2.7 $~pm of error in 100 s of integration for the uncertainty in the ISS orbit, its attitude and the extended structure vibrations.

\subsection{Mission Errors}

Although one could in principle set up complicated engineering models to predict all or each of the systematics, often the uncertainty of the models is too large to make them useful, despite the significant effort required.  A different approach is to accept our ignorance about a non-gravitational acceleration and assess to what extent these can be assumed a constant bias over the time scale of all or part of the mission. (In fact, a constant acceleration produces a linear frequency drift that can be  accounted for in the data analysis by a single unknown parameter.)  In fact, we will use both approaches. 

In  most orbit determination programs some effects, like the
solar radiation pressure, are  included in the set of  routinely 
estimated parameters.  Nevertheless we  want to demonstrate their
influence on LATOR's navigation from the general physics standpoint. This is not only to validate our results,  but also to be a model as to how to study  the influence of the other physical phenomena that are not yet included in the standard navigational packages  for future more demanding missions.  Such missions will involve either spacecraft that will be distant or spacecraft  at shorter distances where high-precision spacecraft navigation will be required. 

In the current design, the LATOR experiment requires that the location of one of the spacecraft with respect to the Sun is known to within 20~m over the duration of the each observing session or $\sim $ 92 min. The major perturbation to the spacecraft trajectory is from local spacecraft disturbances, such as gas leaks for thruster valves and solar radiation pressure. The spacecraft can be designed to eliminate spacecraft errors leaving solar radiation pressure as the major source for the position noise. Other disturbances such as solar wind, magnetic fields, cosmic rays, etc. have been identified and are at least three orders of magnitude lower than solar radiation pressure. 

In this section we will discuss possible systematics generated  external to the spacecraft which might affect the LATOR's mission accuracy.  
 
\subsubsection{Direct Solar Radiation Pressure}
\label{sec:solar_plasma}

There is an exchange of momentum when solar photons impact the
spacecraft and are either absorbed or reflected.  Models for 
this solar pressure effect are usually developed before a mission is launched. The models take into account various parts of the spacecraft exposed to solar radiation; they compute the acceleration directed away from the Sun as a function of spacecraft orientation and solar distance: 
\begin{equation}
a_{\tt s.p.}(r)=\frac{2 f_\odot A }{c~m}
\frac{ \cos\theta(r)}{r^2},
 \label{eq:srp}
\end{equation}
\noindent where $f_\odot=1367 ~{\rm W/m}^{2}$(AU)$^2$ is the 
(effective-temperature Stefan-Boltzmann) 
``solar radiation constant'' at 1 AU from the Sun and $A$ is the effective  size of the craft as seen by the Sun.   $\theta$ is  the angle between the axis of the antenna and the direction of the Sun, $c$ is the speed of light, $M$ is the mass of the spacecraft, and $r$ is the distance from  the Sun to the spacecraft in AU.   
For expected spacecraft values of $A= 1.0 ~{\rm m}^2$ and $m = 150 ~{\rm kg}$, gives an acceleration of $a_{\tt s.p.} \simeq 6.1 \times 10^{-8}~{\rm  m}/{\rm s}^2$ at $r=1$~AU from the Sun.

This acceleration will produce an unmodeled force, which ultimately may result in the error in the radial position of the spacecraft. Over a time $t$ this error is 
$\delta r = \frac{1}{2}\,\delta a\,t^2$ 
where  ${\delta a}$ is the unmodeled acceleration. In turn, this error would lead to a transverse position
error of $\delta x = \frac{1}{4}\,\delta a \, n \, t^3$ 
where $n$ is the spacecraft velocity about the Sun, $n\sim 2 \times 10^{-7}$~rad/s.
If the effect of solar radiation pressure were completely unmodeled, over a period of 21 days, the transverse position error due to solar pressure would be $\sim$~18.2 km, potentially resulting in a 61 nrad  astrometric error. (This is one of the reasons to consider a drag-free spacecraft for the experiment as suggested in \citep{hellings_2005}.) However, if one conduct laser ranging with position knowledge of 60 cm, the transverse position uncertainty over a period of 21 day would only be $\sim$10 cm.  Consequently, it is necessary to use the laser ranging information to predict the transverse position of the spacecraft. 

\begin{figure}[h!]
 \begin{center}
\noindent  \vskip -5pt   
\psfig{figure=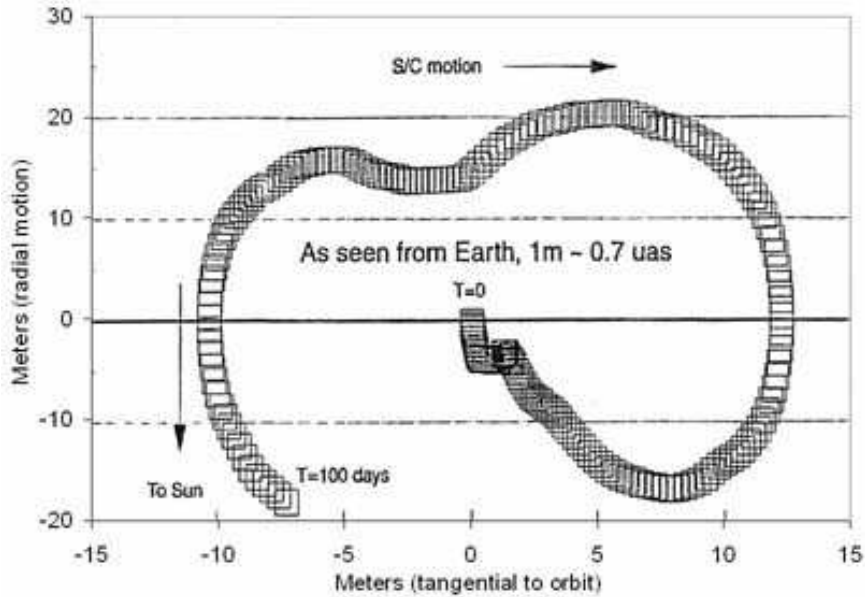,width=115mm}
\end{center}
\vskip -15pt 
  \caption{Simulation of spacecraft motion due to random solar pressure fluctuations.
 \label{fig:solar-rad}}
\end{figure} 

The laser ranging information will be used to solve for the slowly varying changes in the solar pressure leaving the random fluctuations of the solar pressure as the dominant source of position error. Figure~\ref{fig:solar-rad} shows one realization of the position error of the spacecraft.  In this simulation, random fluctuations correspond to 1\% {\small RMS} of the total solar radiation pressure, the spacecraft wanders $\sim$~1~m in the radial direction and 8~cm in the transverse direction in a day.
The use of a redundant optical truss offers an excellent alternative to an ultra-precise orbit determination. This feature also makes LATOR insensitive to spacecraft buffeting from solar wind and solar radiation pressure. This is why, as opposed to other gravitational missions, there is no need for a drag-free spacecraft to enable the high accuracy of the LATOR experiment. (The drag-free option was studied in \citep{hellings_2005} resulting in several interesting design considerations that will be further investigated.) 

\subsubsection{Effect of the Solar Corona}
\label{sec:solar_corona}

The electron density and density gradient in the solar atmosphere  influence the propagation of light and radio-waves through the medium. In fact, radiometric observables will experience effect of the electron density in the solar corona \citep{Brynjolfsson-05}, however, effect on laser ranging and astrometric observations will be significantly less. The use of optical wavelengths offer a significant advantage for the spacecraft communication in the solar system as opposed to the microwave radiation -- the current navigation standard. Such a choice makes the deep space communication effectively free from the solar corona noise. Indeed, the solar plasma effects on wave propagation decrease as $\lambda^2$ and there is a factor of $10^{10}$ reduction in the solar plasma optical path fluctuations by simply moving from the S-band microwave signal $\lambda=10$~cm ($f= 3$~GHz) to the optical wavelengths of $\lambda\sim1 ~\mu$m ($f= 300$ THz). This $10^{10}$ reduction of the dispersive media effects offers tremendous gain in the quality of both spacecraft navigation (increased pointing precision and timing) and deep space communication (very high data transmission rates). LATOR will utilize design capable of rejecting background solar noise in combination with optical  wavelengths for precision navigation; this combination will lead to a significant reduction of the solar corona effect, making its contribution harmless to the mission.

For the navigation purposes, both LATOR spacecraft will be equipped with X-band transponders with both Doppler and range capabilities. The electron density and density gradient in the solar atmosphere will influence the propagation of radio waves through the medium. So, both range and Doppler observations at X-band will affected by the electron density in the interplanetary medium and outer solar corona. This would result in the spacecraft not being able to communicate with the ground when the impact parameter will be less than $\sim 2.5R_\odot$.  This is why the current mission plan includes provision that the radio-navigation will not be conducted for the solar impact parameters smaller than $2.5R_\odot$. The most of the important navigational, instrumental and experimental information will be stored on-board until the time of clear communication link with the NASA Deep Space Network. This is when the mission will step-up to its full potential by utilizing its optical communication capabilities from the extreme solar environment to enable high precision navigation of the spacecraft. 

\subsubsection{Earth's Orbit Velocity}
 
The knowledge of the Earth's orbital velocity, $v_E$, puts requirements on the correction for the stellar aberration.  The error associated with the accuracy of knowledge of the Earth's motion is given as:
{}
\begin{equation}
\Delta v_E \simeq c\Delta\gamma \,\frac{\mu_\odot}{c^2}\frac{r_{3}^2}{p_1p_2(p_2+p_1)}= 2.15~{\rm cm/s} \,\Big(\frac{\Delta\gamma}{2\times 10^{-8}}\Big)\Big(\frac{r_3}{{\rm AU}}\Big)^2\Big(\frac{R_\odot}{p_1}\Big)^3\,\Big[\frac{1}{(1+p_2/p_1)p_2/p_1}\Big].
\label{eq:vel}
\end{equation}
Therefore, the experiment will require a knowledge of the earth's barycentric velocity of 2.2 cm/s, resulting in an angular error of 0.03 prad. Note that a similar uncertainty is tolerable for the ISS's orbit about the Earth.

\subsection{Astrophysical Errors}

Physical phenomena of an astrophysical origin that are external to the LATOR triangle, but do not affect the mission navigation accuracy, are treated as the sources of astrophysical errors.  These errors would  nominally influence both the mission planning and observing scenario, they would be due to non-stationary behavior of the gravity field in the solar system (gravity effects of planets and  asteroids),  unmodeled motion of the fiducial stars, optical properties of the Sun and solar corona near the limb, the properties of the solar surface and etc. We will discuss these sources in some detail. 

\subsubsection{Knowledge of Solar Interior}

Laser ranging between the ISS and  spacecraft will be used to measure the orbits of the flight segments with $\sim$1 cm accuracy. This implies that the solar impact parameter should be measured to $8\times10^{-9}$, a scaling error for the measurement of parameter $\gamma$, but an insignificant error for the other measurements. 

Along with the impact parameter, other solar parameters such as its mass, angular momentum and quadrapole moment must be also known (or will be solved for directly from the data). 
The LATOR instrument may actually be used to gain additional knowledge on the Sun by observing its surface with a Doppler imager. This information may than be used to study the propagation of the sound waves through the solar interior. The resulted data may be used to bootstrap the gravitational solution for the solar oblateness and the higher spherical multipoles of the solar interior.  The instrumental implication of this possibility are currently being investigated and, if feasible, it may be included for the mission proposal. 

\subsubsection{Solar System Gravity}
Since the solar system is not static and the spacecraft are in the orbits
around the Sun, many large solar system bodies, such as the Sun itself, planets, asteroids, and even the galaxy, would have a significant effect on the measurement of $\gamma$ at the eighth decimal place.  Fortunately the ephemerides for the solar system objects are known to sufficient accuracy and the motion of the solar system about the galactic center is sufficiently smooth during the 92 min of each observing session.  Earth orbit crossing asteroids may cause a significant disturbance if they come within $\sim$10,000 km of one of the arms of the triangle. The relativity measurement may either have to be delayed or conducted with a slightly higher sampling rate if one of these are nearby. 

The change in the first order relativistic time delay due to other bodies  in the solar system has to be known to $\sim 10^{-9}$ of the effect from the Sun. The final observational model would have to account for the effects due to all the major solar system bodies. A similar theoretical and algorithmic work is currently being conducted for both SIM and Gaia astrometric missions and may well be used for this mission \citep{Turyshev02,Klioner05}. 

\subsection {Instrument Errors}

In our design considerations we address two types of instrumental errors, namely the offset and scale errors. Thus, in some cases, when a measured value has a systematic offset of a few pm, there may be instrumental errors that lead to further offset errors.  There are many sources of offset (additive)  errors caused by imperfect optics or imperfectly aligned optics at a pm level; there also many sources for scale errors. We take a comfort in the fact that, for the space-based stellar interferometry, we have an ongoing technology program at JPL;  not only has this program already demonstrated metrology accurate to a sub-pm level, but is also has identified a number of the error sources and developed methods to either eliminate them or to minimize their effect at the required level.

The second type of error is a scale error. For instance, in order to measure $\gamma$ in a single measurement with accuracy of two parts in  $10^{8}$ the laser frequency also must be stable to at least to $10^{-8}$ long term; lower accuracy would result in a scale error. The measurement strategy adopted for LATOR would require the laser stability to only $\sim$1\% to achieve accuracy the needed to measure the second order gravity effect. Absolute laser frequency must be known to $10^{-9}$ in order for the scaling error to be negligible. Similarly robust solutions were developed to address the effects of other known sources of scale errors. 

There is a considerable effort currently underway at JPL to evaluate a number of potential errors sources for the LATOR mission, to understand their properties and to establish methods to mitigate their contributions. (A careful strategy is needed to isolate the instrumental effects of the second order of smallness; however, our experience with SIM \citep{mct, Turyshev03,MilmanTuryshev03} is critical in helping us to properly capture its contribution in the instrument models.)  The work is ongoing, this is why the discussion below serves for illustration purposes only. We intend to publish the corresponding analysis and simulations in the subsequent publications.

The final error would have contributions from multiple measurements of the light triangle's four attributes (to enable the redundancy) taken by range and interferometer observations at a series of times. The corresponding errors will be combined with those from orbital position and velocity coordinate uncertainty.  These issues are currently being investigated in the mission covariance analysis; the detailed results of this analysis will be reported elsewhere. However, our current understanding of the  expected mission and instrumental accuracies suggests that LATOR will offer a very significant improvement compare to any other available techniques. This conclusion serves as the strongest experimental motivation to conduct the LATOR experiment.  

\section{Discussion}
\label{sec:conc}

The LATOR mission aims to carry out a test of the curvature of the solar system's gravity  field with an accuracy better than 1 part in 10$^{9}$. In spite of previous space missions exploiting radio waves for tracking the spacecraft, this mission manifests an actual breakthrough in the relativistic gravity experiments as it allow one to take full advantage of the optical techniques that recently became available.  
The LATOR experiment benefits from a number of advantages over techniques that use radio waves to study the light propagation in the solar vicinity.  The use of monochromatic light enables the observation of the spacecraft almost at the limb of the Sun, as seen from the ISS.  The use of narrowband filters, coronagraph optics, and heterodyne detection will suppress background light to a level where the solar background is no longer the dominant noise source.  The short wavelength allows much more efficient links with smaller apertures, thereby eliminating the need for a deployable antenna.  Advances in optical communications technology allow low bandwidth telecommunications with the LATOR spacecraft without having to deploy high gain radio antennae needed to communicate through the solar corona.  Finally, the use of the ISS not only makes the test affordable, but also allows conducting the experiment above the Earth's atmosphere---the major source of astrometric noise for any ground based interferometer.  This fact justifies the placement of LATOR's interferometer node in space. 

The concept is technologically sound; the required technologies have been demonstrated as part of the international laser ranging activities and optical interferometry programs at JPL (i.e. Space Interferometry Mission (SIM) and Keck Interferometer developments. Accuracy of 5 pm was already demonstrated in our SIM-related studies.) The LATOR concept arose from several developments at NASA and JPL that initially enabled optical astrometry and metrology, and also led to developing expertize needed for the precision gravity experiments. Technology that has become available in the last several years such as low cost microspacecraft, medium power highly efficient solid state and fiber lasers, and the development of long range interferometric techniques make possible an unprecedented factor of 30,000 improvement in this test of general relativity. This mission is unique and is the natural next step in solar system gravity experiments that fully exploit modern technologies.

LATOR uses geometric redundancy of the optical truss to achieve a very precise determination of the interplanetary distances between the two micro-spacecraft and a beacon station on the ISS. The experiment takes advantage of the existing space-qualified optical technologies, leading to an outstanding performance in a reasonable mission development time. In addition, the issues of the extended structure vibrations on the ISS, interferometric fringe ambiguity, and signal acquisition on the solar backgrounds have all been analyzed, and do not compromise mission goals.  The ISS is the default location for the interferometer, however, ground- and free-flying versions have also been studied.  While offering programmatic benefits, these options differ in cost, reliability and performance. The  availability of the ISS (via European collaboration) makes presented concept realizable in the near future. A recent JPL Team X study \citep{teamx} confirmed the feasibility of LATOR as a NASA Medium Explorer (MIDEX) class mission; the current mission concept calls for a launch as early as 2014. 

\subsection{LATOR vs Other Gravity Experiments}

Tests of fundamental gravitational physics feature prominently among NASA and ESA goals, missions, and programs.  Prediction of possible deviation of PPN parameters from the general relativistic values provides a robust theoretical paradigm and constructive guidance for experiments that would push beyond the present empirical upper bound on  the PPN parameter $\gamma$  of $\gamma-1=(2.1\pm2.3)\times10^{-5}$ obtained by recent conjunction experiments with Cassini spacecraft \citep{cassini_ber}.\footnote{In addition, any experiment pushing the present upper bounds on $\beta$ (i.e. $\beta - 1 = (1.2 \pm 1.1) \times 10^{-4}$ from \citep{LLR_beta_2004,pescara05} will also be of interest.} Among the future missions that will study the nature of gravity, we discuss here the missions most relevant to LATOR science:

\begin{itemize}
\item 
Configuration similar to the geometry of the Cassini conjunction experiments may be utilized for the microwave ranging between the Earth and a lander on Mars. If the lander were to be equipped with a Cassini-class dual X- and Ka-band communication system, the measurement of the PPN parameter $\gamma$ is possible with accuracy of $\sim$1 part in 10$^6$. However, as oppose to any scenario involving accurate ranging out to the Martian vicinity, the LATOR operations will be conducted at $\sim1$ AU heliocentric distances (well within the asteroid belt) and, thus, will not be affected by the damaging effects of the asteroid belt \citep{Ken_asteroids97,Konopliv_etal_2005}.

\item An ambitious test of the Equivalence Principle -- one of the foundations of general relativity -- is proposed for the STEP (Space Test of Equivalence Principle) mission. The experiment will test the composition independence of gravitational acceleration for laboratory-sized bodies by searching for a violation of the EP with a fractional acceleration accuracy of $\Delta a/a\sim 10^{-18}$ \citep{step,step2}. STEP will be able to test very precisely for any non-metric, long range interactions in physical law, however the results of this mission will say nothing about the metric component of gravity itself. The LATOR mission is designed specifically to test the metric nature of the gravitational interaction.

\item 
The SORT (Solar Orbit Relativity Test) mission concept proposes to use laser pulses and a drag-free spacecraft aided with a precision clock orbiting around the Sun to precisely measure $\gamma$ and $J_2$ (solar quadrupole moment)  \citep{veillet93,veillet94,sort}. SORT would combine a time-delay experiment (via laser signals sent from the Earth and recorded by precise clocks on board two satellites orbiting the Sun) with a light-deflection experiment (interferometric measurement on Earth of the angle between the two light signals emitted from the satellites) \citep{Re1999,sort}. As such, SORT would attempt to measure parameter $\gamma$ with accuracy of 1 part in 10$^6$. In its basic configuration, the LATOR experiment will relay on the redundant geometry formed by the three flight segments (two spacecraft and the ISS) and will not depend on either ultra-stable clocks nor ground-based interferometry that is severely limited by the atmosphere \citep{Shao_Colavita_1992}.

\item 
The ESA's  BepiColombo mission will explore the planet Mercury with equipment allowing an extremely accurate tracking.  This mission will conduct relativity experiments including the study of Mercury's perihelion advance and the relativistic light propagation near the Sun. The BepiColombo mission will enable achievement of the following accuracies: $\sigma_\gamma\simeq 2\times10^{-6}$, $\sigma_\beta\simeq 2\times10^{-6}$ and $\sigma_{J_2}\simeq 2\times10^{-9}$ in measuring the main post-Newtonian parameters \citep{BepiColombo2002}. While a very impressive mission design, its expected accuracy is at least two orders of magnitude worse than that expected from LATOR.  The LATOR mission is a designated relativity mission and it is designed to test solar gravity with accuracy at the level of 1 part in a billion.  

\item 
We stress that the future optical interferometers in space
such as NASA's SIM (Space Interferometry Mission) and ESA's Gaia (formerly known as, Global Astrometric Interferometer for Astrophysics \citep{gaia1995}) would provide improvement in measurement of relativistic parameters  as a  by-product of their astrometric  program. Thus, SIM will be able to reach accuracy of $\sim 10^{-6}$ in measuring PPN parameter $\gamma$. Gaia may potentially reach the accuracy of $10^{-5}-6\times 10^{-7}$ in measuring the $\gamma$ \citep{gaia2003}. However, both of these missions will have rather large exclusion angles and will not be able to test gravity effects on light near the sun.

\item 
A mission concept aiming to reach comparable accuracies in the tests of relativistic gravity in the solar system had been studied in \citep{astrod02} (see also references therein), and \citep{astrod04}. 
The Astrodynamical Space Test of Relativity using Optical Devices (ASTROD) is an ambitious mission concept that utilizes three drag-free spacecraft — one near L1/L2 point, one with an inner solar orbit and one with an outer solar orbit, ranging coherently with one another using lasers to test relativistic gravity and to detect low frequency gravitational waves. The mission may improve the accuracy of determination of the PPN parameter $\gamma$ to $\sim 10^{-7}$ for mini-ASTROD and to $\sim 5\times 10^{-9}$ for a full-scale version \citep{astrod02}. Because of the technological and programmatic complexities, the launch of an ASTROD-like mission is not expected before 2025.
\end{itemize}

A clear advantage of the LATOR mission concept is its independence on both -- the drag-free spacecraft environment and ultra-precise phase-coherent laser transponding techniques. In fact, LATOR will utilize the photon-counting laser ranging methods and redundant optical truss provided by the long-baseline optical multi-chanelled interferometer on the ISS. The LATOR experiment is optimized for its primary science goal -- to measure gravitational deflection of light in the solar gravity to 1 part in 10$^9$ (or at the level of the effects of the second post-Newtonian order of light deflection resulting from gravity's intrinsic non-linearity). There is no major technological breakthroughs needed to satisfy the LATOR mission requirements. All the required technologies already exist and most are space-qualified as a part of our on-going interferometry program at JPL (SIM, TPF (Terrestrial Planet Finder), and Palomar Testbed and Keck Interferometers).

Concluding this section, we point out that the recent progress in relativistic gravity research resulted in a significant tightening of the existing bounds on the PPN parameters obtained at the first post-Newtonian level of accuracy. However, this improvement is not sufficient to lead to groundbreaking tests of fundamental physical laws addressed above. This is especially true, if the cosmological attractor discovered in \citep{Damour_Nordtvedt_93b,DPV02a,DPV02b} is more robust, time variation in the fine structure constant will be confirmed in other experiments and various general relativity extensions will demonstrate feasibility of these methods for cosmology and relativistic gravity. The LATOR mission is proposed to directly address the challenges discussed above. 

\subsection{Conclusions and Further Considerations}

Concluding, we would like to summarize the most significant results of our LATOR mission study. The most natural question is ``Why is LATOR potentially orders of magnitude more sensitive and less expensive?''

First of all, there is a significant advantage in using optical wavelengths as opposed to the microwaves, the present navigational standard. This is based on the fact that solar plasma effects decrease as $\lambda^2$ and, in the case of LATOR, we gain a factor of $10^{10}$ reduction in the solar plasma optical path fluctuations by simply moving from $\lambda=10$~cm to $\lambda=1 ~\mu$m. Another LATOR advantage is its independence of a drag-free technology. In addition, the use of a redundant optical truss offers an excellent alternative to an ultra-precise orbit determination. This feature also makes LATOR insensitive to spacecraft buffeting from solar wind and solar radiation pressure.

Furthermore, the use of existing technologies, laser components and spacecraft make this mission a low cost experiment. Thus, 1~W lasers with sufficient frequency stability and $>10$ years lifetime already developed for optical telecom and also are flight qualified for SIM. Additionally, small optical apertures $\sim$10--20~cm are sufficient and provide this experiment with a high signal-to-noise ratio. There also a significant advantage in using components with no motorized moving parts. This all makes LATOR an excellent candidate for the next flight experiment in fundamental physics. Table \ref{tab:summ_science} summarizes the science objectives for this mission. 

LATOR is envisaged as a partnership between NASA and ESA wherein both partners are essentially equal contributors, while focusing on different mission elements: NASA provides the deep space mission components and interferometer design, while building and servicing infrastructure on the ISS is an ESA contribution. The NASA focus is on mission management, system engineering, software management, integration (both of the payload and the mission), the launch vehicle for the deep space component, and operations. The European focus is on interferometer components, the initial payload integration, optical assemblies and testing of the optics in a realistic ISS environment. The proposed arrangement would provide clean interfaces between familiar mission elements.

This mission may become a 21st century version of the Michelson-Morley experiment in the search for a  cosmologically evolved scalar field in the solar system. As such, LATOR will lead to very robust advances in the tests of fundamental physics: it could discover a violation or extension of general relativity, and/or reveal the presence of an additional long range interaction in the physical law. With this mission testing theory to several orders of magnitude higher precision, finding a violation of general relativity or discovering a new long range interaction could be one of this era's primary steps forward in fundamental physics. There are no analogs to the LATOR experiment; it is unique and is a natural culmination of solar system gravity experiments.

\section*{\large Acknowledgments}
The work described here was carried out at the Jet Propulsion Laboratory, California Institute of Technology, under a contract with the National Aeronautics and Space Administration.


\end{document}